\documentclass{amsart}
\usepackage{amsmath}
\usepackage{amssymb}

\theoremstyle{plain}

\numberwithin{equation}{section}
\input{tcilatex}

\begin{document}
\title{Quantum Field Theory Without Divergence A}
\author{Shi-Hao Chen}
\address{Institute of Theoretical Physics, Northeast Normal University \\
Changchun 130024, China. ~~}
\email{shchen@nenu.edu.cn }
\date{23 Otc. 2002}
\dedicatory{}
\thanks{}

\begin{abstract}
On the basis a new conjecture, we present a new Lagrangian density and a new
quantization method for QED, construct coupling operators and mass
operators, derive scattering operators $S_{f}$ ~and $S_{w}$ \ which are
dependent on each other and supplement new Feynman rules. $S_{f}$ ~and $%
S_{w} $ together determine a Fenman integral. Hence all Feynman integrals
are convergent and it is unnecessary to introduce regularization and
counterterms. That the energy of the vacuum state is equal to zero is
naturally obtained. From this we can easily determine the cosmological
constant according to data of astronomical observation, and it is possible
to correct nonperturbational methods which depend on the energy of the
ground state in quantum field theory. On the same basis as the new QED, we
obtain naturally a new SU(2)XU(1) electroweak unified model whose $\mathcal{%
L=L}_{F}\mathcal{+L}_{W}$ , here $\mathcal{L}$ is left-right symmetric. Thus
the world is left-right symmetric in{\Large \ }principle, but the part
observed by us is asymmetric because $\mathcal{L}_{W}$ and $\mathcal{L}_{F}$
are all asymmetic. This model do not contain any unknown particle with a
massive mass. A conjecture that there is repulsion or gravitation between
the W-particles and the F-particles is presented. If the new interaction is
gravitation, W-matter is the candidate for dark matter. If the new
interaction is repulsion, W-matter is the origin of universe expansion.
\end{abstract}

\maketitle

\section{Quantization for Free Fields\protect\bigskip}

\subsection{Introduction}

There are the following five problems to satisfactorily solve in the
conventional quantum field theory (QFT).

1. The issue of the cosmological constant.

2. The problem of divergence of Feynman integrals with loop diagrams.

3. The problem of the origin of asymmetry in the electroweak\ unified theory.

4. The problem of triviality of $\varphi ^{4}-$theory.

5. The problems of dark matter and the origin of some cosmic phenomena.

In brief, we present a new conjecture and a new quantization method, on the
basis construct a self-consistent QFT without divergence, give a fully
method evaluating Feynman integrals (see the second and third papers) and
possible solutions to the five problems.

Divergence of Feynman integrals with loop diagrams seems to have been solved
by introducing the bare mass and the bare charge or the concepts equivalent
to them. But both bare mass and charge are divergent and unmeasured, thus
QFT is still not perfect. In order to overcome the shortcomings, people have
tried many methods. For example, G. Scharf attempted to solve the difficulty
by the causal approach$^{[1]}.$Feynman integrals with loop diagrams are not
divergent~in some supersymmetric theories. But the supersymmetry theory
lacks experiment foundations. In fact, there should be no divergence and all
physical quantities should be measurable in a self-consistent theory.

According to the given generalized electroweak unified models which are
left-right symmetric before symmetry spontaneously breaking, asymmetry is
caused by symmetry spontaneously breaking. In such models there must be many
unknown particles with massive masses. Such models are troublesome and
causes many new problems. Hence the origin of asymmetry in the electroweak\
unified theory should still be explored.

More than 90\% of matter in cosmos is not composed of the conventional
baryons. Thus the problem of dark matter comes into being. Asymmetry of
left-right and existence of dark matter imply that there is an unknown world.

It is difficult on the basis of the conventional QFT to solve the problems
above. In order to solve the problems it is necessary to present a new
conjecture. The new conjecture of the present theory is the conjecture.

1. A new symmetry and a new Lagrangian density.

\begin{description}
\item[Conjecture] Any particle can exist in two sorts of states ------$%
F-particle$ described by $\mathcal{L}_{F}$\ and $W-particle$ described by $%
\mathcal{L}_{W}$. $\mathcal{L=L}_{F}+\mathcal{L}_{W}$ , $\mathcal{L}_{F}$\
and $\mathcal{L}_{W}$ are independent of each other before quantization and
dependent on each other after quantization. That the particles described by\ 
$\mathcal{L}_{F}$\ (F-particles) and the particles described by $\mathcal{L}%
_{W}$\ (W-particles) are symmetric.
\end{description}

We explain the conjecture as follows.

That the F-particles and the W-particles are symmetric imply that every
particle in $\mathcal{L}_{F}$ is accordant with a particle in $\mathcal{L}%
_{W}$ and the properties of the two particles are the same, e.g., there are
two sorts of electrons, i.e., $F-$electron and $W-$electron. That $\mathcal{L%
}_{F}$ and $\mathcal{L}_{W}$ are independent of each other implies that
there is no coupling between the fields in $\mathcal{L}_{F}$ and the fields
in $\mathcal{L}_{W},$ hence the energy determined by $\mathcal{L}_{F}$ \ and
the energy determined by $\mathcal{L}_{W}$ \ are respectively conservational
and a real F-particle and a real W-particle cannot transform from one to
another. But after quantization, $\mathcal{L}_{F}$ and $\mathcal{L}_{W}$ \
will be dependent on each other and the two sorts of virtual particles can
transform from one to another. The relativistic theory is very perfect, and
existence of negative energies is its essential character. We think that
positive energies and negative energies depend on each other, not only are
negative energies not a difficulty, but have profound physical meanings.
Existence of antiparticles is only, in fact, a result of particle-inversion
symmetry, and do not reveal the essence of negative energies. In frame of
relativistic quantum mechanics, on the basis to reexplain the physical
meanings of negative-energy states we can illuminate necessity and
self-consistency of the conjecture in another paper.

We call the conjecture the F-W (fire-water) symmetry conjecture. We may also
call conjecture the L-R (left-right) symmetry conjecture since $\mathcal{L}%
_{F}$\ and $\mathcal{L}_{W}$ describe respectively the left-hand world
(matter world) and right-hand world (dark-matter world) and $\mathcal{L}%
_{F}\ +\mathcal{L}_{W}$ is left-right symmetry.

2. Transformation operators and a new method to quantize fields.

Because particles can exist in the two sorts of states, we can define
transformation operators which transform a F-particle into a W-particle or a
W-particle into a F-particle, and can quantize fields by the transformation
operators replacing creation and annihilation operators in the conventional
QFT. Thus it is necessary that $g_{f}$ and $m_{ef\text{ }}$ respectively
become operators $G_{F}$ \ and $M_{F}$ \ to be determined by $\mathcal{S}%
_{w} $ ,\ and $g_{w\text{ \ }}$and $m_{ew\text{ }}$ respectively become
operators $G_{W}$ \ and $M_{W}$ to be determined by $\mathcal{S}_{f}$, here $%
\mathcal{S}_{w}$ and $\mathcal{S}_{f}$ are the scattering operators
respectively determined by $\mathcal{L}_{W}$ and $\mathcal{L}_{F}$. $G_{F}$
\ and $M_{F}$ multiplied by field operators $\psi $ \ and $A_{\mu }$\ become
the coupling coefficient $g_{f}(p_{2},p_{1})$ \ and mass $m_{ef\text{ }}(p)$
determined by scattering amplitude $\langle W_{f}\mid S_{w}\mid W_{i}$ $%
\rangle ,$ and $G_{W}$ \ and $M_{W}$ multiplied by field operators $%
\underline{\psi }$ \ and $\underline{A}_{\mu }$ \ become $g_{w}(p_{2},p_{1})$
\ and $m_{ew\text{ }}(p) $ determined by scattering amplitude $\langle
F_{f}\mid S_{f}\mid F_{i}$ $\rangle $\ . Thus after quantization, $\mathcal{L%
}_{F}$ \ and $\mathcal{L}_{W}$ will be dependent on each other.

3. Two sorts of corrections.

In the conventional QED, there are two sorts of parameters, e.g., the
physical charge and the bare charge, and one of corrections originating $%
\mathcal{S}$ \ equivalent to $\mathcal{S}_{f}.$ In contrast with the given
QED, there is only one sort of parameters defined at so-called subtraction
point $q_{2},$ $q_{1}$ and $q\prime ,$ i.e., $%
g_{f}(q_{2},q_{1})=g_{w}(q_{2},q_{1})=g_{0}$ \ and $%
m_{ef}(q)=m_{ew}(q)=m_{e0},$ and there are two sorts of corrections
originating from $\mathcal{S}_{w}$ and $\mathcal{S}_{f}$ to scattering
amplitudes, $g_{0}$ \ and $m_{e0}.$Thus $\mathcal{L}_{F}$ and $\mathcal{L}%
_{W}$ together determine the loop-diagram corrections. When n-loop
corrections originating from $\mathcal{S}_{f}$ and $\mathcal{S}_{w}$ \ are
simultaneously considered, the integrands causing divergence in $\langle
F_{f}\mid S_{f}\mid F_{i}$ $\rangle $ or $\langle W_{f}\mid S_{w}\mid W_{i}$ 
$\rangle $ will cancel each other out, hence all Feynman integrals are
convergent. For example (see third chapter), the one-loop correction to mass
of a free F-electron is

\begin{eqnarray}
\delta m &=&m_{ef}^{\left( 1\right) }\left( q\right)
=m_{eff}^{(1)}(q)+m_{efw}^{(1)}(q)  \notag \\
&=&\frac{3m\alpha }{4i\pi ^{3}}\int d^{4}k\left\{ \frac{1}{(k^{2}+m^{2})^{2}}%
-\frac{1}{(k^{2}+m^{2})^{2}}\right\} =0,  \TCItag{1.1.1}
\end{eqnarray}%
where $m_{eff}^{(1)}$ originates from $\mathcal{S}_{f}$ ,\ $m_{efw}^{(1)}$
originates from $\mathcal{S}_{w}$, and the superscript (1) denotes 1-loop
correction$.$ Thus it is unnecessary to introduce counterterms and
regularization. We give a set of complete Feynman rules to evaluate Feynman
integrals by the new concepts (see the second and third papers).

It should be pointed that\ in the meaning of perturbation theory, for
initial states and final states with given momenta, e.g., the momemta at the
subtraction point, we can give the absolutely precise coupling coefficients
and masses, and from this give absolutely precise $\mathcal{L}_{F}^{(0)}$ , $%
\mathcal{L}_{W}^{(0)},$ $\ \mathcal{H}_{F}^{(0)},$ $\mathcal{H}_{W}^{(0)},$ $%
\mathcal{S}_{f}^{(0)}$ and $\mathcal{S}_{w}^{(0)}.$ But for arbitrary
initial states and final states, we cannot give the absolutely precise
coupling coefficients, masses, $\mathcal{L}_{F}$ , $\mathcal{L}_{W},$ $\ 
\mathcal{H}_{F},$ $\mathcal{H}_{W},$ $\mathcal{S}_{f}$ and $\mathcal{S}_{w}.$%
\ In fact the coupling coefficients and masses\ will be corrected by n-loop
diagrams.\ For arbitrary initial states and final states $\mathcal{S}%
_{f}^{(0)}$ and $\mathcal{S}_{w}^{(0)}$\ etc., are only approximate. Of
course, by $\mathcal{S}_{f}^{(0)}$ and $\mathcal{S}_{w}^{(0)}$ we can obtain
scattering amplitudes approximate to arbitrary n-loop diagrams.

4. $\langle 0\mid H\mid 0\rangle \equiv E_{0}=E_{0F}+E_{0W}=0$ is naturally
derived, thereby we can easily determine the cosmological constant according
to data of astronomical observation, and it is possible to correct
nonperturbational methods which depend on the energy of the ground state in
QFT.

5. Generalizing the present theory to the electroweak unified theory, we
will see a possible origin of symmetry breaking. Accoding to this model, the
world is symmetric on principle since $\mathcal{L=L}_{W}$ $+$ $\mathcal{L}%
_{F}$ is symmetric, but the world observed by us is asymmetric since $%
\mathcal{L}_{W}$ or $\mathcal{L}_{F}$ is asymmetric. In this model there is
no unknown particle with a massive mass (see the third paper).

6. Because there is no interaction between the two sorts of matter by a
given quantizable field, the only possibility is that there is repulsion or
gravitation between the two sorts of matter. If the new interaction is
gravitation, W-matter is the candidate for dark matter$^{[3]}$. \ If the new
interaction is repulsion, it is possible that the new interaction is the
reason for the expansion of cosmos. It is also possible that there is new
and more important relationship between the two sorts of matter.

By the conventional creation and annihilation operators in the conventional
QFT we can also obtain the similar results, provided we suppose $\mathcal{L}%
= $ $\mathcal{L}_{F}$ $+\mathcal{L}_{W}$ and that $g_{f}$ and $m_{f}$ are
determined by $\mathcal{S}_{w}$ $\ $and $g_{w}$ and $m_{w}$ are determined
by $\mathcal{S}_{f}$ (of course, in this case this conjecture is not
natural). It is also possible to obtain the same results, but that
F-particles possess positive energies and W-particles possess negative
energies, provided $\mathcal{L}_{F}$ and $\mathcal{L}_{W}$ \ are independent
of each other in classical meanings.

The present theory contains four parts. The first part takes QED as an
example to illuminate the method to reconstruct QFT, and give the solutions
to the issue of the cosmological constant and the problem of divergent
Feynman integrals in QED. The first part is the present paper. The second
part discuss the problems on dark matter$^{[3]}$. The third part discusses
the problem of the origin of asymmetry in the electroweak\ unified theory in
detail. The fourth part discusses the problem of triviality of $\varphi
^{4}- $theory.

The outline of this chapter is as follows. 2. Lagrangian density; 3.
Quantization for free fields; \ \ \ 4. Energies and charges of particles;\
5. Subsidiary condition; 6. The equations of motion; 7. The physical
meanings of that the energy of the vacuum state is equal to zero; 8. Summary.

The contents of the rest are as follows.

Section 2. Coupling Operators and Feynman Rules

\ \ \ 1. Introduction; 2.\ Interacting Lagrangian density, Hamiltonian
density and equations of motion; \ \ \ 3.\ Coupling operators and mass
operators; 4. Expansion of the Hamiltonian;\ 5. Scattering operators and
Feynman rules; 6. Summary.

Section 3. One-loop Correction and Supplementary Feynman Rules

\ \ \ 1. Introduction; 2. Two sorts of correction; 3. The first sort of
1-loop corrections; 4. The second sort of one-loop corrections and the total
one-loop corrections; 5. $n-$loop corrections for the coupling constants and
the masses;

Section 4. Generalize to the $SU(2)\times U(1)$ electroweak unified model
and interaction between W-matter and F-matter

\ \ \ 1. Generalize to the $SU(2)\times U(1)$ electroweak unified model; 2.
Interaction between W-matter and F-matter.

Section 5. Conclusions and prospects.

Caption for figures.

Appendix A; Appendix B; Appendix C.\ \ 

\subsection{\protect\bigskip Lagrangian density}

We suppose the Lagrangian density for the free Dirac fields and the Maxwell
fields to be

\begin{equation}
\mathcal{L}_{0}=\mathcal{L}_{F0}+\mathcal{L}_{W0},  \tag{1.2.1}
\end{equation}

\begin{equation}
\mathcal{L}_{F0}=-\overline{\psi }(x)(\gamma _{\mu }\partial _{\mu }+m)\psi
(x)-\frac{1}{2}\partial _{\mu }A_{\nu }\partial _{\mu }A_{\nu },  \tag{1.2.2}
\end{equation}

\begin{equation}
\mathcal{L}_{W0}=\overline{\underline{\psi }}(x)(\gamma _{\mu }\partial
_{\mu }+m)\underline{\psi }(x)-\frac{1}{2}\partial _{\mu }\underline{A}_{\nu
}\partial _{\mu }\underline{A}_{\nu },  \tag{1.2.3}
\end{equation}%
(1.2.2) and (1.2.3) imply the Lorentz gauge to be already fixed . The
difference between (1.2.1) and the corresponding Lagrangian density in the
convertional $QED$ is $\mathcal{L}_{W0}$ which describes motion of particles
existing in the other form. We call $\psi $, $A_{\mu ,}$ $\underline{\psi }$
and $\underline{A}_{\mu }$ the F-electron field, the F-photron field, the
W-electron field and the W-photron field, respectively. The signs of the
fermion parts in $\mathcal{L}_{F}$ and $\mathcal{L}_{W}$ are opposite, the
signs of the boson parts are the same. The difference comes from the
difference of property of fermion and boson. This difference is not
essential. The key of matter is that F-particles and W-particles\ are
symmetric. In fact, redefining field operators by transformation operators,
we can also have $\mathcal{L}_{F}$\ and $\mathcal{L}_{W}$\ to be symmetric.

The conjugate fields corresponding to the fields are respectively

\begin{equation}
\pi _{\psi }=\frac{\partial \mathcal{L}}{\partial \dot{\psi}}=i\psi ^{+},\,\
\ \ \ \ \ \pi _{\mu }=\frac{\partial \mathcal{L}}{\partial \dot{A}_{\mu }}=%
\dot{A}_{\mu },  \tag{1.2.4}
\end{equation}

\begin{equation}
\pi _{\underline{\psi }}=\frac{\partial \mathcal{L}}{\partial \underline{%
\dot{\psi}}}=-i\underline{\psi }^{+},\;\ \ \ \ \ \ \underline{\pi }_{\mu }=%
\frac{\partial \mathcal{L}}{\partial \underline{\dot{A}}_{\mu }}=\underline{%
\dot{A}}_{\mu }.  \tag{1.2.5}
\end{equation}%
From the Noether$^{\prime }$s theorem, we obtain the conservational
quantities

\begin{equation}
H_{0}=H_{F0}+H_{W0},  \tag{1.2.6}
\end{equation}

\begin{equation}
H_{F0}=\int d^{3}x[\psi ^{+}\overset{\wedge }{H}_{0}\cdot \psi +\frac{1}{2}%
\left( \overset{\cdot }{A}_{\mu }\cdot \overset{\cdot }{A}_{\mu }+\partial
_{j}A_{\nu }\cdot \partial _{j}A_{\nu }\right) ]\ ,  \tag{1.2.7}
\end{equation}

\begin{equation}
H_{W0}=-\int d^{3}x[(\underline{\psi }^{+}\overset{\wedge }{H}_{0}\cdot 
\underline{\psi }-\frac{1}{2}(\underline{\overset{.}{A}}_{\mu }\cdot 
\underline{\overset{.}{A}}_{\mu }+\partial _{j}\underline{A}_{\nu }\cdot
\partial _{j}\underline{A}_{\nu })],  \tag{1.2.8}
\end{equation}

\begin{equation}
Q=Q_{F}+Q_{W},\;\ \ \ \ Q_{F}=\int d^{3}x\psi _{0}^{+}\cdot \psi ,\ \ \ \
Q_{W}=\int d^{3}x\underline{\psi }_{0}^{+}\cdot \underline{\psi }_{0}. 
\tag{1.2.9}
\end{equation}%
where $\overset{\wedge }{H}_{0}=\gamma _{4}\left( \gamma _{j}\partial
_{j}+m\right) .$ The global gauge transformations corresponding to (1.2.9)
are%
\begin{equation}
\psi \rightarrow \psi ^{\prime }=e^{i\alpha }\psi ,\;\ \underline{\psi }%
\rightarrow \underline{\psi }^{\prime }=e^{-i\alpha }\underline{\psi }\ .\  
\tag{1.2.10}
\end{equation}%
The Euler-Lagrange equations of motion being derived from the Hamilton$^{,}$%
s variational principle are

\begin{equation}
i\frac{\partial }{\partial t}\psi =\overset{\wedge }{H}_{0}\psi ,\;\ \ \ \
\square A_{\mu }=0,  \tag{1.2.11}
\end{equation}

\begin{equation}
i\frac{\partial }{\partial t}\underline{\psi }=\overset{\wedge }{H}_{0}%
\underline{\psi },\;\ \ \ \ \ \ \ \square \underline{A}_{\mu }=0. 
\tag{1.2.12}
\end{equation}%
It is seen from (1.2.11)-(1.2.12) that although $\mathcal{L}_{F0}\neq 
\mathcal{L}_{W0},$ the equations satisfied by $\underline{\psi }$ and $%
\underline{A}_{\mu }$ are the same as those satisfied by $\psi $ and $A_{\mu
},$ respectively. This implies that for a relativistic physical system, only
equations of motion are insufficient for corrective description of all
properties of the system. A complete Lagrangian density is very necessary.

When $\psi $, etc., are regarded as the classical fields and

\begin{equation}
\partial _{\mu }A_{\mu }=\partial _{\mu }\underline{A}_{\mu }=0, 
\tag{1.2.13}
\end{equation}%
$\psi $ and $\underline{\psi }$ can be expanded in terms of the complete set
of plane-wave solutions

\begin{equation}
\frac{1}{\sqrt{V}}u_{\mathbf{p}s}e^{ipx},\qquad \frac{1}{\sqrt{V}}v_{\mathbf{%
p}s}e^{-ipx},\quad \text{ \hspace{0in}}s=1,2,  \tag{1.2.14}
\end{equation}%
where $px=\mathbf{px}-E_{\mathbf{p}}t,$ $E_{\mathbf{p}}=\sqrt{\mathbf{p}%
^{2}+m^{2}},$ and $A_{\mu }$ and $\underline{A}_{\mu }$ can be expanded in
terms of the complete set

\begin{equation}
\frac{1}{\sqrt{2\omega _{\mathbf{k}}V}}e_{\mathbf{k}\mu }^{\lambda }e^{\pm
ikx},  \tag{1.2.15}
\end{equation}%
where $kx=\mathbf{kx}-\omega _{\mathbf{k}}t,\omega _{\mathbf{k}}=\mid 
\mathbf{k}\mid $ , $\lambda =1,2,.$ To get a completeness relation, it is
necessary to form a guartet of orthonormal 4-vectors$^{\left[ 4\right] }.$

\begin{eqnarray*}
e_{k}^{1} &=&\left( \mathbf{\varepsilon }_{\mathbf{k}}^{1},0\right)
,\;e_{k}^{2}=\left( \mathbf{\varepsilon }_{\mathbf{k}}^{2},0\right)
,\;e_{k}^{3}=-\left[ k+\eta \left( k\eta \right) \right] \diagup k\eta
,\;\eta =\left( 0,0,0,i\right) , \\
e_{k}^{4} &=&i\eta ,\;\ \ \ \mathbf{\varepsilon }_{\mathbf{k}}^{1,2}\cdot 
\mathbf{k=0.}
\end{eqnarray*}%
Moreover, all four vectors are normalized to 1, i.e.,

\begin{equation*}
e_{\mathbf{k}}^\lambda e_{\mathbf{k}}^{\lambda ^{^{\prime }}}=\delta
_{\lambda \lambda ^{^{\prime }}},\text{ \qquad }\sum_{\lambda =1}^4e_{%
\mathbf{k}\mu }^\lambda e_{\mathbf{k}\nu }^\lambda =\delta _{\mu \nu .}
\end{equation*}

\subsection{Quantization for free fields}

\subsubsection{Field operators and equations of motion}

We now regard $\psi $ etc., as quantum fields. $\psi $, $A_{\mu ,}$ $%
\underline{\psi }$ \hspace{0in}and $\hspace{0in}\underline{A}_{\mu }$ as the
solutions of the equations of the quantum fields $\left( 1.2.11\right)
-\left( 1.2.12\right) $ can also be expanded in terms of the complete sets $%
\left( 1.2.14\right) $ and $\left( 1.2.15\right) ,$ respectively, only the
expanding coefficients are all operators. Thus we have

\begin{equation}
\psi _{0}\left( x\right) =\frac{1}{\sqrt{V}}\sum_{\mathbf{p}s}\left( 
\underline{I}_{\mathbf{p}}\eqslantless a_{\mathbf{p}s}(t)\mid u_{\mathbf{p}%
s}e^{i\mathbf{px}}+\mid b_{\mathbf{p}s}(t)\eqslantgtr \underline{I}_{(%
\mathbf{-p)}}v_{\mathbf{p}s}e^{-i\mathbf{px}}\right) ,  \tag{1.3.1}
\end{equation}

\begin{eqnarray}
A_{0\mu }\left( x\right) &=&\frac{1}{\sqrt{V}}\sum_{\mathbf{k}}\frac{1}{%
\sqrt{2\omega _{\mathbf{k}}}}  \notag \\
&&\cdot \sum_{\lambda =1}^{4}e_{\mathbf{k}\mu }^{\lambda }\left( \underline{j%
}_{\mathbf{k}}\eqslantless c_{\mathbf{k}\lambda }(t)\mid e^{i\mathbf{kx}%
}+\mid \overline{c}_{\mathbf{k}\lambda }(t)\eqslantgtr \underline{j}_{%
\mathbf{(-k)}}e^{-i\mathbf{kx}}\right) ,  \TCItag{1.3.2}
\end{eqnarray}%
\begin{equation}
\underline{\psi }_{0}\left( x\right) =\frac{1}{\sqrt{V}}\sum_{\mathbf{p}%
s}\left( \mid \underline{b}_{\mathbf{p}s}(t)\eqslantgtr I_{\mathbf{p}}u_{%
\mathbf{p}s}e^{i\mathbf{px}}+I_{(-\mathbf{p)}}\eqslantless \underline{a}_{%
\mathbf{p}s}(t)\mid v_{\mathbf{p}s}e^{-i\mathbf{px}}\right) ,  \tag{1.3.3}
\end{equation}

\begin{eqnarray}
\underline{A}_{0\mu }\left( x\right) &=&\frac{1}{\sqrt{V}}\sum_{\mathbf{k}}%
\frac{1}{\sqrt{2\omega _{\mathbf{k}}}}  \notag \\
&&\cdot \sum_{\lambda =1}^{4}e_{\mathbf{k}\mu }^{\lambda }\left( \mid 
\underline{\overline{c}}_{\mathbf{k}\lambda }(t)\eqslantgtr j_{(-\mathbf{k)}%
}e^{i\mathbf{kx}}+j_{\mathbf{k}}\eqslantless \underline{c}_{\mathbf{k}%
\lambda }(t)\mid e^{-i\mathbf{kx}}\right) ,  \TCItag{1.3.4}
\end{eqnarray}

\begin{eqnarray}
\pi _{0\psi } &=&i\psi _{0}^{+}\left( x\right)  \notag \\
&\equiv &\frac{i}{\sqrt{V}}\sum_{\mathbf{p}s}\left( \mid a_{\mathbf{p}%
s}(t)\eqslantgtr \underline{I}_{\mathbf{p}}^{+}u_{\mathbf{p}s}^{+}e^{-i%
\mathbf{px}}+\underline{I}_{(\mathbf{-p)}}^{+}\eqslantless b_{\mathbf{p}%
s}(t)\mid v_{\mathbf{p}s}^{+}e^{i\mathbf{px}}\right) ,  \TCItag{1.3.5}
\end{eqnarray}

\begin{eqnarray}
\pi _{0\mu } &=&\dot{A}_{0\mu }\left( x\right) \equiv \frac{-i}{\sqrt{V}}%
\sum_{\mathbf{k}}\sqrt{\frac{\omega _{\mathbf{k}}}{2}}  \notag \\
&&\cdot \sum_{\lambda =4}^{4}e_{\mathbf{k}\mu }^{\lambda }\left( \underline{j%
}_{\mathbf{k}}\eqslantless c_{\mathbf{k}\lambda }(t)\mid e^{i\mathbf{kx}%
}-\mid \overline{c}_{\mathbf{k}\lambda }(t)\eqslantgtr \underline{j}_{(-%
\mathbf{k)}}e^{-i\mathbf{kx}}\right) ,  \TCItag{1.3.6}
\end{eqnarray}%
\begin{equation}
\underline{\pi }_{0\psi }=-i\underline{\psi }_{0}^{+}(x)\equiv \frac{-i}{%
\sqrt{V}}\sum_{\mathbf{p}s}\left( I_{\mathbf{p}}^{+}\eqslantless \underline{b%
}_{\mathbf{p}s}(t)\mid u_{\mathbf{p}s}^{+}e^{-i\mathbf{px}}+\mid \underline{a%
}_{\mathbf{p}s}(t)\eqslantgtr I_{-\mathbf{p}}^{+}v_{\mathbf{p}s}^{+}e^{i%
\mathbf{px}}\right) ,  \tag{1.3.7}
\end{equation}

\begin{eqnarray}
\underline{\pi }_{0\mu } &=&\underline{\dot{A}}_{0\mu }\left( x\right)
\equiv \frac{-i}{\sqrt{V}}\sum_{\mathbf{k}}\sqrt{\frac{\omega _{\mathbf{k}}}{%
2}}  \notag \\
&&\cdot \sum_{\lambda =1}^{4}e_{\mathbf{k}\mu }^{\lambda }\left( \mid 
\underline{\overline{c}}_{\mathbf{k}\lambda }\eqslantgtr j_{(-\mathbf{k)}%
}e^{i\mathbf{kx}}-j_{\mathbf{k}}\eqslantless \underline{c}_{\mathbf{k}%
\lambda }\mid e^{-i\mathbf{kx}}\right) ,  \TCItag{1.3.8}
\end{eqnarray}

\begin{eqnarray}
&\mid &\underline{\overline{c}}_{\mathbf{k}\lambda }\eqslantgtr =\left\{ 
\begin{array}{c}
\mid \underline{c}_{\mathbf{k}\lambda }\eqslantgtr ,\lambda =1,2,3, \\ 
-\mid \underline{c}_{\mathbf{k}\lambda }\eqslantgtr ,\lambda =4%
\end{array}%
\right. ,  \TCItag{1.3.9a} \\
&\mid &\overline{c}_{\mathbf{k}\lambda }\eqslantgtr =\left\{ 
\begin{array}{c}
\mid c_{\mathbf{k}\lambda }\eqslantgtr ,\lambda =1,2,3, \\ 
-\mid c_{\mathbf{k}\lambda }\eqslantgtr ,\lambda =4%
\end{array}%
\right. ,  \TCItag{1.3.9b}
\end{eqnarray}%
\begin{eqnarray}
&\eqslantless &a_{\mathbf{p}s}(t)\mid =\eqslantless a_{\mathbf{p}s}\mid
e^{-i\omega _{\mathbf{p}}t},\;\ \ \ \ \ \mid b_{\mathbf{p}s}(t)\eqslantgtr
=\mid b_{\mathbf{p}s}\eqslantgtr e^{i\omega _{\mathbf{p}}t},  \notag \\
&\mid &a_{\mathbf{p}s}(t)\eqslantgtr =\mid a_{\mathbf{p}s}\eqslantgtr
e^{i\omega _{\mathbf{p}}t},\;\ \ \ \ \ \ \ \eqslantless b_{\mathbf{p}%
s}(t)\mid =\eqslantless b_{\mathbf{p}s}\mid e^{-i\omega _{\mathbf{p}}t}, 
\notag \\
&\eqslantless &c_{\mathbf{k}\lambda }(t)\mid =\eqslantless c_{\mathbf{k}%
\lambda }\mid e^{-i\omega _{\mathbf{k}}t},\;\ \ \ \ \mid \overline{c}_{%
\mathbf{k}\lambda }(t)\eqslantgtr =\mid \overline{c}_{\mathbf{k}\lambda
}\eqslantgtr e^{i\omega _{\mathbf{k}}t}  \notag \\
&\mid &\underline{b}_{\mathbf{p}s}(t)\eqslantgtr =\mid \underline{b}_{%
\mathbf{p}s}\eqslantgtr e^{-i\omega _{\mathbf{p}}t},\;\ \ \ \ \ \ \
\eqslantless \underline{a}_{\mathbf{p}s}(t)\mid =\eqslantless \underline{a}_{%
\mathbf{p}s}\mid e^{i\omega _{\mathbf{p}}t}  \notag \\
&\eqslantless &\underline{b}_{\mathbf{p}s}(t)\mid =\eqslantless \underline{b}%
_{\mathbf{p}s}\mid e^{i\omega _{\mathbf{p}}t},\;\ \ \ \ \ \mid \underline{a}%
_{\mathbf{p}s}(t)\eqslantgtr =\mid \underline{a}_{\mathbf{p}s}\eqslantgtr
e^{-i\omega _{\mathbf{p}}t},  \notag \\
&\mid &\overline{\underline{c}}_{\mathbf{k}\lambda }(t)\eqslantgtr =\mid 
\underline{\overline{c}}_{\mathbf{k}\lambda }\eqslantgtr e^{-i\omega _{%
\mathbf{k}}t},\;\ \ \ \ \eqslantless \underline{c}_{\mathbf{k}\lambda
}(t)\mid =\eqslantless \underline{c}_{\mathbf{k}\lambda }\mid e^{i\omega _{%
\mathbf{k}}t}.  \TCItag{1.3.10}
\end{eqnarray}%
We call such operators as $I_{\mathbf{p}}\eqslantless \underline{a}_{\mathbf{%
p}s}(t)\mid $ and $\mid \underline{c}_{\mathbf{k}\lambda }(t)\eqslantgtr j_{%
\mathbf{k}}$ transformation operators. Such an operator as $\eqslantless 
\underline{a}_{\mathbf{p}s}(t)\mid $ changes as time, and $\underline{I}_{%
\mathbf{p}}$ and $j_{\mathbf{k}}$ etc. do not change. In the Heisenberg
picture the evolution of a quantum field system as time is carried by the
field operators according to the equations of motion

\begin{equation}
\dot{F}=-i\left[ F,H_{0}\right] =-i\left[ F,H_{0F}\right] ,  \tag{1.3.11}
\end{equation}

\begin{equation}
\dot{W}=i\left[ W,H_{0}\right] =i\left[ W,H_{0W}\right] ,  \tag{1.3.12}
\end{equation}%
where $F=\psi _{0}(x),$ $\psi _{0}^{+}(x),$ $A_{0\mu }(x),$ $\pi _{0\mu
}(x), $ $\mid a_{\mathbf{p}s}(t)\eqslantgtr ,$ $\mid b_{\mathbf{p}%
s}(t)\eqslantgtr ,$ $\mid c_{\mathbf{k}\lambda }(t)\eqslantgtr ,$ $%
\eqslantless a_{\mathbf{p}s}(t)\mid ,$ $\eqslantless b_{\mathbf{p}s}(t)\mid $
and $\eqslantless c_{\mathbf{k}\lambda }(t)\mid ,$ $W=$ $\underline{\psi }%
_{0}(x),$ $\underline{\psi }_{0}^{+}(x),$ $\underline{A}_{0\mu }(x),$ $%
\underline{\pi }_{0\mu }(x), $ $\mid \underline{a}_{\mathbf{p}%
s}(t)\eqslantgtr ,$ $\mid \underline{b}_{\mathbf{p}s}(t)\eqslantgtr ,$ $\mid 
\underline{c}_{\mathbf{k}\lambda }(t)\eqslantgtr ,\eqslantless \underline{a}%
_{\mathbf{p}s}(t)\mid ,$ $\eqslantless \underline{b}_{\mathbf{p}s}(t)\mid $
and $\eqslantless \underline{c}_{\mathbf{k}\lambda }(t)\mid .$ The equation
(1.3.11) is well-known as the Heisenberg equation, while (1.3.12) is a new
equation of motion. We will see 
\begin{equation}
\left[ H_{0W},H_{0F}\right] =\left[ H_{0F},H_{0}\right] =\left[ H_{0W},H_{0}%
\right] =0,  \tag{1.3.13}
\end{equation}%
hence $H_{0F}$ and $H_{0W}$ are the constants of motion. Thus, from $%
(1.3.11)-\left( 1.3.13\right) $ we have 
\begin{equation}
F\left( t\right) =e^{iH_{0}t}F\left( 0\right)
e^{-iH_{0}t}=e^{iH_{F0}t}F\left( 0\right) e^{-iH_{F0}t},  \tag{1.3.14}
\end{equation}

\begin{equation}
W\left( t\right) =e^{-iH_{0}t}W\left( 0\right)
e^{iH_{0}t}=e^{-iH_{W0}t}W\left( 0\right) e^{iH_{W0}t},  \tag{1.3.15}
\end{equation}%
$\left( 1.3.11\right) $ $-\left( 1.3.12\right) $ are consistent with $\left(
1.2.11\right) -\left( 1.2.12\right) $, and $\left( 1.3.14\right) $ $-\left(
1.3.15\right) $ are consistent with $\left( 1.3.10\right) .$

\subsubsection{Properties and multiplication rules of the transformation
operators.}

1. A transformation operator as $I_{\mathbf{p}}\eqslantless \underline{a}_{%
\mathbf{p}s}\mid $ is regarded a whole, hence the order of its two parts
cannot be reversed, though $I_{\mathbf{p}}$ $\ $and $\eqslantless \underline{%
a}_{\mathbf{p}s}\mid $ can be be separated, say, $I_{\mathbf{p}}\eqslantless 
\underline{a}_{\mathbf{p}s}\mid $ and $\mid \overline{c}_{\mathbf{k}\lambda
}\eqslantgtr \underline{j}_{\left( -\mathbf{k}\right) }$ \ cannot be written
as $\eqslantless \underline{a}_{\mathbf{p}s}\mid I_{\mathbf{p}}$ and $%
\underline{j}_{\left( -\mathbf{k}\right) }\mid \overline{c}_{\mathbf{k}%
\lambda }\eqslantgtr $, respectively. When $I_{\mathbf{p}}$ $\ $and $%
\eqslantless \underline{a}_{\mathbf{p}s}\mid $ do not constitute a
transformation operator, they can exchange their order. Hence this property
should be regarded as a definition for transformation operators.

2. Commutators or anticommutators of such operators as $\eqslantless a_{%
\mathbf{p}s}\mid $ and $\mid a_{\mathbf{p}s}\eqslantgtr $ .

Such an operator in the form $\eqslantless a_{\mathbf{p}s}\mid $ is
equivalent to an annihilation operator $a_{\mathbf{p}s},$ and such an
operator in the form $\mid a_{\mathbf{p}s}\eqslantgtr $ is equivalent to an
creation operator $a_{\mathbf{p}s}^{+}$ in the conventional $QED$. Thus, let 
$\alpha =a,$ $b,$ $\underline{a}$ and $\underline{b}$, $\gamma =c$ and $%
\underline{c},$ we have

\begin{equation}
\{\eqslantless \alpha _{\mathbf{p}s}(t)\mid ,\mid \alpha _{\mathbf{p}%
^{^{\prime }}s^{^{\prime }}}(t)\eqslantgtr \}=\delta _{\alpha \alpha
^{\prime }}\delta _{\mathbf{pp}^{\prime }}\delta _{ss^{^{\prime }}} 
\tag{1.3.16}
\end{equation}

\begin{equation}
\lbrack \eqslantless \gamma _{\mathbf{k}\lambda }(t)\mid ,\mid \gamma _{%
\mathbf{k}\ ^{^{\prime }}\lambda ^{^{\prime }}}^{^{\prime }}(t)\eqslantgtr
]=\left\{ 
\begin{array}{c}
\delta _{\gamma \gamma ^{^{\prime }}}\delta _{\mathbf{kk}^{^{\prime
}}}\delta _{\lambda \lambda ^{^{\prime }}},\lambda =1,2,3, \\ 
-\delta _{\gamma \gamma ^{^{\prime }}}\delta _{\mathbf{kk}^{^{\prime
}}}\delta _{\lambda \lambda ^{^{\prime }}},\lambda =4%
\end{array}%
,\right.  \tag{1.3.17}
\end{equation}%
The other commutators or anticommutators are all zero.

3. States and inner products of states.

(1.3.16) and (1.3.17) are the same as the anticommutation relations and the
commutation relations of the conventional $QED,$ respectiveky. As the
conventional $QED$, from (1.3.16)-(1.3.17) we see free fermion states and
n-photon states to be%
\begin{equation}
\mid \alpha _{\mathbf{p}s}\rangle =\mid \alpha _{\mathbf{p}s}\eqslantgtr
\mid 0\rangle ,\;\ \ \ \ \ \ \ \langle \alpha _{\mathbf{p}s}\mid =\langle
0\mid \eqslantless \alpha _{\mathbf{p}s}\mid ,  \tag{1.3.18}
\end{equation}%
\begin{eqnarray}
&\mid &n_{\mathbf{k}\lambda }\rangle \equiv \frac{1}{\sqrt{n!}}(\mid c_{%
\mathbf{k}\lambda }\eqslantgtr )^{n}\mid 0\rangle ,  \notag \\
\langle n_{\mathbf{k}\lambda } &\mid &\equiv \langle 0\mid (\eqslantless c_{%
\mathbf{k}\lambda }\mid )^{n}\frac{1}{\sqrt{n!}},  \TCItag{1.3.19a} \\
&\mid &\underline{n}_{\mathbf{k}\lambda }\rangle \equiv \frac{1}{\sqrt{n!}}%
(\mid \underline{c}_{\mathbf{k}\lambda }\eqslantgtr )^{n}\mid 0\rangle , 
\notag \\
\langle \underline{n}_{\mathbf{k}\lambda } &\mid &\equiv \langle 0\mid
(\eqslantless \underline{c}_{\mathbf{k}\lambda }\mid )^{n}\frac{1}{\sqrt{n!}}%
.  \TCItag{1.3.19b}
\end{eqnarray}%
The other sort of representation is in appendix A. Considering
(1.3.16)-(1.3.17) and 
\begin{eqnarray}
&\eqslantless &\alpha _{\mathbf{p}s}\mid 0\rangle =\eqslantless \gamma _{%
\mathbf{k}\lambda }\mid 0\rangle =\langle 0\mid \alpha _{\mathbf{p}%
s}\eqslantgtr =\langle 0\mid \gamma _{\mathbf{k}\lambda }\eqslantgtr =0, 
\notag \\
\langle 0 &\mid &0\rangle =1,  \TCItag{1.3.20}
\end{eqnarray}%
we obtain the inner products of states to be 
\begin{equation}
\langle \alpha _{\mathbf{p}s}\mid \cdot \mid \alpha _{\mathbf{p}^{\prime
}s^{\prime }}^{\prime }\rangle =\langle \alpha _{\mathbf{p}s}\mid \alpha _{%
\mathbf{p}^{\prime }s^{\prime }}^{\prime }\rangle =\delta _{\alpha \alpha
^{\prime }}\delta _{\mathbf{pp}^{\prime }}\delta _{ss^{^{\prime }}}, 
\tag{1.3.21}
\end{equation}

\begin{equation}
\langle \gamma _{\mathbf{k}\lambda }\mid \cdot \mid \gamma _{\mathbf{k}%
^{\prime }\ \lambda ^{^{\prime }}}^{\prime }\rangle =\langle \gamma _{%
\mathbf{k}\lambda }\mid \gamma _{\mathbf{k}^{\prime }\ \lambda ^{^{\prime
}}}^{\prime }\rangle =\left\{ 
\begin{array}{c}
\delta _{\gamma \gamma ^{^{\prime }}}\delta _{\mathbf{kk}^{^{\prime
}}}\delta _{\lambda \lambda ^{^{\prime }}},\lambda =1,2,3, \\ 
-\delta _{\gamma \gamma ^{\prime }}\delta _{\mathbf{kk}^{^{\prime }}}\delta
_{\lambda \lambda ^{^{\prime }}},\lambda =4%
\end{array}%
\right.  \tag{1.3.22}
\end{equation}

\begin{equation}
\langle \beta _{\mathbf{p}s}\mid \cdot \mid \gamma _{\mathbf{k}\lambda
}\rangle =0.  \tag{1.3.23}
\end{equation}

4. Inner products of $I_{\mathbf{p}}$ , $J_{\mathbf{p}}$ ,$\underline{I}_{%
\mathbf{p}},$ and $\underline{J}_{\mathbf{p}}.$

The original form of transformation operators is the same as $\mid 
\underline{a}_{\mathbf{p}s}\rangle \eqslantless a_{\mathbf{p}s}\mid ,$ $\mid
b_{_{\mathbf{p}s}}\eqslantgtr \langle \underline{b}_{\mathbf{p}s}\mid $ or $%
\mid c_{\mathbf{k}\lambda }\rangle \eqslantless \underline{c}_{\mathbf{k}%
\lambda }\mid ,$ and their essential meanings are 
\begin{eqnarray}
&\mid &\underline{a}_{\mathbf{p}s}\rangle \eqslantless a_{\mathbf{p}s}\mid
\mid a_{\mathbf{p}s}\rangle =\mid \underline{a}_{\mathbf{p}s}\rangle
,\;\langle b_{\mathbf{p}s}\mid \mid b_{\mathbf{p}s}\eqslantgtr \langle 
\underline{b}_{\mathbf{p}s}\mid =\langle \underline{b}_{\mathbf{p}s}\mid 
\notag \\
&\mid &c_{\mathbf{k}\lambda }\rangle \eqslantless \underline{c}_{\mathbf{k}%
\lambda }\mid \mid \underline{c}_{\mathbf{k}\lambda }\rangle =\mid c_{%
\mathbf{k}\lambda }\rangle .  \TCItag{1.3.24}
\end{eqnarray}%
By such transformation operators we can construct a new QFT from which the
same results as the present papers can be derived with more complicated
discussion. In order to compare the new QFT with the conventional QFT, we
write the transformation operators as the following form.%
\begin{eqnarray}
&\mid &\underline{a}_{\mathbf{p}s}\rangle \eqslantless a_{\mathbf{p}%
s}(t)\mid =\mid \underline{a}_{\mathbf{p}s}\rangle \underline{I}_{\mathbf{p}%
}^{+}\cdot \underline{I}_{\mathbf{p}}\eqslantless a_{\mathbf{p}s}(t)\mid , 
\notag \\
&\mid &b_{_{\mathbf{p}s}}(t)\eqslantgtr \langle \underline{b}_{\mathbf{p}%
s}\mid =\mid b_{_{\mathbf{p}s}}(t)\eqslantgtr \underline{I}_{(-\mathbf{p)}%
}\cdot \underline{I}_{(-\mathbf{p)}}^{+}\langle \underline{b}_{\mathbf{p}%
s}\mid ,  \notag \\
&\mid &\underline{c}_{\mathbf{k}\lambda }\rangle \eqslantless c_{\mathbf{k}%
\lambda }(t)\mid =\mid \underline{c}_{\mathbf{k}\lambda }\rangle \underline{J%
}_{\mathbf{k}}^{+}\cdot \underline{J}_{\mathbf{k}}\eqslantless c_{\mathbf{k}%
\lambda }(t)\mid ,  \notag \\
&\mid &b_{\mathbf{p}s}\rangle \eqslantless \underline{b}_{\mathbf{p}%
s}(t)\mid =\mid b_{\mathbf{p}s}\rangle I_{\mathbf{p}}^{+}\cdot I_{\mathbf{p}%
}\eqslantless \underline{b}_{\mathbf{p}s}(t)\mid ,  \notag \\
&\mid &a_{\mathbf{p}s}\rangle \eqslantless \underline{a}_{\mathbf{p}%
s}(t)\mid =\mid a_{\mathbf{p}s}\rangle I_{(-\mathbf{p)}}^{+}\cdot I_{(-%
\mathbf{p)}}\eqslantless \underline{a}_{\mathbf{p}s}(t)\mid ,  \notag \\
&\mid &c_{\mathbf{k}\lambda }\rangle \eqslantless \underline{c}_{\mathbf{k}%
\lambda }(t)\mid =\mid c_{\mathbf{k}\lambda }\rangle J_{\mathbf{k}}^{+}\cdot
J_{\mathbf{k}}\eqslantless \underline{c}_{\mathbf{k}\lambda }(t)\mid . 
\TCItag{1.3.25}
\end{eqnarray}%
where $\underline{I}_{\mathbf{p}}^{+}\cdot \underline{I}_{\mathbf{p}}=%
\underline{J}_{\mathbf{k}}^{+}\cdot \underline{J}_{\mathbf{k}}=I_{\mathbf{p}%
}^{+}\cdot I_{\mathbf{p}}=J_{\mathbf{k}}^{+}\cdot J_{\mathbf{k}}=1.$ For the
cause of mathematics we introduce $\underline{I}_{\mathbf{p}}$ etc.. In
orter to easily deal with problems, now we divide such operators into two
parts. For example, we divide $\mid \underline{a}_{\mathbf{p}s}\rangle 
\underline{I}_{\mathbf{p}}^{+}\cdot \underline{I}_{\mathbf{p}}\eqslantless
a_{\mathbf{p}s}(t)\mid $ into $\mid \underline{a}_{\mathbf{p}s}\rangle 
\underline{I}_{\mathbf{p}}^{+}$ $\ $and $\underline{I}_{\mathbf{p}%
}\eqslantless a_{\mathbf{p}s}(t)\mid $, leave such a part as $\mid 
\underline{\alpha }_{\mathbf{p}s}\rangle \underline{I}_{\mathbf{p}}^{+}$ \
to belong a coupling operator or a mass operator (see below), and leave such
a part as $\underline{I}_{\mathbf{p}}\eqslantless \alpha _{\mathbf{p}s}\mid $
\ to belong a field operator and also call it a transformation operator.

Let $K=I_{\mathbf{p}},$ $I_{\mathbf{p}}^{+},$ $\underline{I}_{\mathbf{p}},$ $%
\underline{I}_{\mathbf{p}}^{+},$ $J_{\mathbf{k}}$ and $\underline{J}_{%
\mathbf{k}}.$ Their properties and the multiplication rules are as follows.

\begin{equation}
\ J_{-\mathbf{k}}=J_{\mathbf{k}}^{+},\;\ \ \ \ \underline{J}_{-\mathbf{k}}=%
\underline{J}_{\mathbf{k}}^{+}  \tag{1.3.26}
\end{equation}%
$I_{\mathbf{p}},$ $I_{\mathbf{p}}^{+},$ $\underline{I}_{\mathbf{p}}^{+},$ $%
\underline{I}_{\mathbf{p}},$ $J_{\mathbf{k}}$, $\underline{J}_{\mathbf{k}}$
\ can be regarded a base vector of $I_{\mathbf{p}}-space,$ $\cdot \cdot
\cdot $ a base vector of $\underline{J}_{\mathbf{k}}-space,$ respectively.
Their inner products are defined as 
\begin{equation}
(I_{\mathbf{p}^{\prime }},\;I_{\mathbf{p}})=I_{\mathbf{p}^{\prime
}}^{+}\cdot I_{\mathbf{p}}=I_{\mathbf{p}}\cdot I_{\mathbf{p}^{\prime
}}^{+}=\delta _{\mathbf{pp}^{\prime }}  \tag{1.3.27a}
\end{equation}%
Similarly, we have 
\begin{eqnarray}
\underline{I}_{\mathbf{p}^{\prime }}^{+}\cdot \underline{I}_{\mathbf{p}} &=&%
\underline{I}_{\mathbf{p}}\cdot \underline{I}_{\mathbf{p}^{\prime
}}^{+}=\delta _{\mathbf{pp}^{\prime }}\;\ \ \ \   \notag \\
\ \ \ J_{\mathbf{k}^{\prime }}^{+}\cdot J_{\mathbf{k}} &=&\underline{J}_{%
\mathbf{k}^{\prime }}^{+}\cdot \underline{J}_{\mathbf{k}}=\delta _{\mathbf{kk%
}^{\prime }},  \notag \\
&&The\,other\;inner\;products\;are\;all\;zero.  \TCItag{1.3.27b}
\end{eqnarray}

$5.$Multiplication rules and commutation or anticommutation relations for
the transformation operators.

When multipling a transformation operator containing a factor $K$ by other
operator containing a factor $K$, we define the product to be such an
operator obtained after achieving multiplication of the two $K^{\prime }s$ ,
i.e.,%
\begin{eqnarray}
AK_{A}\cdot K_{B}B &\equiv &A(K_{A}\cdot K_{B})B,  \TCItag{1.3.28a} \\
\ \ K_{B}B\cdot \ AK_{A} &\equiv &\pm A(K_{A}\cdot K_{B})B,  \TCItag{1.3.28b}
\\
K_{B}B\cdot K_{A}A &=&(K_{B}\cdot K_{A})BA,\;BK_{B}\cdot AK_{A}\
=BA(K_{B}\cdot K_{A}),  \TCItag{1.3.28c} \\
\ [A,K] &=&0.  \TCItag{1.3.28d}
\end{eqnarray}%
where $A$, $B=$ $\eqslantless a_{\mathbf{p}s}(t)\mid $ etc. or $\mid b_{%
\mathbf{p}s}\rangle $ \ etc., when bath $A$ and $B$ are fermion operators,
(1.3.28b) takes `-', and otherwise takes `+'. $K$ can be constructed by
states (see appendix B), but this is not necessary. It is obvious that only
transformation operators can not form a closed algebra. A product of two
F-transformation operators (two W-transformation operators) is a F-operators
which transform a F-state into other F-state (a W-operators which transform
a W-state into other W-state ). The F-transformation operators and the
F-operators form a closed algebra. The W-transformation operators and the
W-operators form a closed algebra. It can be seen from (1.3.27 ) that the
product of a F-transformation operator and \hspace{0in}a W-transformation
operator must be equal to zero.

When many operators containing $K$ multiply, associative law does no longer
hold water. Hence we should appoint the associative order of such the
operators. But the complicacy cannot appear in fact, since after Lagrangian
or Hamiltonian density is construced by completing the products of two field
operators and the products of field operators and coupling operators or mass
operators (see second paper), thus Lagrangian or Hamiltonian density does no
longer contain $K$ so that multiplication of many operators containing $K$
will not appear. In interaction parts of Lagrangian or Hamiltonian density
that every field operator directly multiplies by a coupling operator or a
mass operator will be defined.

When a transformation operator and a operator not containing $K$ multiply,
the multiplication rule is the same as usual.

It is obvious that an essential difference between transformation operators
and the creation or annihilation operators is the factor $K$ .

From (1.3.16)-(1.3.17), (1.3.27)-(1.3.28) and (1.3.1)-(1.3.8), we easily
derive the commutation \ or anticommutation relations of the transformation
operators and the field operators.%
\begin{eqnarray}
\{\underline{I}_{\mathbf{p}} &\eqslantless &a_{\mathbf{p}s}(t)\mid ,\mid a_{%
\mathbf{p}^{\prime }s^{\prime }}(t)\eqslantgtr \underline{I}_{\mathbf{p}%
^{\prime }}^{+}\}  \notag \\
&=&\{\underline{I}_{-\mathbf{p}}^{+}\eqslantless b_{\mathbf{p}s}(t)\mid
,\mid b_{\mathbf{p}^{\prime }s^{\prime }}(t)\eqslantgtr \underline{I}_{-%
\mathbf{p}^{\prime }}\}  \notag \\
&=&[\underline{J}_{\mathbf{k}}\eqslantless c_{\mathbf{k}\lambda }(t)\mid
,\mid \overline{c}_{\mathbf{k}^{\prime }\lambda ^{\prime }}(t)\eqslantgtr 
\underline{J}_{\mathbf{k}^{\prime }}^{+}]=0,  \TCItag{1.3.29}
\end{eqnarray}%
\begin{eqnarray}
\{I_{-\mathbf{p}}^{+} &\eqslantless &\underline{a}_{\mathbf{p}s}(t)\mid
,\mid \underline{a}_{\mathbf{p}^{\prime }s^{\prime }}(t)\eqslantgtr I_{-%
\mathbf{p}^{\prime }}\}  \notag \\
&=&\{I_{\mathbf{p}}\eqslantless \underline{b}_{\mathbf{p}s}(t)\mid ,\mid 
\underline{b}_{\mathbf{p}^{\prime }s^{\prime }}(t)\eqslantgtr I_{\mathbf{p}%
^{\prime }}^{+}  \notag \\
&=&[J_{\mathbf{k}}\eqslantless \underline{c}_{\mathbf{k}\lambda }(t)\mid
,\mid \underline{\overline{c}}_{\mathbf{k}^{\prime }\lambda ^{\prime
}}(t)\eqslantgtr J_{\mathbf{k}^{\prime }}^{+}]=0,  \TCItag{1.3.30}
\end{eqnarray}%
\begin{equation}
\{\psi _{\alpha }(\mathbf{x},t),\psi _{\beta }^{+}(\mathbf{y},t)\}=\{\psi
_{\alpha }(\mathbf{x},t),\psi _{\beta }(\mathbf{y},t)\}=\{\psi _{\alpha
}^{+}(\mathbf{x},t),\psi _{\beta }^{+}(\mathbf{y},t)\}=0,  \tag{1.3.31}
\end{equation}%
\begin{equation}
\lbrack A_{\mu }(\mathbf{x,t),}\pi _{\nu }(\mathbf{y,t)]}\mathbf{=}[A_{\mu }(%
\mathbf{x,t),}A_{\nu }(\mathbf{y,t)]}\mathbf{=}[\pi _{\mu }(\mathbf{x,t),}%
\pi _{\nu }(\mathbf{y,t)]=}0,  \tag{1.3.32}
\end{equation}%
\begin{equation}
\{\underline{\psi }_{\alpha }(\mathbf{x},t),\underline{\psi }_{\beta }^{+}(%
\mathbf{y},t)\}=\{\underline{\psi }_{\alpha }(\mathbf{x},t),\underline{\psi }%
_{\beta }(\mathbf{y},t)\}=\{\underline{\psi }_{\alpha }^{+}(\mathbf{x},t),%
\underline{\psi }_{\beta }^{+}(\mathbf{y},t)\}=0,  \tag{1.3.33}
\end{equation}%
\begin{equation}
\lbrack \underline{A}_{\mu }(\mathbf{x,t),}\underline{\pi }_{\nu }(\mathbf{%
y,t)]}\mathbf{=}[\underline{A}_{\mu }(\mathbf{x,t),}\underline{A}_{\nu }(%
\mathbf{y,t)]}\mathbf{=}[\underline{\pi }_{\mu }(\mathbf{x,t),}\underline{%
\pi }_{\nu }(\mathbf{y,t)]=}0.  \tag{1.3.34}
\end{equation}%
The others are all zero as well. The commutation or anticommutation
relations are different from those of the convertional QED.

The new QFT seem to be more complicated than the conventional QFT from the
rules above, in fact it is not true since regularization and counterterms
are no longer necessary in the new QFT.

\subsection{The energies and charges of particles}

From $\left( 1.2.7\right) -\left( 1.2.9\right) ,$ $(1.3.1)-(1.3.8)$ and $%
\left( 1.3.26\right) -\left( 1.3.28\right) $ we obtain

\begin{eqnarray}
H_{F0} &=&\sum_{\mathbf{p}s}E_{\mathbf{p}}\left( \mid a_{\mathbf{p}%
s}\eqslantgtr \eqslantless a_{\mathbf{p}s}\mid +\mid b_{\mathbf{p}%
s}\eqslantgtr \eqslantless b_{\mathbf{p}s}\mid \right)  \notag \\
+\sum_{\mathbf{k}}\omega _{\mathbf{k}}(\sum_{\lambda =1}^{3} &\mid &c_{%
\mathbf{k}\lambda }\eqslantgtr \eqslantless c_{\mathbf{k}\lambda }\mid -\mid
c_{\mathbf{k}4}\eqslantgtr \eqslantless c_{\mathbf{k}4}\mid ), 
\TCItag{1.4.1}
\end{eqnarray}

\begin{eqnarray}
H_{W0} &=&\sum_{\mathbf{p}s}E_{\mathbf{p}}\left( \mid \underline{b}_{\mathbf{%
p}s}\eqslantgtr \eqslantless \underline{b}_{\mathbf{p}s}\mid +\mid 
\underline{a}_{\mathbf{p}s}\eqslantgtr \eqslantless \underline{a}_{\mathbf{p}%
s}\mid \right)  \notag \\
+\sum_{\mathbf{k}}\omega _{\mathbf{k}}(\sum_{\lambda =1}^{3} &\mid &%
\underline{c}_{\mathbf{k}\lambda }\eqslantgtr \eqslantless \underline{c}_{%
\mathbf{k}\lambda }\mid -\mid \underline{c}_{\mathbf{k}4}\eqslantgtr
\eqslantless \underline{c}_{\mathbf{k}4}\mid ),  \TCItag{1.4.2}
\end{eqnarray}

\begin{equation}
Q_{F}=\sum_{\mathbf{p}s}\left( \mid a_{\mathbf{p}s}\eqslantgtr \eqslantless
a_{\mathbf{p}s}\mid -\mid b_{\mathbf{p}s}\eqslantgtr \eqslantless b_{\mathbf{%
p}s}\mid \right) ,  \tag{1.4.3}
\end{equation}

\begin{equation}
Q_{W}=\sum_{\mathbf{p}s}\left( -\mid \underline{b}_{\mathbf{p}s}\eqslantgtr
\eqslantless \underline{b}_{\mathbf{p}s}\mid +\mid \underline{a}_{\mathbf{p}%
s}\eqslantgtr \eqslantless \underline{a}_{\mathbf{p}s}\mid \right) . 
\tag{1.4.4}
\end{equation}%
From (1.4.1) and (1.4.2) we see that energies are positive-definite and
(1.3.11) and (1.3.12) are consistent with (1.3.10), respectively. It is
easily seen from $\left( 1.4.1\right) -\left( 1.4.4\right) $ that

\begin{equation}
\langle \underline{\sigma }\mid H_{0}\mid \underline{\sigma }\rangle
=\langle \sigma \mid H_{0}\mid \sigma \rangle ,  \tag{1.4.5}
\end{equation}

\begin{equation}
\langle \underline{\sigma }\mid Q\mid \underline{\sigma }\rangle =\langle
\sigma \mid Q\mid \sigma \rangle ,  \tag{1.4.6}
\end{equation}%
where $\sigma =a_{\mathbf{p}s,}$ $b_{\mathbf{p}s,}$ $c_{\mathbf{k}\lambda .}$

The Hamiltonian and charge operators can also be written as

\begin{equation}
H_{F0}=\int d^{3}x:[\psi ^{\prime +}\overset{\wedge }{H}_{0}\psi ^{\prime }+%
\frac{1}{2}\left( \overset{\cdot }{A^{\prime }}_{\mu }\overset{\cdot }{%
A^{\prime }}_{\mu }+\partial _{j}A_{\nu }^{^{\prime }}\partial _{j}A_{\nu
}^{^{\prime }}\right) ]\ :,  \tag{1.4.7}
\end{equation}

\begin{equation}
H_{W0}=-\int d^{3}x:[(\underline{\psi }^{^{\prime }+}\overset{\wedge }{H}_{0}%
\underline{\psi }^{^{\prime }}-\frac{1}{2}(\underline{\overset{.}{A}}_{\mu
}^{\prime }\underline{\overset{.}{A}}_{\mu }^{\prime }+\partial _{j}%
\underline{A}_{\nu }^{\prime }\partial _{j}\underline{A}_{\nu }^{\prime })]:,
\tag{1.4.8}
\end{equation}

\begin{equation}
Q_{F}=\int d^{3}x:\psi ^{\prime +}\psi ^{^{\prime }}:,\;\ \ \ \ Q_{F}=\int
d^{3}x:\underline{\psi }^{\prime +}\underline{\psi }^{^{\prime }}:, 
\tag{1.4.9}
\end{equation}%
where the double-dot notation : $\cdots $: is known as normal ordering. An
operator product is in normal ordered form if all operators as $\mid \alpha
\eqslantgtr $ stand to the left of all operators as $\eqslantless \alpha
\mid .$ In (1.4.7)-(1.4.9),

\begin{equation}
\psi _{0}^{\prime }\left( x\right) =\frac{1}{\sqrt{V}}\sum_{\mathbf{p}%
s}\left( \eqslantless a_{\mathbf{p}s}(t)\mid u_{\mathbf{p}s}e^{i\mathbf{px}%
}+\mid b_{\mathbf{p}s}(t)\eqslantgtr v_{\mathbf{p}s}e^{-i\mathbf{px}}\right)
,  \tag{1.4.10}
\end{equation}

\begin{equation}
A_{0\mu }^{\prime }\left( x\right) =\frac{1}{\sqrt{V}}\sum_{\mathbf{k}}\frac{%
1}{\sqrt{2\omega _{\mathbf{k}}}}\sum_{\lambda =1}^{4}e_{\mathbf{k}\mu
}^{\lambda }\left( \eqslantless c_{\mathbf{k}\lambda }(t)\mid e^{i\mathbf{kx}%
}+\mid \overline{c}_{\mathbf{k}\lambda }(t)\eqslantgtr e^{-i\mathbf{kx}%
}\right) ,  \tag{1.4.11}
\end{equation}

\begin{equation}
\underline{\psi ^{\prime }}_{0}\left( x\right) =\frac{1}{\sqrt{V}}\sum_{%
\mathbf{p}s}\left( \mid \underline{b}_{\mathbf{p}s}(t)\eqslantgtr u_{\mathbf{%
p}s}e^{i\mathbf{px}}+\eqslantless \underline{a}_{\mathbf{p}s}(t)\mid v_{%
\mathbf{p}s}e^{-i\mathbf{px}}\right) ,  \tag{1.4.12}
\end{equation}

\begin{equation}
\underline{A}_{\mu }^{\prime }\left( x\right) =\frac{1}{\sqrt{V}}\sum_{%
\mathbf{k}}\frac{1}{\sqrt{2\omega _{\mathbf{k}}}}\sum_{\lambda =1}^{4}e_{%
\mathbf{k}\mu }^{\lambda }\left( \mid \underline{\overline{c}}_{\mathbf{k}%
\lambda }(t)\eqslantgtr e^{i\mathbf{kx}}+\eqslantless \underline{c}_{\mathbf{%
k}\lambda }(t)\mid e^{-i\mathbf{kx}}\right) ,  \tag{1.4.13}
\end{equation}

\begin{equation}
\pi _{0\psi }^{\prime }=i\psi _{0\left( x\right) }^{^{\prime }+}=\frac{i}{%
\sqrt{V}}\sum_{\mathbf{p}s}\left( \mid a_{\mathbf{p}s}(t)\eqslantgtr u_{%
\mathbf{p}s}^{+}e^{-i\mathbf{px}}+\eqslantless b_{\mathbf{p}s}(t)\mid v_{%
\mathbf{p}s}^{+}e^{i\mathbf{px}}\right)  \tag{1.4.14}
\end{equation}

\begin{eqnarray}
\pi _{0\mu }^{\prime } &=&\dot{A}_{0\mu }^{\prime }\left( x\right)  \notag \\
&=&\frac{-i}{\sqrt{V}}\sum_{\mathbf{k}}\sqrt{\frac{\omega _{\mathbf{k}}}{2}}%
\sum_{\lambda =1}^{4}e_{\mathbf{k}\mu }^{\lambda }\left( \eqslantless c_{%
\mathbf{k}\lambda }(t)\mid e^{i\mathbf{kx}}-\mid \overline{c}_{\mathbf{k}%
\lambda }(t)\eqslantgtr e^{-i\mathbf{kx}}\right) ,  \TCItag{1.4.15}
\end{eqnarray}

\begin{equation}
\pi _{0\psi }^{\prime }=-i\underline{\psi }_{0}^{^{\prime }+}(x)=\frac{-i}{%
\sqrt{V}}\sum_{\mathbf{p}s}\left( \eqslantless \underline{b}_{\mathbf{p}%
s}(t)\mid u_{\mathbf{p}s}^{+}e^{-i\mathbf{px}}+\mid \underline{a}_{\mathbf{p}%
s}(t)\eqslantgtr v_{\mathbf{p}s}^{+}e^{i\mathbf{px}}\right) ,  \tag{1.4.16}
\end{equation}

\begin{equation}
\underline{\pi }_{0\mu }^{\prime }=\underline{\dot{A}}_{0\mu }^{\prime }=%
\frac{-i}{\sqrt{V}}\sum_{\mathbf{k}}\sqrt{\frac{\omega _{\mathbf{k}}}{2}}%
\sum_{\lambda =1}^{4}e_{\mathbf{k}\mu }^{\lambda }\left( \mid \underline{%
\overline{c}}_{\mathbf{k}\lambda }(t)\eqslantgtr e^{i\mathbf{kx}%
}-\eqslantless \underline{c}_{\mathbf{k}\lambda }(t)\mid e^{-i\mathbf{kx}%
}\right) ,  \tag{1.4.17}
\end{equation}%
where the operators $\mid a_{\mathbf{p}s}(t)\eqslantgtr $ etc., are the same
as (1.3.10). From (1.3.16)-(1.3.17) and (1.4.10)-(1.4.17) we have%
\begin{equation}
\{\psi _{\alpha }^{\prime }(\mathbf{x,}t),\psi _{\beta }^{\prime +}(\mathbf{%
y,}t)\}=\{\underline{\psi }_{\alpha }^{\prime }(\mathbf{x,}t),\underline{%
\psi }_{\beta }^{\prime +}(\mathbf{y,}t)\}=\delta (\mathbf{x}-\mathbf{y}%
)\delta _{\alpha \beta },  \tag{1.4.18}
\end{equation}%
\begin{equation}
\lbrack A_{\mu }^{\prime }(\mathbf{x,}t),\pi _{\nu }^{\prime }(\mathbf{y,}%
t)]=[\underline{A}_{\mu }^{\prime }(\mathbf{x,}t),\underline{\pi }_{\nu
}^{\prime }(\mathbf{y,}t)]=i\delta (\mathbf{x-y})\delta _{\mu \nu }. 
\tag{1.4.19}
\end{equation}%
All other anticommutators and commutators are zero. (1.4.18)-(1.4.19) are
the same as those in the convintional $QED$. In contrast with the
conventional QFT, (1.4.1)-(1.4.4) or (1.4.7)-(1.4.9) are the deductions from
the multiplication rules (1.3.28), and propagators can be deduce from
(1.3.16)-(1.3.17) or (1.4.18)-(1.4.19) (see (2.5.9)-(2.5.12)). Hence in
present theory, (1.4.1)-(1.4.4) or (1.4.7)-(1.4.9) (thereby $\langle 0\mid
H\mid 0\rangle =0)$ and (1.3.16)-(1.3.17) or (1.4.18)-(1.4.19) are
self-consistent. In the conventional QFT, both propagators and Hamiltonian
are deduced from (1.4.18)-(1.4.19), hence it is inevitable that $\langle
0\mid H\mid 0\rangle $ is divergent or $\langle 0\mid H\mid 0\rangle =0$ and
to introduce the definition of normal products which is equivalent to%
\begin{eqnarray*}
\{\psi _{\alpha }^{\prime }(\mathbf{x,}t),\psi _{\beta }^{\prime +}(\mathbf{%
y,}t)\} &=&\{\underline{\psi }_{\alpha }^{\prime }(\mathbf{x,}t),\underline{%
\psi }_{\beta }^{\prime +}(\mathbf{y,}t)\}=0, \\
\lbrack A_{\mu }^{\prime }(\mathbf{x,}t),\pi _{\nu }^{\prime }(\mathbf{y,}%
t)] &=&[\underline{A}_{\mu }^{\prime }(\mathbf{x,}t),\underline{\pi }_{\nu
}^{\prime }(\mathbf{y,}t)]=0.
\end{eqnarray*}%
Thus the conventional QFT is not self-consistent.  

\subsection{Subsidiary condition}

After the Maxwell field is quantized, the Lorentz condition $\left(
1.2.13\right) $ is no longer applicable. From $\left( 1.4.11\right) $ and $%
\left( 1.4.13\right) $ we have

\begin{equation}
\left( \partial _{\mu }A_{\mu }^{\prime }\right) ^{+}=\frac{i}{\sqrt{V}}%
\sum_{\mathbf{k}}\frac{\mid \mathbf{k\mid }}{\sqrt{2\omega _{\mathbf{k}}}}%
(\eqslantless c_{\mathbf{k}3}\mid -i\eqslantless c_{\mathbf{k}4}\mid
)e^{ikx},  \tag{1.5.1}
\end{equation}

\begin{equation}
\left( \partial _{\mu }\underline{A}_{\mu }^{\prime }\right) ^{-}=-\frac{i}{%
\sqrt{V}}\sum_{\mathbf{k}}\frac{\mid \mathbf{k\mid }}{\sqrt{2\omega _{%
\mathbf{k}}}}(\eqslantless \underline{c}_{\mathbf{k}3}\mid -i\eqslantless 
\underline{c}_{\mathbf{k}4}\mid )e^{-ikx},  \tag{1.5.2}
\end{equation}%
Thus we define the subsidiary condition to be

\begin{equation}
\left( \partial _{\mu }A_{\mu }^{\prime }\right) ^{+}\mid c_{p}\rangle =0, 
\tag{1.5.3}
\end{equation}

\begin{equation}
\left( \partial _{\mu }\underline{A}_{\mu }^{\prime }\right) ^{-}\mid 
\underline{c}_{p}\rangle =0.  \tag{1.5.4}
\end{equation}%
$\mid c_{p}\rangle $ and $\mid \underline{c}_{p}\rangle $ are known as
F-physics state ket and W-physics state ket, respectively. From $\left(
1.5.1\right) -\left( 1.5.4\right) $ we obtain

\begin{equation}
\mid c_{p}\rangle =\mid c_{T}\rangle \{1+\sum_{\mathbf{k}}f\left( \mathbf{k}%
\right) \mid c_{p\mathbf{k}}\rangle +\cdots +\sum_{\mathbf{k}_{1}\cdot \cdot
\cdot \mathbf{k}_{n}}f\left( \mathbf{k}_{1}\cdots \mathbf{k}_{n}\right) \mid
c_{p\mathbf{k}_{1}}\cdots c_{p\mathbf{k}_{n}}\rangle \},  \tag{1.5.5}
\end{equation}

\begin{eqnarray}
&\mid &\underline{c}_{p}\rangle =\mid \underline{c}_{T}\rangle \{1+\sum_{%
\mathbf{k}}\underline{f}\left( \mathbf{k}\right) \mid \underline{c}_{p%
\mathbf{k}}\rangle +\cdots   \notag \\
+\sum_{\mathbf{k}_{1}\cdot \cdot \cdot \mathbf{k}_{n}}\underline{f}\left( 
\mathbf{k}_{1}\cdots \mathbf{k}_{n}\right)  &\mid &\underline{c}_{p\mathbf{k}%
_{1}}\cdots \underline{c}_{p\mathbf{k}_{n}}\rangle \},  \TCItag{1.5.6}
\end{eqnarray}%
where $\mid c_{T}\rangle $ and $\mid \underline{c}_{T}\rangle $ are states
containing only transverse photrons, and

\begin{equation}
\mid c_{p\mathbf{k}}\rangle =\mid c_{\mathbf{k}3}\rangle +i\mid c_{\mathbf{k}%
4}\rangle ,  \tag{1.5.7}
\end{equation}

\begin{equation}
\mid \underline{c}_{p\mathbf{k}}\rangle =\mid \underline{c}_{\mathbf{k}%
3}\rangle +i\mid \underline{c}_{\mathbf{k}4}\rangle .  \tag{1.5.8}
\end{equation}%
From (1.4.1)-(1.4.2) and (1.5.5)-(1.5.8) we obtain

\begin{eqnarray}
\langle c_{p\mathbf{k}} &\mid &c_{p\mathbf{k}^{^{\prime }}}\rangle =\langle 
\underline{c}_{p\mathbf{k}}\mid \underline{c}_{p\mathbf{k}^{^{\prime
}}}\rangle  \notag \\
&=&\langle c_{p\mathbf{k}}\mid H_{0}\mid c_{p\mathbf{k}}\rangle =\langle 
\underline{c}_{p\mathbf{k}}\mid H_{0}\mid \underline{c}_{p\mathbf{k}}\rangle
=0,  \TCItag{1.5.9}
\end{eqnarray}

\begin{equation}
\langle c_{p}\mid c_{p^{^{\prime }}}\rangle =\langle c_{T}\mid
c_{T^{^{\prime }}}\rangle ,\text{ }\langle \underline{c}_{p}\mid \underline{c%
}_{p^{^{\prime }}}\rangle =\langle \underline{c}_{T}\mid \underline{c}%
_{T^{^{\prime }}}\rangle ,  \tag{1.5.10}
\end{equation}

\begin{eqnarray}
\langle c_{p} &\mid &H_{0}\mid c_{p}\rangle =\langle c_{T}\mid H_{0}\mid
c_{T}\rangle ,  \notag \\
\langle \underline{c}_{p} &\mid &H_{0}\mid \underline{c}_{p}\rangle =\langle 
\underline{c}_{T}\mid H_{0}\mid \underline{c}_{T}\rangle ,\hspace{0pt} 
\TCItag{1.5.11}
\end{eqnarray}

\subsection{The equations of motion}

From $\left( 1.3.1\right) $ -$\left( 1.3.4\right) ,$ $\left( 1.3.6\right)
,\left( 1.3.8\right) ,$ $\left( 1.4.1\right) $ and $\left( 1.4.2\right) $ ,
we have

\begin{equation}
i\frac{\partial \psi _{0}}{\partial t}=\left[ \psi _{0},H_{F0}\right] =%
\overset{\wedge }{H}_{0}\psi _{0},  \tag{1.6.1}
\end{equation}

\begin{equation}
i\frac{\partial \underline{\psi }_{0}}{\partial t}=-[\underline{\psi }%
_{0},H_{W0}]=\overset{\wedge }{H}_{0}\underline{\psi }_{0},  \tag{1.6.2}
\end{equation}

\begin{equation}
\frac{\partial A_{0\mu }}{\partial t}=-i\left[ A_{0\mu },H_{F0}\right] ,%
\text{ \hspace{0in} \hspace{0in} \hspace{0in} \hspace{0in} \hspace{0in} 
\hspace{0in} \hspace{0in} \hspace{0in} \hspace{0in} \hspace{0in} \hspace{0in}
\hspace{0in} \hspace{0in} \hspace{0in} }\frac{\partial \dot{A}_{0\mu }}{%
\partial t}=-i[\dot{A}_{0\mu }H_{F0}]=\triangledown ^{2}A_{0\mu }, 
\tag{1.6.3}
\end{equation}

\begin{equation}
\frac{\partial \underline{A}_{0\mu }}{\partial t}=i[\underline{A}_{0\mu
},H_{W0}],\text{\hspace{0in} \hspace{0in} \hspace{0in} \hspace{0in} \hspace{%
0in} \hspace{0in} \hspace{0in} \hspace{0in} \hspace{0in} \hspace{0in} 
\hspace{0in} }\frac{\partial \underline{\dot{A}}_{0\mu }}{\partial t}=i[%
\underline{\dot{A}}_{0\mu },H_{W0}]=\triangledown ^{2}\underline{A}_{0\mu }.
\tag{1.6.4}
\end{equation}%
\begin{equation}
i\frac{\partial \psi _{0}^{\prime }}{\partial t}=\left[ \psi _{0}^{\prime
},H_{F0}\right] =\overset{\wedge }{H}_{0}\psi _{0}^{\prime },  \tag{1.6.5}
\end{equation}%
\begin{equation}
i\frac{\partial \underline{\psi }_{0}^{\prime }}{\partial t}=-[\underline{%
\psi }_{0}^{\prime },H_{W0}]=\overset{\wedge }{H}_{0}\underline{\psi }%
_{0}^{\prime },  \tag{1.6.6}
\end{equation}%
\begin{equation}
\frac{\partial A_{0\mu }^{\prime }}{\partial t}=-i\left[ A_{0\mu }^{\prime
},H_{F0}\right] ,\text{ \hspace{0in} \hspace{0in} \hspace{0in} \hspace{0in} 
\hspace{0in} \hspace{0in} \hspace{0in} \hspace{0in} \hspace{0in} \hspace{0in}
\hspace{0in} \hspace{0in} \hspace{0in} \hspace{0in} }\frac{\partial \dot{A}%
_{0\mu }^{\prime }}{\partial t}=-i[\dot{A}_{0\mu }^{\prime
}H_{F0}]=\triangledown ^{2}A_{0\mu }^{\prime },  \tag{1.6.7}
\end{equation}%
\begin{equation}
\frac{\partial \underline{A}_{0\mu }^{\prime }}{\partial t}=i[\underline{A}%
_{0\mu }^{\prime },H_{W0}],\text{\hspace{0in} \hspace{0in} \hspace{0in} 
\hspace{0in} \hspace{0in} \hspace{0in} \hspace{0in} \hspace{0in} \hspace{0in}
\hspace{0in} \hspace{0in} }\frac{\partial \underline{\dot{A}}_{0\mu
}^{\prime }}{\partial t}=i[\underline{\dot{A}}_{0\mu }^{\prime
},H_{W0}]=\triangledown ^{2}\underline{A}_{0\mu }^{\prime }.  \tag{1.6.8}
\end{equation}%
Since $\left[ H_{0,}H_{F0}\right] $ =$\left[ H_{0,}H_{W0}\right] =0$ , $%
H_{F0}$ and $H_{W0}$ are the constants of motion. Thus we obtain
(1.3.11)-(1.3.12) and

\begin{equation}
\psi _{0}^{\prime }\left( \mathbf{x,}t\right) =e^{iH_{F0}t}\psi _{0}^{\prime
}\left( \mathbf{x,0}\right) e^{-iH_{F0}t},  \tag{1.6.9}
\end{equation}

\begin{equation}
\ \underline{\psi }_{0}^{\prime }\left( \mathbf{x,}t\right)
=e_{0}^{-iH_{W0}t}\underline{\psi }_{0}^{\prime }\left( \mathbf{x,0}\right)
e^{iH_{W0}t},  \tag{1.6.10}
\end{equation}

\begin{equation}
\ A_{0\mu }^{\prime }\left( \mathbf{x,}t\right) =e^{iH_{F0}t}A_{0\mu
}^{\prime }\left( \mathbf{x,0}\right) e^{-iH_{F0}t},  \tag{1.6.11}
\end{equation}

\begin{equation}
\underline{A}_{0\mu }^{\prime }\left( \mathbf{x,}t\right) =e^{-iH_{W0}t}%
\underline{A}_{0\mu }^{\prime }\left( \mathbf{x,0}\right) e^{iH_{W0}t}, 
\tag{1.6.12}
\end{equation}%
As seen the equations $\left( 1.6.1\right) -\left( 1.6.4\right) $ are
consistent with $\left( 1.2.11\right) -\left( 1.2.12\right) ,$ respectively.

\subsection{The physical meanings of that the energy of the vacuum state is
equal to zero}

From (1.2.6), (1.4.1) and (1.4.2) we obtain that the energy of the vacuum
state,

\begin{equation}
E_{0}=\langle 0\mid H_{0}\mid 0\rangle =0.  \tag{1.7.1}
\end{equation}%
It is easily seen that (1.7.1) is not relative to the definition for
multiplication of transformation operators. (1.7.1) holds water provided $%
H_{0}$ is composed of transformation operators. The result is in contrast
with the given QED. According to the given QED, before redefining $H_{0}$ as
normal-ordered products $E_{0}\neq 0.$ After redefining $H_{0}$ as
normal-ordered products, $E_{0}=0.$This definition is, in fact, equivalent
to demand 
\begin{equation*}
\{a_{\mathbf{p}s},a_{\mathbf{p}s}^{+}\}=\{b_{\mathbf{p}s},b_{\mathbf{p}%
s}^{+}\}=[c_{\mathbf{k}\lambda },c_{\mathbf{k}\lambda }^{+}]=0,
\end{equation*}%
in the conventional $QED$. But in fact these commutation relations are equal
to 1, and in other cases, e.g. in propagators, they must also be 1. Thus the
conventional $QFT$ is not consistent. In fact, this definition only
transfers the divergence difficulty of the energy of the ground state. We
may arbitrarily choose the zero point of energy in quantum field theory. \
But in the theory of gravitation, if $E_{0}\neq 0$, $E_{0}$ will have
gravitational effect. Hence we are not at liberty to redefine $E_{0}$ $=0$ .
Thus the knotty problem of the cosmological constant arises in the
conventional $QFT$ and the relativistic theory of gravitation$^{[5]}$. In
the present theory $E_{0}=0,$ hence the density of the energy of the vacuum
state $\rho _{vac}=0.$ Thus, it is seem that from the equation of
gravitation field 
\begin{equation*}
R_{\mu \nu }-\frac{1}{2}g_{\mu \nu }R-\lambda g_{\mu \nu }=-8\pi G\left(
T_{\mu \nu }-\rho _{vac}g_{\mu \nu }\right) ,
\end{equation*}%
and astronomical observation values we can easily determine the cosmological
constant $\lambda .$ But when to generalize\ the new QED to the standard
electroweak model, expectation valves of Higgs fields are not equal to zero.
Thus in order to solve the cosmological constant problem, we must consider
contribution of Higgs fields. Considering Higgs fields, in another form of
the same idea we can also obtain $E_{0}=E_{F0}+E_{W0}=0$ (see hep-th/0203230)%
$.$ Thus the cosmological constant $\lambda $ can be determined according to
astronomical observation values. We will discuss the cosmological constant
problem in another paper.

We second simply discuss the correction originating from $E_0=0$ to a
nonperturtational method in quantum field theory.

When one evaluates the energy of a system by a nonperturtational method,
e.g., a Hartree-type approximation$^{\left[ 6\right] },$ it is necessary to
subtract the zero-point energy $E_{0}^{\left[ 7\right] }.$ According to the
given quantum field theory $E_{0}\neq 0,$ while according to the present
theory $E_{0}=0,$ hence we will obtain different results in nature.

We will discuss the two knotty problems above in detail in other papers.

\subsection{Summary}

We have presented a new conjecture. According to the conjectures, a particle
can exist in two forms which are symmetric. From this we have presented a
new Lagrangian density and a new quantization method for QED. That the
energy of the vacuum state is equal to zero is naturally obtained. From this
the cosmological constant is easily determined by astronomical observation
values and it is possible to correct nonperturbational methods which depend
on the energy of the ground state in quantum field theory.

\section{\protect\bigskip Coupling Operators and Feynman Rules}

\subsection{Introduction}

\ In the preceding chapter, we have presented a new Lagrangian density $%
\mathcal{L=}$ $\mathcal{L}_{F}+\mathcal{L}_{W}$ , have defined the
transformation operators and have quantized free fields by the
transformation operators. Thus, in order to quantize interacting fields, it
is necessary to transform the coupling coefficient $g_{f}$ and the
electromagnetic mass $m_{ef\text{ }}$ in $\mathcal{L}_{F}$ respectively into
a coupling operator $G_{F}$ \ and a mass operator $M_{F}$,\ and to transform 
$g_{w\text{ \ }}$and $m_{ew\text{ }}$ in $\mathcal{L}_{W}$ respectively into
operators $G_{W}$ \ and $M_{W}$ . In the present chapter we will construct
the coupling operators and mass operators, derive scattering operators $%
\mathcal{S}_{w},$ $\mathcal{S}_{f}$ and Feynman rules. We will see that $%
G_{F}$ \ and $M_{F}$ \ are determined by $\mathcal{S}_{w}$ ,\ and $G_{W}$ \
and $M_{W}$ are determined by $\mathcal{S}_{f}$. $G_{F}$ \ and $M_{F}$
multiplied by field operators $\psi $ \ and $A_{\mu }$ \ become the coupling
coefficient $g_{f}(p_{2},p_{1})$ \ and the mass $m_{ef\text{ }}(p)$
determined by the scattering amplitude $\langle W_{f}\mid S_{w}\mid
W_{i}\rangle $\ , and $G_{W}$ \ and $M_{W}$ multiplied by field operators $%
\underline{\psi }$ \ and $\underline{A}_{\mu }$ \ become $g_{w}(p_{2},p_{1})$
\ and $m_{ew\text{ }}(p)$ determined by the scattering amplitude $\langle
F_{f}\mid S_{f}\mid F_{i}$ $\rangle $\ . It is seen that after quantization, 
$\mathcal{L}_{F}$ and $\mathcal{L}_{W}$ \ are dependent on each other.

We think that for a self-consistent theory, there should be only one sort of
physical parameters which are measurable and finite, and there is no other
sort of parameters which are unmeasurable and divergent as the bare mass and
the bare charge in the convertional QED. The mass $m_{e0}$ and the coupling
constant $g_{0}$ defined at the so-called subtraction point are the
parameters of the sole sort. From this we can determine $\mathcal{L}%
_{F}^{(0)}$ and $\mathcal{L}_{W}^{(0)},$ thereby determine scattering
operators $S_{f}^{(0)}$ and $S_{w}^{(0)}$. $S_{f}^{(0)}$ and $S_{w}^{(0)}$
determine together the $one-loop$ corrections to the coupling constants and
the masses, thereby determine the $one-loop$ corrections of scattering
amplitudes as well. Thus we can derive $S_{f}^{(1-loop)}$, $%
S_{w}^{(1-loop)}, $ and $S_{f}^{(n-loop)}$ and $S_{w}^{(n-loop)}$ order by
order. We will see that the integrands causing divergence in a total
correction cancel each other out. Thus all Feynman integrals will be
convergent, and it is unnecessary to introduce regularization and
counterterms.\ 

\subsection{Interaction Lagrangian density, Hamiltonian density and
equations of motion}

From $(1.2.1)-(1.2.3)$ we can construct a Lagrangian density

\begin{equation}
\mathcal{L=L}_{F0}+\mathcal{L}_{W0}+\mathcal{L}_{F1}+\mathcal{L}_{W1,} 
\tag{2.2.1}
\end{equation}%
\begin{equation}
\mathcal{L}_{F0}=-\overline{\psi }(x)(\gamma _{\mu }\partial _{\mu }+m)\psi
(x)-\frac{1}{2}\partial _{\mu }A_{\nu }\partial _{\mu }A_{\nu },\text{ 
\hspace{0in} \hspace{0in} \hspace{0in}\hspace{0in} \hspace{0in} }\mathcal{L}%
_{F1}=ig\overline{\psi }\gamma _{\mu }A_{\mu }\psi ,  \tag{2.2.2}
\end{equation}

\begin{equation}
\mathcal{L}_{W0}=\underline{\overline{\psi }}(x)(\gamma _{\mu }\partial
_{\mu }+m)\underline{\psi }(x)-\frac{1}{2}\partial _{\mu }\underline{A}_{\nu
}\partial _{\mu }\underline{A}_{\nu },\text{ \hspace{0in} \hspace{0in} 
\hspace{0in} \hspace{0in}}\mathcal{L}_{W1}=ig\overline{\underline{\psi }}%
\gamma _{\mu }\underline{A}_{\mu }\underline{\psi },  \tag{2.2.3}
\end{equation}%
which is invariant under the following gauge transformation.

\begin{equation}
\psi \rightarrow \psi ^{\prime }=e^{i\theta }\psi ,\text{ \hspace{0in} 
\hspace{0in} \hspace{0in} \hspace{0in} \hspace{0in} \hspace{0in} \hspace{0in}
\hspace{0in}}\overline{\psi }\rightarrow \overline{\psi }^{\prime }=%
\overline{\psi }e^{-i\theta },  \tag{2.2.4}
\end{equation}

\begin{equation}
A_{\mu }\rightarrow A_{\mu }^{^{\prime }}=A_{\mu }+\partial _{\mu }\theta
\left( x\right) ,  \tag{2.2.5}
\end{equation}

\begin{equation}
\underline{\psi }\rightarrow \underline{\psi }^{\prime }=e^{-i\theta }%
\underline{\psi },\text{ \hspace{0in} \hspace{0in} \hspace{0in} \hspace{0in} 
\hspace{0in} \hspace{0in} \hspace{0in} \hspace{0in} \hspace{0in} }\underline{%
\overline{\psi }}\rightarrow \text{ }\underline{\overline{\psi }}^{\prime }=%
\text{ }\underline{\overline{\psi }}e^{i\theta },  \tag{2.2.6}
\end{equation}

\begin{equation}
\underline{A}_{\mu }\rightarrow \underline{A^{^{\prime }}}_{\mu }=\underline{%
A}_{\mu }+\partial _{\mu }\theta \left( x\right) ,  \tag{2.2.7}
\end{equation}%
where $\psi ,A_{\mu },\underline{\psi }$ and $\underline{A}_{\mu }$ are
regarded as the classical fields, and $g$ and $m$ are constants. After
quantization, as mentioned in the first chapter, field operators are
composed of transformation operators, hence $g$ and $m$ must become
respectively a coupling operator and a mass operator, i.e., in $\left(
2.2.2\right) ,$%
\begin{equation*}
m\rightarrow M_{F}^{^{\prime }}=m+\left( M_{F}^{^{\prime }}-m\right) \equiv
m+M_{F},\;\ \ \ g\rightarrow G_{F\text{ }};
\end{equation*}%
in $\left( 2.2.3\right) ,$%
\begin{equation*}
m\rightarrow M_{W}^{^{\prime }}=m+\left( M_{W}^{^{\prime }}-m\right) \equiv
m+M_{W};\;\ \ \ g\rightarrow G_{W};
\end{equation*}%
here all $M_{F},G_{F\text{ }},M_{W}$ and $G_{W}$ are operators. $G_{F,}$ $%
G_{W,}$ $M_{F\text{ }}$ and $M_{W}$ are determined on the basis of the gauge
and Lorentz invariance of the Lagrangian density $\left( 2.2.1\right) .$
Thus we have

\begin{equation}
\mathcal{L}_{FI}=i\overline{\psi }\gamma _{\mu }\cdot G_{F}\cdot A_{\mu
}\psi -\overline{\psi }\cdot M_{F}\cdot \psi ,  \tag{2.2.8}
\end{equation}

\begin{equation}
\mathcal{L}_{W1}=i\overline{\underline{\psi }}\gamma _{\mu }\cdot G_{W}\cdot 
\underline{A}_{\mu }\underline{\psi }+\underline{\overline{\psi }}\cdot
M_{W}\cdot \underline{\psi },  \tag{2.2.9}
\end{equation}%
In (2.2.8)-(2.2.9) every field operator directly multiplies by a coupling
operator or a mass operator. From $\left( 2.2.1\right) $, $\left(
2.2.8\right) $ and $\left( 2.2.9\right) $ we obtain the Hamiltonian to be

\begin{equation}
H=H_{F}+H_{W},  \tag{2.2.10}
\end{equation}

\begin{equation}
H_{F}=H_{F0}+H_{FI},\text{ \hspace{0in} \hspace{0in} \hspace{0in}\hspace{0in}
}H_{W}=H_{W0}+H_{WI},  \tag{2.2.11}
\end{equation}

\begin{equation*}
H_{F0}=\int d^{3}x[\psi ^{+}\gamma _{4}\left( \gamma _{j}\partial
_{j}+m\right) \cdot \psi +\frac{1}{2}\left( \overset{\cdot }{A}_{\mu }\cdot 
\overset{\cdot }{A}_{\mu }+\partial _{j}A_{\nu }\partial _{j}\cdot A_{\nu
}\right) ]\ ,
\end{equation*}

\begin{equation*}
H_{W0}=\int d^{3}x[-(\underline{\psi }^{+}\gamma _{4}(\gamma _{j}\partial
_{j}+m)\cdot \underline{\psi }+\frac{1}{2}(\underline{\dot{A}}_{\mu }\cdot 
\underline{\dot{A}}_{\mu }+\partial _{j}\underline{A}_{\mu }\partial
_{j}\cdot \underline{A}_{\mu })],
\end{equation*}

\begin{equation}
H_{FI}=\int d^{3}x\mathcal{H}_{FI}\equiv -\int d^{3}x\mathcal{L}_{FI}, 
\tag{2.2.12}
\end{equation}

\begin{equation}
H_{WI}=\int d^{3}x\mathcal{H}_{WI}\equiv -\int d^{3}x\mathcal{L}_{W1}. 
\tag{2.2.13}
\end{equation}

The Euler-Lagrange equations of motion can be derived from $\left(
2.2.1\right) ,$ $\left( 2.2.8\right) $ and $\left( 2.2.9\right) .$

\begin{equation}
i\frac{\partial }{\partial t}\psi =\gamma _{4}\left( \gamma _{j}\partial
_{j}-i\gamma _{\mu }G_{F}\cdot A_{\mu }+m+M_{F}\cdot \right) \psi , 
\tag{2.2.14}
\end{equation}

\begin{equation}
\square A_{\mu }=-i\overline{\psi }\gamma _{\mu }\cdot G_{F}\cdot \psi , 
\tag{2.2.15}
\end{equation}

\begin{equation}
i\frac{\partial }{\partial t}\underline{\psi }=\gamma _{4}\left( \gamma
_{j}\partial _{j}+i\gamma _{\mu }G_{W}\cdot \underline{A}_{\mu
}+m+M_{W}\cdot \right) \underline{\psi },  \tag{2.2.16}
\end{equation}

\begin{equation}
\square \underline{A}_{\mu }=-i\underline{\overline{\psi }}\gamma _{\mu
}\cdot G_{W}\cdot \underline{\psi },  \tag{2.2.17}
\end{equation}%
In the Heisenberg picture the evolution of a quantum field system as time is
carried by the field operators according to the equations of motion as well,

\begin{equation}
\dot{F}=-i\left[ F,H\right] =-i\left[ F,H_{F}\right] ,  \tag{2.2.18}
\end{equation}

\begin{equation}
\dot{W}=i\left[ W,H\right] =i\left[ W,H_{W}\right] ,  \tag{2.2.19}
\end{equation}%
where $F=\psi (x),$ $\psi ^{+}(x),$ $A_{\mu }(x),$ $\pi _{\mu }(x),$ $\mid
a_{\mathbf{p}s}(t)\eqslantgtr ,$ $\mid b_{\mathbf{p}s}(t)\eqslantgtr ,$ $%
\mid c_{\mathbf{k}\lambda }(t)\eqslantgtr ,$,$\eqslantless a_{\mathbf{p}%
s}(t)\mid ,$ $\eqslantless b_{\mathbf{p}s}(t)\mid $ and $\eqslantless c_{%
\mathbf{k}\lambda }(t)\mid ,$ and $W=$ $\underline{\psi }(x),$ $\underline{%
\psi }^{+}(x),$ $\underline{A}_{\mu }(x),$ $\underline{\pi }_{\mu }(x),$ $%
\mid \underline{a}_{\mathbf{p}s}(t)\eqslantgtr ,$ $\mid \underline{b}_{%
\mathbf{p}s}(t)\eqslantgtr ,$ $\mid \underline{c}_{\mathbf{k}\lambda
}(t)\eqslantgtr ,\eqslantless \underline{a}_{\mathbf{p}s}(t)\mid ,$ $%
\eqslantless \underline{b}_{\mathbf{p}s}(t)\mid $ and $\eqslantless 
\underline{c}_{\mathbf{k}\lambda }(t)\mid .$ It can be proven that $\left(
2.2.18\right) $ and $\left( 2.2.19\right) $ are consistent with $\left(
2.2.14\right) -\left( 2.2.17\right) ,$ respectively (see appendix C). The
equal-time anticommutation and commutation relations are the same as
(1.3.16)-(1.3.17) and (1.3.29)-(1.3.34), respectively. The equation (2.2.18)
is well-known as the Heisenberg equation, while (2.2.19) is a new equation
of motion. It is easily seen from $\left( 2.2.10\right) -\left(
2.2.13\right) $ that 
\begin{equation}
\left[ H_{F},H\right] =\left[ H_{W},H\right] =0,  \tag{2.2.20}
\end{equation}%
hence $H_{F}$ and $H_{W}$ are the constants of motion. Thus, from $\left(
2.2.18\right) $ and $\left( 2.2.19\right) $ we have

\begin{equation}
F\left( \mathbf{x,}t\right) =e^{iHt}F\left( \mathbf{x,}0\right)
e^{-iHt}=e^{iH_{F}t}F\left( \mathbf{x,}0\right) e^{-iH_{F}t},  \tag{2.2.21}
\end{equation}

\begin{equation}
W\left( \mathbf{x,}t\right) =e^{-iHt}W\left( \mathbf{x,}0\right)
e^{iHt}=e^{-iH_{W}t}W\left( \mathbf{x,}0\right) e^{iH_{W}t}.  \tag{2.2.22}
\end{equation}%
At a given time, say $t=0$, we can still expand the field operators in terms
of their Fourier components as $\left( 1.3.1\right) -\left( 1.3.8\right) $,
provided let $t=0$ in $\left( 1.3.1\right) -\left( 1.3.8\right) .$ Of
course, in this case we cannot interpret $\eqslantless \underline{a}_{%
\mathbf{p}s}(t)\mid ,$ $\mid b_{\mathbf{p}s}(t)\eqslantgtr ,$ etc., as
single-particle annihilation or creation operators. Passage from time $t=0$
to $t$ involves the replacements $\left( 1.3.1\right) -\left( 1.3.8\right) $
by $\left( 2.2.21\right) -\left( 2.2.22\right) .$ After completing the
products of two field operators and the products of field operators and
coupling operators or mass operators, Lagrangian and Hamiltonian densities
does no longer contain $K-\func{factor}$. Thus, that many operators
containing $K$ \ multiply will not appear.

\subsection{Coupling operators and mass operators}

The coupling operators $G_{F},$ $G_{W}$ \ and the mass operators $M_{F}$ and 
$M_{W}$ should satisfy the following demands.

1. After multipying by field operators, a coupling operator and a mass
operator should become a coupling coefficient and a mass operator,
respectively.

2. The coupling operators and the mass operators possess the same symmetry
as those satisfied by the Lagrangian density containing them, e.g., the
Lorentz invariance, the symmetry between F-particles and W-particles,
particle-antiparticle symmetry, etc..

In order to contruct the coupling operators, we first define the coupling
coefficients. Let $A_{gf}\left( \mid a_{\mathbf{p}_{1}s_{1}}c_{\mathbf{k}%
\lambda }\rangle \rightarrow \mid a_{\mathbf{p}_{2}s_{2}}\rangle \right) $
be a transition amplitude of an initial state $\mid a_{\mathbf{p}%
_{1}s_{1}}c_{\mathbf{k}\lambda }\rangle $ to a final state $\mid a_{\mathbf{p%
}_{2}s_{2}}\rangle ,$ here it is not necessary that $p_{1,}$ $p_{2}$ and $k$
satisfy the mass shell restrications. We define the F-coupling constant as 
\begin{eqnarray}
&&g_{f}(\mid a_{\mathbf{p}_{1}s_{1}}c_{\mathbf{k}\lambda }\rangle
\rightarrow \mid a_{\mathbf{p}_{2}s_{2}}\rangle )  \notag \\
&=&g_{f}\left( p_{2},p_{1}\right) =\frac{A_{gf}\left( \mid a_{\mathbf{p}%
_{1}s_{1}}c_{\mathbf{k}\lambda }\rangle \rightarrow \mid a_{\mathbf{p}%
_{2}s_{2}}\rangle \right) }{A_{gf}^{^{\prime }}\left( \mid a_{\mathbf{p}%
_{1}s_{1}}c_{\mathbf{k}\lambda }\rangle \rightarrow \mid a_{\mathbf{p}%
_{2}s_{2}}\rangle \right) }  \notag \\
&=&\frac{\langle a_{\mathbf{p}_{2}s_{2}}\mid S_{f}\mid a_{\mathbf{p}%
_{1}s_{1}}c_{\mathbf{k}\lambda }\rangle }{A_{gf}^{^{\prime }}\left( \mid a_{%
\mathbf{p}_{1}s_{1}}c_{\mathbf{k}\lambda }\rangle \rightarrow \mid a_{%
\mathbf{p}_{2}s_{2}}\rangle \right) },  \TCItag{2.3.1}
\end{eqnarray}%
where $S_{f}$ is the scattering operator for F-states. 
\begin{eqnarray}
&&A_{gf}^{\prime }\left( \mid a_{\mathbf{p}_{1}s_{1}}c_{\mathbf{k}\lambda
}\rangle \rightarrow \mid a_{\mathbf{p}_{2}s_{2}}\rangle \right)   \notag \\
&=&\langle a_{\mathbf{p}_{2}s_{2}}\mid \left( -\int d^{4}x:\overline{\psi }%
_{0}^{\prime }\gamma _{\mu }A_{\mu }^{\prime }\psi _{0}^{\prime }:\right)
\mid a_{\mathbf{p}_{1}s_{1}}c_{\mathbf{k}\lambda }\rangle   \notag \\
&=&-\left( 2\pi \right) ^{4}\delta ^{4}\left( p_{2}-p_{1}-k\right) \frac{1}{V%
}\overline{u}_{\mathbf{p}_{2}s_{2}}\gamma _{\mu }u_{\mathbf{p}_{1}s_{1}}%
\frac{1}{\sqrt{2\omega _{k}V}}e_{\mathbf{k}\mu }^{\lambda }.  \TCItag{2.3.2}
\end{eqnarray}%
where $k=p_{2}-p_{1},$ $\psi ^{\prime }$ and $A_{\mu }^{\prime }$ are the
same as (1.4.10)-(1.4.11) in form, but do not satisfy the equations for free
particle when $p_{2}$, $p_{1}$ and $k$ are not in mass shell. The act of $%
A_{gf}^{\prime }$ is only to eliminate the external-line factors of $A_{gf}.$
Considering the Lorentz invariance and that an outgoing $\mid b_{\left( -%
\mathbf{p}\right) }\rangle $ is equivalent to an ingoing $\mid a_{\mathbf{p}%
}\rangle $, and an outgoing $\mid c_{\left( -\mathbf{k}\right) }\rangle $ is
equivalent to an ingoing $\mid c_{\mathbf{k}}\rangle ,$ we have 
\begin{eqnarray}
&&g_{f}\left( p_{2},p_{1}\right)   \notag \\
&=&g_{f}\left( \mid c_{\mathbf{k}\lambda }\rangle \rightarrow \mid b_{(-%
\mathbf{p}_{1}s_{1})}a_{\mathbf{p}_{2}s_{2}}\rangle \right) =g_{f}\left(
\mid a_{\mathbf{p}_{1}s_{1}}b_{(-\mathbf{p}_{2}s_{2})}\rangle \mid c_{%
\mathbf{k}\lambda }\rangle \rightarrow \mid 0\rangle \right)   \notag \\
&=&g_{f}\left( \mid b_{(-\mathbf{p}_{2}s_{2})}c_{\mathbf{k}\lambda }\rangle
\rightarrow \mid b_{(-\mathbf{p}_{1}s_{1})}\rangle \right) =g_{f}\left( \mid
a_{\mathbf{p}_{1}s_{1}}\rangle \rightarrow \mid c_{(-\mathbf{k}\lambda )}a_{%
\mathbf{p}_{2}s_{2}}\rangle \right)   \notag \\
&=&g_{f}\left( \mid 0\rangle \rightarrow \mid c_{(-\mathbf{k}\lambda )}b_{(-%
\mathbf{p}_{1}s_{1})}a_{\mathbf{p}_{2}s_{2}}\rangle \right) =g_{f}\left(
\mid a_{\mathbf{p}_{1}s_{1}}b_{(-\mathbf{p}_{2}s_{2})}\rangle \rightarrow
\mid c_{(-\mathbf{k}\lambda )}\rangle \right)   \notag \\
&=&g_{f}\left( \mid b_{(-\mathbf{p}_{2}s_{2})}\rangle \rightarrow \mid c_{(-%
\mathbf{k}\lambda )}b_{(-\mathbf{p}_{1}s_{1})}\rangle \right) =g_{f}[\left(
p_{2}-p_{1}\right) ^{2}]=g_{f}\left( -\mu ^{2}\right) .  \TCItag{2.3.3}
\end{eqnarray}%
Similarly, we can define the W-coupling constant

\begin{eqnarray}
g_{w}\left( p_{2},p_{1}\right)  &=&g_{w}\left( \mid \underline{a}_{\mathbf{p}%
_{1}s_{1}}\underline{c}_{\mathbf{k}\lambda }\rangle \rightarrow \mid 
\underline{a}_{\mathbf{p}_{2}s_{2}}\rangle \right)   \notag \\
&=&\frac{A_{gw}\left( \mid \underline{a}_{\mathbf{p}_{1}s_{1}}\underline{c}_{%
\mathbf{k}\lambda }\rangle \rightarrow \mid \underline{a}_{\mathbf{p}%
_{2}s_{2}}\rangle \right) }{A_{gw}^{\prime }\left( \mid \underline{a}_{%
\mathbf{p}_{1}s_{1}}\underline{c}_{\mathbf{k}\lambda }\rangle \rightarrow
\mid \underline{a}_{\mathbf{p}_{2}s_{2}}\rangle \right) }  \notag \\
&=&\frac{\langle \underline{a}_{p_{2}s_{2}}\mid S_{w}\mid \underline{a}_{%
\mathbf{p}_{1}s_{1}}\underline{c}_{\mathbf{k}\lambda }\rangle }{%
A_{gw}^{\prime }\left( \mid \underline{a}_{\mathbf{p}_{2}s_{2}}\underline{c}%
_{\mathbf{k}\lambda }\rangle \rightarrow \mid \underline{a}_{\mathbf{p}%
_{2}s_{2}}\rangle \right) }  \TCItag{2.3.4}
\end{eqnarray}

\begin{eqnarray}
&&A_{gw}^{\prime }\left( \mid \underline{a}_{\mathbf{p}_{1}s_{1}}\underline{c%
}_{\mathbf{k}\lambda }\rangle \rightarrow \mid \underline{a}_{\mathbf{p}%
_{2}s_{2}}\rangle \right)   \notag \\
&=&\langle \underline{a}_{\mathbf{p}_{2}s_{2}}\mid \int d^{4}x:\underline{%
\overline{\psi }}_{0}^{\prime }\gamma _{\mu }\underline{A}_{\mu }^{\prime }%
\underline{\psi }_{0}^{\prime }:\mid \underline{a}_{\mathbf{p}_{1}s_{1}}%
\underline{c}_{\mathbf{k}\lambda }\rangle   \notag \\
&=&\left( 2\pi \right) ^{4}\delta ^{4}\left( p_{2}-p_{1}-k\right) \frac{1}{V}%
\overline{v}_{\mathbf{p}_{1}s_{1}}\gamma _{\mu }v_{\mathbf{p}_{2}s_{2}}\frac{%
1}{\sqrt{2\omega _{k}V}}e_{\mathbf{k}\mu }^{\lambda },  \TCItag{2.3.5}
\end{eqnarray}%
Considering the symmetry of the F-particles and the W-particles, we have

\begin{equation}
g_{f}\left( p_{2},p_{1}\right) =g_{w}\left( p_{2,}p_{1}\right) \equiv
g\left( p_{2},p_{1}\right) =g[\left( p_{2}-p_{1}\right) ^{2}]=g\left( -\mu
^{2}\right) .  \tag{2.3.6}
\end{equation}

What relation is there between $g_{f}\left( p_{2},p_{1}\right) $ and $G_{F}$%
? We analyse $G_{F}$ as follows.

Replacing (1.3.10) or (1.3.14)-(1.3.15) by (2.2.21)-(2.2.22), we obtain the
expansions of $\psi ,$ $A_{\mu },$ $\underline{\psi }$ and $\underline{A}%
_{\mu }$ which are the same as (1.3.1)-(1.3.4) in form. Substituting the
expansions into (2.2.12), (2.2.13), we obtain the expansions of the coupling
terms $\mathcal{H}_{FI}$ \ and $\mathcal{H}_{WI}$. Without loss of
generality, we take a term in (2.2.12)

\begin{equation}
\mid a_{\mathbf{p}_{2}s_{2}}(t)\eqslantgtr I_{\mathbf{p}_{2}}^{+}\cdot
G_{F}\mid \cdot I_{\mathbf{p}_{1}}\eqslantless a_{\mathbf{p}%
_{1}s_{1}}(t)\mid \underline{J}_{\mathbf{k}}\eqslantless c_{\mathbf{k}%
\lambda }(t)\mid   \tag{2.3.7}
\end{equation}%
as an example to illuminate the structure of $G_{F}.$The c-number part
relating to external lines in (2.3.7) is ignored. Recalling
(1.3.25)-(1.3.28),\ from the factor $I_{\mathbf{p}_{2}}^{+}G_{F}I_{\mathbf{p}%
_{1}}\underline{J}_{\mathbf{k}}$ we can see that the corresponding part in $%
G_{F}$ can be written as%
\begin{equation}
\underline{I}_{\mathbf{p}_{2}}\langle \underline{a}_{\mathbf{p}%
_{2}s_{2}}\mid G_{F}^{\prime }\mid \underline{a}_{\mathbf{p}_{1}s_{1}}%
\underline{c}_{\mathbf{k}\lambda }\rangle \underline{I}_{\mathbf{p}_{1}}^{+}%
\underline{J}_{\mathbf{k}}^{+}.  \tag{2.3.8}
\end{equation}%
Because $\langle \underline{a}_{\mathbf{p}_{2}s_{2}}\mid $ and $\mid 
\underline{a}_{\mathbf{p}_{1}s_{1}}\underline{c}_{\mathbf{k}\lambda }\rangle 
$ \ can be \ regarded as a final W-state and an initial\ W-state, $\langle 
\underline{a}_{\mathbf{p}_{2}s_{2}}\mid G_{F}^{\prime }\mid \underline{a}_{%
\mathbf{p}_{1}s_{1}}\underline{c}_{\mathbf{k}\lambda }\rangle $ must be
directly proportional to the scattering amplitude $A_{gw}\left( \mid 
\underline{a}_{\mathbf{p}_{1}s_{1}}\underline{c}_{\mathbf{k}\lambda }\rangle
\rightarrow \mid \underline{a}_{\mathbf{p}_{2}s_{2}}\rangle \right) .$
Comparing with (2.3.4), we define 
\begin{eqnarray}
\langle \underline{a}_{\mathbf{p}_{2}s_{2}} &\mid &G_{F}^{\prime }\mid 
\underline{a}_{\mathbf{p}_{1}s_{1}}\underline{c}_{\mathbf{k}\lambda }\rangle
\equiv g_{f}\left( p_{2,}p_{1}\right)   \notag \\
&\equiv &\frac{\langle \underline{a}_{\mathbf{p}_{2}s_{2}}\mid S_{w}\mid 
\underline{a}_{\mathbf{p}_{1}s_{1}}\underline{c}_{\mathbf{k}\lambda }\rangle 
}{A_{gw}^{\prime }\left( \mid \underline{a}_{\mathbf{p}_{2}s_{2}}\underline{c%
}_{\mathbf{k}\lambda }\rangle \rightarrow \mid \underline{a}_{\mathbf{p}%
_{2}s_{2}}\rangle \right) },\;  \TCItag{2.3.9}
\end{eqnarray}%
as a coupling coefficient. From (2.3.4) and (2.3.9) we can anew obtain
(2.3.6). This implies (2.3.9) is consistent with (2.3.6). Thus (2.3.8) can
be written as 
\begin{equation}
\underline{I}_{\mathbf{p}_{2}}g_{f}\left( p_{2,}\;p_{1}\right) \underline{I}%
_{\mathbf{p}_{1}}^{+}\underline{J}_{\mathbf{k}}^{+}.  \tag{2.3.9a}
\end{equation}%
It is possible that in $g_{f}\left( p_{2,}\;p_{1}\right) $ $p_{2}\neq m^{2}\;
$or $p_{1}^{2}\neq m^{2},$ i.e., $p_{0}$ \ is not solely determined by $%
\mathbf{p}$. We determine $p_{20}$ and $p_{10}$ according to the following
method. It is easily seen that%
\begin{eqnarray}
\langle a_{\mathbf{p}_{2}^{\prime }s_{2}} &\mid &(-i\dint d^{4}x\mathcal{H}%
_{F})\mid a_{\mathbf{p}_{1}^{\prime }s_{1}}c_{\mathbf{k}^{\prime }\lambda
}\rangle   \notag \\
&\varpropto &\langle a_{\mathbf{p}_{2}^{\prime }s_{2}}\mid \cdot \mid a_{%
\mathbf{p}_{2}s_{2}}(t)\eqslantgtr \langle \underline{a}_{\mathbf{p}%
_{2}s_{2}}\mid G_{F}^{\prime }\mid \underline{a}_{\mathbf{p}_{1}s_{1}}%
\underline{c}_{\mathbf{k}\lambda }\rangle   \notag \\
&\eqslantless &a_{\mathbf{p}_{1}s_{1}}(t)\mid \eqslantless c_{\mathbf{k}%
\lambda }(t)\mid \cdot \mid a_{\mathbf{p}_{1}^{\prime }s_{1}}c_{\mathbf{k}%
^{\prime }\lambda }\rangle   \notag \\
&=&\langle a_{\mathbf{p}_{2}^{\prime }s_{2}}\mid \cdot \mid a_{\mathbf{p}%
_{2}s_{2}}(t)\eqslantgtr g_{f}\left( p_{2,}p_{1}\right)   \notag \\
&\eqslantless &a_{\mathbf{p}_{1}s_{1}}(t)\mid \eqslantless c_{\mathbf{k}%
\lambda }(t)\mid \cdot \mid a_{\mathbf{p}_{1}^{\prime }s_{1}}\underline{c}_{%
\mathbf{k}^{\prime }\lambda }\rangle   \TCItag{2.3.9b}
\end{eqnarray}%
determines a scattering amplitude. Substituting (2.3.9b) into (2.3.1), we
can obtain a coupling coeffecient $g_{f}(p_{20}^{\prime },p_{10}^{\prime }).$
In the state $\mid a_{\mathbf{p}_{1}^{\prime }s_{1}}\rangle $ the relation
between $p_{10}^{\prime }$ and $\mathbf{p}_{1}^{\prime }$ is definite.
Similarly, the relation between $p_{20}^{\prime }$ and $\mathbf{p}%
_{2}^{\prime }$ and the relation between $k_{0}^{\prime }$ and $\mathbf{k}%
^{\prime }$ are also definite. It is obviously necessary that $\mathbf{p}%
_{2}^{\prime }=\mathbf{p}_{2},$ $\mathbf{p}_{1}^{\prime }=\mathbf{p}_{1}$
and $\mathbf{k}^{\prime }=\mathbf{k}.$ On the basis of the W-F-symmetry, we
demand%
\begin{equation}
p_{20}^{\prime }=p_{20},\;\ \ \ p_{10}^{\prime }=p_{10},\;\ \ \
k_{0}^{\prime }=k_{0}.  \tag{2.3.9c}
\end{equation}%
Thus we have $p_{2}^{\prime }=p_{2},\;p_{1}^{\prime }=p_{1}$,$\ k^{\prime }=k
$ and 
\begin{equation}
g_{f}(p_{20}^{\prime },p_{10}^{\prime })=g_{f}\left( p_{2,}\;p_{1}\right) . 
\tag{2.3.9d}
\end{equation}%
Even in interaction picture the momenta meeting at a vertex cannot also all
satisfy the relation $p^{2}=m^{2}.$ In the case $p_{20}$ and $p_{10}$ in $%
g_{f}\left( p_{2,}\;p_{1}\right) $ at a vertex should respectively be equal
to the momenta of the propagators joining the vertex. $g_{f}\left(
p_{2,}\;p_{1}\right) $ is determined according to the scattering amplitude
(2.3.9b) and is accurate. But we cannot accurately determined all $%
g_{f}\left( p_{2,}\;p_{1}\right) .$ When $p_{20}^{\prime }\neq p_{20}$ or$\
\ p_{10}^{\prime }\neq p_{10},$ $g_{f}\left( p_{2,}\;p_{1}\right) $ is
approximate and should be corrected.

From (2.3.9a), considering the expansions of $\overline{\psi },$ $\psi $ and$%
\ A_{\mu },$ we obtain%
\begin{eqnarray}
G_{F} &=&\sum_{\mathbf{p}_{2}}\sum_{\mathbf{p}_{1}}g_{f}\left(
p_{2},p_{1}\right) \{\underline{I}_{\mathbf{p}_{2}}\underline{I}_{\mathbf{p}%
_{1}}^{+}+\underline{I}_{\mathbf{p}_{2}}\underline{I}_{(-\mathbf{p}%
_{1})}^{+}+\underline{I}_{(-\mathbf{p}_{2})}\underline{I}_{\mathbf{p}%
_{1}}^{+}+\underline{I}_{(-\mathbf{p}_{2})}\underline{I}_{(-\mathbf{p}%
_{1})}^{+}\}\underline{J}_{\mathbf{k}}^{+}  \notag \\
&=&\sum_{\mathbf{p}_{2}}\sum_{\mathbf{p}_{1}}g_{f}\left( p_{2},p_{1}\right) 
\underline{I}_{\mathbf{p}_{2}}\underline{I}_{\mathbf{p}_{1}}^{+}\underline{J}%
_{(-\mathbf{k)}}.  \TCItag{2.3.10}
\end{eqnarray}%
For the initial\ W-state $\mid a_{\mathbf{p}_{1}s_{1}}c_{\mathbf{k}\lambda
}\rangle $ and the final W-state $\mid a_{\mathbf{p}_{2}s_{2}}\rangle ,$ $%
g_{f}(p_{2},p_{1})$ is precise. If for arbitrary $\mathbf{p}_{2}$ and $%
\mathbf{p}_{1}$ we can also determine $g(p_{2},p_{1})$, (2.3.10) is the
precise couplng operator. In fact it is impossible to determine all $%
g_{f}\left( p_{2},p_{1}\right) .$ We can only accurately determine some $%
g_{f}\left( p_{1},p_{2}\right) ,$ e.g., $g_{f}\left( q_{2},q_{1}\right) ,$
where $q_{2}$ \ and $q_{1}$ are the momenta at the so-called subtraction
point which satisfy the restrictions

\begin{equation}
i\gamma q_{1}+m_{0}=i\gamma q_{2}+m_{0}=0,\;\ q^{\prime }=q_{2}-q_{1},\text{ 
\hspace{0in} }q^{\prime 2}=0,  \tag{2.3.11}
\end{equation}%
We define

\begin{equation}
g_{f}\left( q_{2},q_{1}\right) =g_{f0}.  \tag{2.3.12}
\end{equation}%
Replacing $g_{f}\left( p_{2},p_{1}\right) $ by $g_{f0},$ we obtain a
approximate coupling operator,%
\begin{equation}
G_{F}^{(0)}=g_{f0}\sum_{\mathbf{p}_{2}}\sum_{\mathbf{p}_{1}}\underline{I}_{%
\mathbf{p}_{2}}\underline{I}_{\mathbf{p}_{1}}^{+}\underline{J}_{\mathbf{k}%
}^{+}.\;\ \   \tag{2.3.13}
\end{equation}%
Analogously to $G_{F}(p_{2},p_{1})$ and $G_{F}^{(0)},$ and considering the
symmetry between $\mathcal{L}_{F}$ $\ $and $\ \mathcal{L}_{W}$, from (2.3.9)
we have%
\begin{eqnarray}
\langle a_{\mathbf{p}_{2}s_{2}} &\mid &G_{W}^{\prime }\mid a_{\mathbf{p}%
_{1}s_{1}}c_{\mathbf{k}\lambda }\rangle \equiv \frac{\langle a_{\mathbf{p}%
_{2}s_{2}}\mid S_{f}\mid a_{\mathbf{p}_{1}s_{1}}c_{\mathbf{k}\lambda
}\rangle }{A_{gf}^{^{\prime }}\left( \mid a_{\mathbf{p}_{1}s_{1}}c_{\mathbf{k%
}\lambda }\rangle \rightarrow \mid a_{\mathbf{p}_{2}s_{2}}\rangle \right) } 
\notag \\
&\equiv &g_{w}(p_{2},p_{1})=g_{f}(p_{2},p_{1})=\langle \underline{a}_{%
\mathbf{p}_{2}s_{2}}\mid G_{F}^{\prime }\mid \underline{a}_{\mathbf{p}%
_{1}s_{1}}\underline{c}_{\mathbf{k}\lambda }\rangle ,  \TCItag{2.3.14}
\end{eqnarray}%
\begin{equation}
G_{W}=\sum_{\mathbf{p}_{2}}\sum_{\mathbf{p}_{1}}g_{w}\left(
p_{2},p_{1}\right) I_{\mathbf{p}_{2}}I_{\mathbf{p}_{1}}^{+}\underline{J}_{%
\mathbf{k}}^{+},\text{\ }  \tag{2.3.15}
\end{equation}%
\begin{equation}
G_{W}^{(0)}=g_{w0}\sum_{\mathbf{p}_{2}}\sum_{\mathbf{p}_{1}}I_{\mathbf{p}%
_{2}}I_{\mathbf{p}_{1}}^{+}J_{\mathbf{k}}^{+},\;\ \ \
g_{w0}(q_{2},q_{1})=g_{f0}=g_{0}.  \tag{2.3.16}
\end{equation}

\ Let $m_{ef}$ be the mass originating from the electromagnetic interaction, 
$m_{f}^{\prime }$ be the mass originating from the other interactions, $%
m_{f}=$ $m_{f}^{\prime }+m_{ef}$ is the total mass of a F-electron.
Analogously to $G_{F},$ we define

\begin{eqnarray}
m_{ef}\left( \mid a_{\mathbf{p}_{1}s_{1}}\rangle \rightarrow \mid a_{\mathbf{%
p}_{2}s_{2}}\rangle \right) &=&\frac{A_{mf}\left( \mid a_{\mathbf{p}%
_{1}s_{1}}\rangle \rightarrow \mid a_{\mathbf{p}_{2}s_{2}}\rangle \right) }{%
A_{mf}^{^{\prime }}\left( \mid a_{\mathbf{p}_{1}s_{1}}\rangle \rightarrow
\mid a_{\mathbf{p}_{2}s_{2}}\rangle \right) }  \notag \\
&=&\frac{\langle a_{\mathbf{p}_{2}s_{2}}\mid S_{f}\mid a_{\mathbf{p}%
_{1}s_{1}}\rangle }{\langle a_{\mathbf{p}_{2}s_{2}}\mid \left( -i\int d^{4}x:%
\overline{\psi }^{^{\prime }}\psi ^{^{\prime }}:\right) \mid a_{\mathbf{p}%
_{1}s_{1}}\rangle },  \TCItag{2.3.17}
\end{eqnarray}

\begin{eqnarray}
A_{mf}^{^{\prime }}\left( \mid a_{\mathbf{p}_{1}s_{1}}\rangle \rightarrow
\mid a_{\mathbf{p}_{2}s_{2}}\rangle \right) &=&\langle a_{\mathbf{p}%
_{2}s_{2}}\mid \left( -i\int d^{4}x:\overline{\psi }^{^{\prime }}\psi
^{^{\prime }}:\right) \mid a_{\mathbf{p}_{1}s_{1}}\rangle  \notag \\
&=&-i\left( 2\pi \right) ^{4}\delta ^{4}\left( p_{2}-p_{1}\right) \frac{1}{V}%
\overline{u}_{\mathbf{p}_{2}s_{2}}u_{\mathbf{p}_{1}s_{1}}.  \TCItag{2.3.18}
\end{eqnarray}%
Because of the symmetry of $\mid a_{\mathbf{p}_{1}s_{1}}\rangle $ and $\mid
b_{\mathbf{p}_{1}s_{1}}\rangle $, we have

\begin{equation}
m_{ef}\left( \mid a_{\mathbf{p}_{1}s_{1}}\rangle \rightarrow \mid a_{\mathbf{%
p}_{2}s_{2}}\rangle \right) =m_{ef}\left( \mid b_{\mathbf{p}%
_{1}s_{1}}\rangle \rightarrow \mid b_{\mathbf{p}_{2}s_{2}}\rangle \right)
=m_{ef}\left( p_{1}\right) .  \tag{2.3.19}
\end{equation}%
Similarly, we can define the electromagnetic mass of a W-electron as

\begin{eqnarray}
m_{ew}\left( p_{1}\right) &=&\frac{A_{mw}\left( \mid \underline{a}_{\mathbf{p%
}_{1}s_{1}}\rangle \rightarrow \mid \underline{a}_{\mathbf{p}%
_{2}s_{2}}\rangle \right) }{A_{mw}^{^{\prime }}\left( \mid \underline{a}_{%
\mathbf{p}_{1}s_{1}}\rangle \rightarrow \mid \underline{a}_{\mathbf{p}%
_{2}s_{2}}\rangle \right) }  \notag \\
&=&\frac{\langle \underline{a}_{\mathbf{p}_{2}s_{2}}\mid S_{w}\mid 
\underline{a}_{\mathbf{p}_{1}s_{11}}\rangle }{\langle \underline{a}_{\mathbf{%
p}_{2}s_{2}}\mid -i\int d^{4}x:\underline{\overline{\psi }}_{0}^{^{\prime }}%
\underline{\psi }_{0}^{^{\prime }}:\mid \underline{a}_{\mathbf{p}%
_{1}s_{11}}\rangle },  \TCItag{2.3.20}
\end{eqnarray}

\begin{eqnarray}
A_{mw}^{^{\prime }}\left( \mid \underline{a}_{\mathbf{p}_{1}s_{1}}\rangle
\rightarrow \mid \underline{a}_{\mathbf{p}_{2}s_{2}}\rangle \right)
&=&\langle \underline{a}_{\mathbf{p}_{2}s_{2}}\mid -i\int d^{4}x:\underline{%
\overline{\psi }}_{0}^{^{\prime }}\underline{\psi }_{0}^{^{\prime }}:\mid 
\underline{a}_{\mathbf{p}_{1}s_{11}}\rangle  \notag \\
&=&-i\left( 2\pi \right) ^{4}\delta ^{4}\left( p_{1}-p_{2}\right) \frac{1}{V}%
\overline{v}_{\mathbf{p}_{1}s_{1}}v_{\mathbf{p}_{1}s_{1}},  \TCItag{2.3.21}
\end{eqnarray}%
Considering the symmetry of $\mathcal{L}_{W}$ and $\mathcal{L}_{F}$, we have

\begin{equation}
m_{ew}\left( p\right) =m_{ef}\left( p\right) \equiv m_{e}\left( p\right) . 
\tag{2.3.22}
\end{equation}

Without loss of generality, we take a term in (2.2.12)

\begin{equation}
\mid a_{\mathbf{p}_{2}s_{2}}(t)\eqslantgtr I_{\mathbf{p}_{2}}^{+}\cdot
M_{F}\cdot I_{\mathbf{p}_{1}}\eqslantless a_{\mathbf{p}_{1}s_{1}}(t)\mid 
\tag{2.3.23}
\end{equation}%
as an example to illuminate the structure of $M_{F}$. Recalling
(1.3.25)-(1.3.28),\ we can know that the part in $M_{F}$ corresponding to $%
I_{\mathbf{p}_{2}}^{+}\cdot M_{F}\cdot I_{\mathbf{p}_{1}}$ should be written
as%
\begin{equation}
I_{\mathbf{p}_{2}}\langle \underline{a}_{\mathbf{p}_{2}s_{2}}\mid
M_{F}^{^{\prime }}\mid \underline{a}_{\mathbf{p}_{1}s_{1}}\rangle I_{\mathbf{%
p}_{1}}^{+}  \tag{2.3.24}
\end{equation}%
Because $\langle \underline{a}_{\mathbf{p}_{2}s_{2}}\mid $ and $\mid 
\underline{a}_{\mathbf{p}_{1}s_{1}}\rangle $ may be regarded a initial
W-state and a final W-state, respectively, $\langle \underline{a}_{\mathbf{p}%
_{2}s_{2}}\mid M_{F}^{\prime }\mid \underline{a}_{\mathbf{p}%
_{1}s_{1}}\rangle $ must be directly proportional to the scattering
amplitude $A_{mw}\left( \mid \underline{a}_{\mathbf{p}_{1}s_{1}}\rangle
\rightarrow \mid \underline{a}_{\mathbf{p}_{2}s_{2}}\rangle \right) .$
Comparing $\langle \underline{a}_{\mathbf{p}_{2}s_{2}}\mid M_{F}^{^{\prime
}}\mid \underline{a}_{\mathbf{p}_{1}s_{1}}\rangle $ with (2.3.20), we can
define $\langle \underline{a}_{\mathbf{p}_{2}s_{2}}\mid M_{F}^{^{\prime
}}\mid \underline{a}_{\mathbf{p}_{1}s_{1}}\rangle $ as the electromagnetic
mass of a F-electron,%
\begin{equation}
\langle \underline{a}_{\mathbf{p}_{2}s_{2}}\mid M_{F}^{\prime }\mid 
\underline{a}_{\mathbf{p}_{1}s_{1}}\rangle \equiv \frac{\langle \underline{a}%
_{\mathbf{p}_{2}s_{2}}\mid S_{w}\mid \underline{a}_{\mathbf{p}%
_{1}s_{11}}\rangle }{\langle \underline{a}_{\mathbf{p}_{2}s_{2}}\mid -i\int
d^{4}x:\underline{\overline{\psi }}_{0}^{^{\prime }}\underline{\psi }%
_{0}^{^{\prime }}:\mid \underline{a}_{\mathbf{p}_{1}s_{11}}\rangle }\equiv
m_{ef}\left( p_{1}\right) ,  \tag{2.3.25}
\end{equation}%
where $p_{2}=p_{1}.$ From (2.3.20) and (2.3.25) we can anew obtain (2.3.22).
It is seen that (2.3.25) is consistent with (2.3.22). Substituting (2.3.25)
into (2.3.24), and considering%
\begin{equation*}
\langle \underline{a}_{\mathbf{p}_{2}s_{2}}\mid M_{F}^{^{\prime }}\mid 
\underline{a}_{\mathbf{p}_{1}s_{1}}\rangle =\langle \underline{b}_{\mathbf{p}%
_{2}s_{2}}\mid M_{F}^{^{\prime }}\mid \underline{b}_{\mathbf{p}%
_{1}s_{1}}\rangle =m_{ef}\left( p_{1}\right)
\end{equation*}%
and the expansions of $\underline{\overline{\psi }}$ \ and $\underline{\psi }%
,$ we have%
\begin{equation}
M_{F}=\sum_{\mathbf{p}_{2}}\sum_{\mathbf{p}_{1}}m_{ef}(p_{1})\underline{I}_{%
\mathbf{p}_{2}}\underline{I}_{\mathbf{p}_{1}}^{+}.  \tag{2.3.26}
\end{equation}%
For some $p,$ e.g., $p=q,$ we can determine $m_{ef}\left( q\right) .$ $\func{%
Re}$placing $m_{ef}\left( p_{1}\right) $ in (2.3.26) by $m_{ef}\left(
q\right) ,$ we obtain a approximate mass operator%
\begin{equation}
M_{F}^{(0)}=m_{ef0}\sum_{\mathbf{p}_{2}}\sum_{\mathbf{p}_{1}}\underline{I}_{%
\mathbf{p}_{2}}\underline{I}_{\mathbf{p}_{1}}^{+},\;\ \ \ \ m_{ef0}\equiv
m_{ef}\left( q\right) \ .  \tag{2.3.27}
\end{equation}

Analogously to (2.3.26) and (2.3.27) and considering the symmetry between $%
\mathcal{L}_{F}$ and $\mathcal{L}_{W},$ we have%
\begin{eqnarray}
\langle a_{\mathbf{p}_{2}s_{2}} &\mid &M_{W}^{\prime }\mid a_{\mathbf{p}%
_{1}s_{1}}\rangle \equiv \frac{\langle a_{\mathbf{p}_{2}s_{2}}\mid S_{f}\mid
a_{\mathbf{p}_{1}s_{1}}\rangle }{\langle a_{\mathbf{p}_{2}s_{2}}\mid \left(
-i\int d^{4}x:\overline{\psi }^{^{\prime }}\psi ^{^{\prime }}:\right) \mid
a_{\mathbf{p}_{1}s_{1}}\rangle }  \notag \\
&\equiv &m_{ew}\left( p_{1}\right) =m_{ef}\left( p_{1}\right) =\langle 
\underline{a}_{p_{2}s_{2}}\mid M_{F}^{\prime }\mid \underline{a}%
_{p_{1}s_{1}}\rangle ,\;\   \TCItag{2.3.28}
\end{eqnarray}%
\begin{equation}
M_{W}=\sum_{\mathbf{p}_{2}}\sum_{\mathbf{p}_{1}}m_{ew}(p_{1})I_{\mathbf{p}%
_{2}}I_{\mathbf{p}_{1}}^{+},\;\   \tag{2.3.29}
\end{equation}%
\begin{equation}
M_{W}^{(0)}=m_{ew0}\sum_{\mathbf{p}_{2}}\sum_{\mathbf{p}_{1}}I_{\mathbf{p}%
_{2}}I_{\mathbf{p}_{1}}^{+},\;\ \ m_{ew0}\equiv m_{ew}\left( q\right)
=m_{ef0}\equiv m_{e0}\ .  \tag{2.3.30}
\end{equation}%
Because we cannot differentiate $m_{ef0}$ from $m_{f0}^{^{\prime }}$ in $%
m_{f0}=$ $m_{f0}^{^{\prime }}+m_{ef0}$, for convenience, we define 
\begin{equation}
m_{ef0}=m_{ew0}=m_{e0}=0,\;\ \ \ \ \ \ m_{f0}=m_{f0}^{^{\prime
}}=m_{w0}=m_{w0}^{^{\prime }}=m_{0}.  \tag{2.3.31}
\end{equation}%
Of course, for an interacting field operator $\mathbf{p}^{2}-p_{0}^{2}\neq
-m^{2}$ and $k^{2}\neq 0.$ When $p_{2}=q_{2}$ and $p_{1}=q_{1},$ both $g_{0}$
and $m_{e0}$ are accurate. When $p_{2}\neq q_{2}$ or $p_{1}\neq q_{1},$ both 
$g_{0}$ and $m_{e0}$ are approximate. If the subtraction point is a
nonphysical point for a coupling constant, ~we assume that $g$ can
analytically expand from the nonphysical point into physical regions.

\subsection{Expansion of the Hamiltonian}

Because when many operators containing $K$ multiply, associative law does no
longer hold water. Hence it is necessary to define the order of
multiplication of a copling or a mass operator and transformation operators.
When we construct the interaction Lagrangian or Hamiltonian density, we
define the orter to be that the coupling operators and the mass operators
derectly multiply by every transformation operator. Thus substituting the
expansions of $\psi ,$ $\overline{\psi },$ $A_{\mu }$, \underline{$\psi $}$,$
\underline{$\overline{\psi }$} \ and \underline{$A$}$_{\mu }$ \ and
(2.3.10), (2.3.15), (2.3.26) and (2.3.29) into (2.2.11)-(2.2.13), we obtain

\begin{eqnarray}
H_{F0} &=&\sum_{\mathbf{p}s}\{\mid a_{\mathbf{p}s}\left( t\right)
\eqslantgtr \eqslantless a_{\mathbf{p}s}\left( t\right) \mid u_{\mathbf{p}%
s}^{+}\gamma _{4}\left( i\gamma _{j}p_{j}+m\right) u_{\mathbf{p}s}  \notag \\
+ &\eqslantless &b_{(-\mathbf{p)}s}\left( t\right) \mid \eqslantless a_{%
\mathbf{p}s}\left( t\right) \mid v_{(-\mathbf{p)}s}^{+}\gamma _{4}\left(
i\gamma _{j}p_{j}+m\right) u_{\mathbf{p}s}  \notag \\
+ &\mid &a_{(-\mathbf{p)}s}\left( t\right) \eqslantgtr \mid b_{\mathbf{p}%
s}\left( t\right) \eqslantgtr u_{(-\mathbf{p)}s}^{+}\gamma _{4}\left(
-i\gamma _{j}p_{j}+m\right) v_{\mathbf{p}s}  \notag \\
- &\mid &b_{\mathbf{p}s}\left( t\right) \eqslantgtr \eqslantless b_{\mathbf{p%
}s}\left( t\right) \mid v_{\mathbf{p}s}^{+}\gamma _{4}\left( -i\gamma
_{j}p_{j}+m\right) v_{\mathbf{p}s}\}  \notag \\
+\sum_{\mathbf{k}}\frac{\omega _{\mathbf{k}}^{2}+\mathbf{k}^{2}}{2\omega _{%
\mathbf{k}}}\{\sum_{\lambda =1}^{3} &\mid &c_{\mathbf{k}\lambda }\left(
t\right) \eqslantgtr \eqslantless c_{\mathbf{k}\lambda }\left( t\right) \mid
-\mid c_{\mathbf{k}4}\left( t\right) \eqslantgtr \eqslantless c_{\mathbf{k}%
4}\left( t\right) \mid  \notag \\
-\sum_{\mathbf{k}}\frac{\omega _{\mathbf{k}}^{2}-\mathbf{k}^{2}}{4\omega _{%
\mathbf{k}}}\{\sum_{\lambda =1}^{3}( &\eqslantless &c_{\mathbf{k}\lambda
}\left( t\right) \mid \eqslantless c_{-\mathbf{k}\lambda }\left( t\right)
\mid +\mid c_{\mathbf{k}\lambda }\left( t\right) \eqslantgtr \mid c_{-%
\mathbf{k}\lambda }\left( t\right) \eqslantgtr )  \notag \\
-( &\eqslantless &c_{\mathbf{k}4}\left( t\right) \mid \eqslantless c_{-%
\mathbf{k}4}\left( t\right) \mid +\mid c_{\mathbf{k}4}\left( t\right)
\eqslantgtr \mid c_{-\mathbf{k}4}\left( t\right) \eqslantgtr )\}, 
\TCItag{2.4.1}
\end{eqnarray}

\begin{eqnarray}
H_{FIM} &=&\sum_{\mathbf{p}s}m_{ef}\left( p\right) \{\mid a_{\mathbf{p}%
s}\left( t\right) \eqslantgtr \eqslantless a_{\mathbf{p}s}\left( t\right)
\mid u_{\mathbf{p}s}^{+}\gamma _{4}u_{\mathbf{p}s}  \notag \\
+ &\eqslantless &b_{(-\mathbf{p)}s}\left( t\right) \mid \eqslantless a_{%
\mathbf{p}s}\left( t\right) \mid v_{(-\mathbf{p)}s}^{+}\gamma _{4}u_{\mathbf{%
p}s}  \notag \\
+ &\mid &a_{(-\mathbf{p)}s}\left( t\right) \eqslantgtr \mid b_{\mathbf{p}%
s}\left( t\right) \eqslantgtr u_{(-\mathbf{p)}s}^{+}\gamma _{4}v_{\mathbf{p}%
s}  \notag \\
- &\mid &b_{\mathbf{p}s}\left( t\right) \eqslantgtr \eqslantless b_{\mathbf{p%
}s}\left( t\right) \mid v_{\mathbf{p}s}^{+}\gamma _{4}v_{\mathbf{p}s}\} 
\TCItag{2.4.2}
\end{eqnarray}%
\begin{eqnarray}
H_{FIG} &=&-\frac{i}{\sqrt{V}}\sum_{\mathbf{p}_{2}s_{2}}\sum_{\mathbf{p}%
_{1}s_{1}}\{\mid a_{\mathbf{p}_{2}s_{2}}\left( t\right) \eqslantgtr
\eqslantless a_{\mathbf{p}_{1}s_{1}}\left( t\right) \mid \overline{u}_{%
\mathbf{p}_{2}s_{2}}\gamma _{\mu }u_{\mathbf{p}_{1}s_{1}}  \notag \\
+ &\eqslantless &b_{(-\mathbf{p}_{2})s_{2}}\left( t\right) \mid \eqslantless
a_{\mathbf{p}_{1}s_{1}}\left( t\right) \mid \overline{v}_{(-\mathbf{p}%
_{2})s_{2}}\gamma _{\mu }u_{\mathbf{p}_{1}s_{1}}  \notag \\
+ &\mid &a_{\mathbf{p}_{2}s_{2}}\left( t\right) \eqslantgtr \mid b_{(-%
\mathbf{p}_{1})s_{1}}\left( t\right) \eqslantgtr \overline{u}_{\mathbf{p}%
_{2}s_{2}}\gamma _{\mu }v_{(-\mathbf{p}_{1})s_{1}}  \notag \\
- &\mid &b_{(-\mathbf{p}_{1})s_{1}}\left( t\right) \eqslantgtr \eqslantless
b_{(-\mathbf{p}_{2})s_{2}}\left( t\right) \mid \overline{v}_{(-\mathbf{p}%
_{2})s_{2}}\gamma _{\mu }v_{(-\mathbf{p}_{1})s_{1}}\}g\left(
p_{2},p_{1}\right) \frac{1}{\sqrt{2\omega _{\mathbf{k}}}}  \notag \\
&&\cdot \sum_{\lambda =1}^{4}\left( e_{\mathbf{k}\mu }^{\lambda
}\eqslantless c_{\mathbf{k}\lambda }\left( t\right) \mid +e_{(-\mathbf{k}\mu
)}^{\lambda }\mid \overline{c}_{(-\mathbf{k}\lambda )}\left( t\right)
\eqslantgtr \right) ,  \TCItag{2.4.3}
\end{eqnarray}

where $\mathbf{k=p}_{2}\mathbf{-p}_{1}.$%
\begin{eqnarray}
H_{W0} &=&\sum_{\mathbf{p}s}\{\mid \underline{b}_{\mathbf{p}s}\left(
t\right) \eqslantgtr \eqslantless \underline{b}_{\mathbf{p}s}\left( t\right)
\mid u_{\mathbf{p}s}^{+}\gamma _{4}\left( i\gamma _{j}p_{j}+m\right) u_{%
\mathbf{p}s}  \notag \\
- &\mid &\underline{a}_{(-\mathbf{p)}s}\left( t\right) \eqslantgtr \mid 
\underline{b}_{\mathbf{p}s}\left( t\right) \eqslantgtr v_{(-\mathbf{p)}%
s}^{+}\gamma _{4}\left( i\gamma _{j}p_{j}+m\right) u_{\mathbf{p}s}  \notag \\
- &\eqslantless &\underline{b}_{(-\mathbf{p)}s}\left( t\right) \mid
\eqslantless \underline{a}_{\mathbf{p}s}\left( t\right) \mid u_{(-\mathbf{p)}%
s}^{+}\gamma _{4}\left( -i\gamma _{j}p_{j}+m\right) v_{\mathbf{p}s}  \notag
\\
- &\mid &\underline{a}_{\mathbf{p}s}\left( t\right) \eqslantgtr \eqslantless 
\underline{a}_{\mathbf{p}s}\left( t\right) \mid v_{\mathbf{p}s}^{+}\gamma
_{4}\left( -i\gamma _{j}p_{j}+m\right) v_{\mathbf{p}s}\}  \notag \\
+\sum_{\mathbf{k}}\frac{\omega _{\mathbf{k}}^{2}+\mathbf{k}^{2}}{2\omega _{%
\mathbf{k}}}\{\sum_{\lambda =1}^{3} &\mid &\underline{c}_{\mathbf{k}\lambda
}\left( t\right) \eqslantgtr \eqslantless \underline{c}_{\mathbf{k}\lambda
}\left( t\right) \mid -\mid \underline{c}_{\mathbf{k}4}\left( t\right)
\eqslantgtr \eqslantless \underline{c}_{\mathbf{k}4}\left( t\right) \mid \} 
\notag \\
-\sum_{\mathbf{k}}\frac{\omega _{\mathbf{k}}^{2}-\mathbf{k}^{2}}{4\omega _{%
\mathbf{k}}}\{\sum_{\lambda =1}^{3}( &\eqslantless &\underline{c}_{\mathbf{k}%
\lambda }\left( t\right) \mid \eqslantless \underline{c}_{-\mathbf{k}\lambda
}\left( t\right) \mid +\mid \underline{c}_{\mathbf{k}\lambda }\left(
t\right) \eqslantgtr \mid \underline{c}_{-\mathbf{k}\lambda }\left( t\right)
\eqslantgtr )  \notag \\
-( &\eqslantless &\underline{c}_{\mathbf{k}4}\left( t\right) \mid
\eqslantless \underline{c}_{-\mathbf{k}4}\left( t\right) \mid +\mid 
\underline{c}_{\mathbf{k}4}\left( t\right) \eqslantgtr \mid \underline{c}_{-%
\mathbf{k}4}\left( t\right) \eqslantgtr )\},  \TCItag{2.4.4}
\end{eqnarray}

\begin{eqnarray}
H_{WIM} &=&\sum_{\mathbf{p}s}m_{ew}\left( p\right) \{\mid \underline{b}_{%
\mathbf{p}s}\left( t\right) \eqslantgtr \eqslantless \underline{b}_{\mathbf{p%
}s}\left( t\right) \mid u_{\mathbf{p}s}^{+}\gamma _{4}u_{\mathbf{p}s}  \notag
\\
- &\mid &\underline{a}_{(-\mathbf{p)}s}\left( t\right) \eqslantgtr \mid 
\underline{b}_{\mathbf{p}s}\left( t\right) \eqslantgtr v_{(-\mathbf{p)}%
s}^{+}\gamma _{4}u_{\mathbf{p}s}  \notag \\
- &\eqslantless &\underline{b}_{(-\mathbf{p)}s}\left( t\right) \mid
\eqslantless \underline{a}_{\mathbf{p}s}\left( t\right) \mid u_{(-\mathbf{p)}%
s}^{+}\gamma _{4}v_{\mathbf{p}s}  \notag \\
- &\mid &\underline{a}_{\mathbf{p}s}\left( t\right) \eqslantgtr \eqslantless 
\underline{a}_{\mathbf{p}s}\left( t\right) \mid v_{\mathbf{p}s}^{+}\gamma
_{4}v_{\mathbf{p}s}\}  \TCItag{2.4.5}
\end{eqnarray}%
\begin{eqnarray}
H_{WIG} &=&-\frac{i}{\sqrt{V}}\sum_{\mathbf{p}_{2}s_{2}}\sum_{\mathbf{p}%
_{1}s_{1}}\{-\mid \underline{b}_{\mathbf{p}_{2}s_{2}}\left( t\right)
\eqslantgtr \eqslantless \underline{b}_{\mathbf{p}_{1}s_{1}}\left( t\right)
\mid u_{\mathbf{p}_{1}s_{1}}^{+}\gamma _{4}\gamma _{\mu }u_{\mathbf{p}%
_{2}s_{2}}  \notag \\
+ &\mid &\underline{a}_{(-\mathbf{p}_{1}s_{1})}\left( t\right) \eqslantgtr
\mid \underline{b}_{\mathbf{p}_{2}s_{2}}\left( t\right) \eqslantgtr v_{-%
\mathbf{p}_{1}s_{1}}^{+}\gamma _{4}\gamma _{\mu }u_{\mathbf{p}_{2}s_{2}} 
\notag \\
+ &\eqslantless &\underline{b}_{\mathbf{p}_{1}s_{1}}\left( t\right) \mid
\eqslantless \underline{a}_{(-\mathbf{p}_{2}s_{2})}\left( t\right) \mid u_{%
\mathbf{p}_{1}s_{1}}^{+}\gamma _{4}\gamma _{\mu }v_{-\mathbf{p}_{2}s_{2}} 
\notag \\
+ &\mid &\underline{a}_{(-\mathbf{p}_{1}s_{1})}\left( t\right) \eqslantgtr
\eqslantless \underline{a}_{(-\mathbf{p}_{2}s_{2})}\left( t\right) \mid v_{-%
\mathbf{p}_{1}s_{1}}^{+}\gamma _{4}\gamma _{\mu }v_{-\mathbf{p}%
_{2}s_{2}}\}\cdot g\left( p_{2},p_{1}\right) \frac{1}{\sqrt{2\omega _{%
\mathbf{k}}}}  \notag \\
\sum_{\lambda =1}^{4}e_{(-\mathbf{k)}\mu }^{\lambda } &\mid &\underline{%
\overline{c}}_{(-\mathbf{k)}\lambda }\left( t\right) \eqslantgtr +e_{\mathbf{%
k}\mu }^{\lambda }\eqslantless \underline{c}_{\mathbf{k}\lambda }\left(
t\right) \mid ).  \TCItag{2.4.6}
\end{eqnarray}%
From (2.4.1)-(2.4.6) we can prove (2.2.18)-(2.2.19) to be consistent with
(2.2.14)-(2.2.17). (see appendix C).

In fact, we can only obtain an approximate Hamiltonian in which the coupling
constants and masses are approximately regarded as the same, i.e., $g\left(
p_{2},p_{1}\right) =g_{0}$ and $m_{e}\left( p\right) =0.$ In this case, we
can write $H_{F}$ and $H_{W}$ as%
\begin{eqnarray}
&&H_{F}^{(0-loop)}\left( g_{0},m_{e0}\right)  \notag \\
&=&\int d^{3}x:[\psi ^{\prime +}\gamma _{4}\left( \gamma _{j}\partial
_{j}-ig_{0}\gamma _{\mu }A_{\mu }^{\prime }+m\right) \psi ^{\prime }+\frac{1%
}{2}\left( \dot{A}_{\mu }^{\prime }\dot{A}_{\mu }^{\prime }+\partial
_{j}A_{\nu }^{\prime }\partial _{j}A_{\nu }^{\prime }\right) ]:, 
\TCItag{2.4.7}
\end{eqnarray}

\begin{eqnarray}
&&H_{W}^{(0-loop)}\left( g_{0},m_{e0}\right)  \notag \\
&=&\int d^{3}x:[-\underline{\psi }^{^{\prime }+}\gamma _{4}(\gamma
_{j}\partial _{j}+ig_{0}\gamma _{\mu }\underline{A}_{\mu }^{\prime }+m)%
\underline{\psi }^{^{\prime }}+\frac{1}{2}(\underline{\dot{A}}^{\prime }%
\underline{\dot{A}}^{\prime }+\partial _{j}^{\prime }\underline{A}_{\nu
}^{\prime }\underline{A}_{\nu }^{\prime })]:  \TCItag{2.4.8}
\end{eqnarray}%
here the expansions of the field operators $\psi ^{^{\prime }}\left( \mathbf{%
x},t\right) $ etc. are the same as $(1.4.10)-(1.4.17)$ in form, but the
operators $\eqslantless a_{\mathbf{p}s}\left( t\right) \mid $ etc. in
(1.4.10) or (1.4.14)-(1.4.15) must be replaced by (2.2.21)-(2.2.22). When $%
p_{2}=q_{2}$ or $p_{1}=q_{1},$both $H_{F}^{(0-loop)}\left(
g_{0},m_{e0}\right) $ and $H_{W}^{(0-loop)}\left( g_{0},m_{e0}\right) $ are
accurate. When $p_{2}\neq q_{2}$ or $p_{1}\neq q_{1},$ both $%
H_{F}^{(0-loop)}\left( g_{0},m_{e0}\right) $ and $H_{W}^{(0-loop)}\left(
g_{0},m_{e0}\right) $ are approximate. The equal-time anticommulation and
commulation relations for the operators $\eqslantless \underline{b}_{\mathbf{%
p}s}\left( t\right) \mid $ and $\psi _{\rho }^{\prime }\left( \mathbf{x}%
,t\right) ,$ etc., are the same as $(1.3.16)-(1.3.17)$ and $\left(
1.4.18\right) -\left( 1.4.19\right) ,$ respectively. In this case, the
Heisenberg equations are

\begin{equation}
\dot{F}=-i\left[ F,H_{F}^{(0-loop)}\right] =-i\left[ F,H^{(0-loop)}\right] ,
\tag{2.4.9}
\end{equation}%
\hspace{0pt}

\begin{equation}
\dot{W}=i[W,H_{W}^{(0-loop)}]=i[W,H^{(0-loop)}],  \tag{2.4.10}
\end{equation}%
where $H^{(0-loop)}=H_{F}^{(0-loop)}+H_{W}^{(0-loop)},$ $F=\psi ^{\prime },$ 
$A_{\mu }^{\prime },$ $\psi ,$ $A_{\mu },$ $\eqslantless a_{\mathbf{p}%
s}(t)\mid ,$ $\eqslantless b_{\mathbf{p}s}(t)\mid ,$ $\eqslantless c_{%
\mathbf{k}\lambda }(t)\mid ;$ $W=$ $\underline{\psi }^{\prime },$ $%
\underline{A}_{\mu }^{\prime },$ $\underline{\psi },$ $\underline{A}_{\mu },$
$\eqslantless \underline{a}_{\mathbf{p}s}(t)\mid ,$ $\eqslantless \underline{%
b}_{\mathbf{p}s}(t)\mid ,$ $\eqslantless \underline{c}_{\mathbf{k}\lambda
}(t)\mid .$ From $(2.4.9)-(2.4.10)$ we obtain $(2.2.14)-(2.2.17)$ (where $%
M_{F}\rightarrow M_{F}^{(0)},$ $G_{F}\rightarrow G_{F}^{(0)},$ $%
M_{W}\rightarrow M_{W}^{(0)},$ $G_{W}\rightarrow G_{W}^{(0)})$ and 
\begin{equation}
i\frac{\partial }{\partial t}\psi ^{\prime }=\gamma _{4}\left( \gamma
_{j}\partial _{j}-i\gamma _{\mu }g_{0}A_{\mu }+m\right) \psi ^{\prime }, 
\tag{2.4.11}
\end{equation}

\begin{equation}
\square A_{\mu }^{\prime }=-i\overline{\psi }^{\prime }\gamma _{\mu
}g_{0}\psi ^{\prime },  \tag{2.4.12}
\end{equation}

\begin{equation}
i\frac{\partial }{\partial t}\underline{\psi }^{\prime }=\gamma _{4}\left(
\gamma _{j}\partial _{j}+i\gamma _{\mu }g_{0}\underline{A}_{\mu }^{\prime
}+m\right) \underline{\psi }^{\prime },  \tag{2.4.13}
\end{equation}

\begin{equation}
\square \underline{A}_{\mu }^{\prime }=-i\underline{\overline{\psi }}%
^{\prime }\gamma _{\mu }g_{0}\underline{\psi }^{\prime },  \tag{2.4.14}
\end{equation}%
\begin{equation}
\psi ^{\prime }\left( \mathbf{x,}t\right) =e^{iH^{(0-loop)}t}\psi ^{\prime
}\left( \mathbf{x,}0\right) e^{-iH^{(0-loop)}t}=e^{iH_{F}^{(0-loop)}t}\psi
^{\prime }\left( \mathbf{x,}0\right) e^{-iH_{F}^{(0-loop)}t},  \tag{2.4.15}
\end{equation}%
\begin{equation}
A_{\mu }^{\prime }\left( \mathbf{x,}t\right) =e^{iH^{(0-loop)}t}A_{\mu
}^{\prime }\left( \mathbf{x,}0\right)
e^{-iH^{(0-loop)}t}=e^{iH_{F}^{(0-loop)}t}A_{\mu }^{\prime }\left( \mathbf{x,%
}0\right) e^{-iH_{F}^{(0-loop)}t},  \tag{2.4.16}
\end{equation}%
\begin{equation}
\underline{\psi }^{\prime }\left( \mathbf{x,}t\right) =e^{-iH^{(0-loop)}t}%
\underline{\psi }^{\prime }\left( \mathbf{x,}0\right)
e^{iH^{(0-loop)}t}=e^{-iH_{W}^{(0-loop)}t}\underline{\psi }^{\prime }\left( 
\mathbf{x,}0\right) e^{iH_{W}^{(0-loop)}t},  \tag{2.4.17}
\end{equation}

\begin{equation}
\underline{A}_{\mu }^{\prime }\left( \mathbf{x,}t\right) =e^{-iH^{(0-loop)}t}%
\underline{A}_{\mu }^{\prime }\left( \mathbf{x,}0\right)
e^{iH^{(0-loop)}t}=e^{-iH_{W}^{(0-loop)}t}\underline{A}_{\mu }^{\prime
}\left( \mathbf{x,}0\right) e^{iH_{W}^{(0-loop)}t},  \tag{2.4.18}
\end{equation}%
(2.4.11)-(2.4.12) and (2.4.15)-(2.4.16) are the same as those of the
conventional $QED$, (2.4.13)-(2.4.14) and (2.4.17)-(2.4.18) are different
from those of the conventional $QED.$ It is seen from the $\mathcal{L}_{F}$
we can also obtain the results of the conventional $QED$.

\subsection{Scattering operators and Feynman rules}

As the conventional $QED$, transforming the Heisenberg picture into the
interaction picture, we can derive the scattering operactior $S_{f}$ and $%
S_{w}$ from (2.4.2)-(2.4.3), (2.4.5)-(2.4.6), (2.4.7) and (2.4.8), 
\begin{equation}
S_{f}=1+\sum_{n=1}^{\infty }\frac{\left( -i\right) ^{n}}{n!}\int_{-\infty
}^{\infty }d^{4}x_{1}..\int_{-\infty }^{\infty }d^{4}x_{n}T:\mathcal{H}%
_{FI}\left( g_{,}m_{e,}x\right) ..\mathcal{H}_{FI}\left( g_{,}m_{e,}x\right)
\tag{2.5.1}
\end{equation}%
\begin{equation}
S_{w}=1+\sum_{n=1}^{\infty }\frac{i^{n}}{n!}\int_{-\infty }^{\infty
}d^{4}x_{1}...\int_{-\infty }^{\infty }d^{4}x_{n}T:\mathcal{H}_{WI}\left(
g_{,}m_{e,}x\right) ...\mathcal{H}_{WI}\left( g_{,}m_{e,}x\right) 
\tag{2.5.2}
\end{equation}%
\begin{equation}
S_{f}^{(0)}\left( g_{0},m_{e0}\right) =1+\sum_{n=1}^{\infty }\frac{\left(
-i\right) ^{n}}{n!}\int_{-\infty }^{\infty }d^{4}x_{1}...\int_{-\infty
}^{\infty }d^{4}x_{n}T:\mathcal{H}_{FI}^{(0)}...\mathcal{H}_{FI}^{(0)} 
\tag{2.5.3}
\end{equation}%
\begin{equation}
S_{w}^{(0)}\left( g_{0},m_{e0}\right) =1+\sum_{n=1}^{\infty }\frac{i^{n}}{n!}%
\int_{-\infty }^{\infty }d^{4}x_{1}...\int_{-\infty }^{\infty }d^{4}x_{n}T:%
\mathcal{H}_{WI}^{(0)}...\mathcal{H}_{WI}^{(0)}  \tag{2.5.4}
\end{equation}%
\begin{equation}
\mathcal{H}_{FI}=-i\overline{\psi }\gamma _{\mu }\cdot G_{F}\cdot A_{\mu
}\psi +\overline{\psi }\cdot M_{F}\cdot \psi ,  \tag{2.5.5}
\end{equation}

\begin{equation}
\mathcal{H}_{WI}=-i\overline{\underline{\psi }}\gamma _{\mu }\cdot G_{W}%
\underline{A}_{\mu }\underline{\psi }-\underline{\overline{\psi }}\cdot
M_{W}\cdot \underline{\psi },  \tag{2.5.6}
\end{equation}%
\begin{equation}
\mathcal{H}_{FI}^{(0)}=\mathcal{H}_{FI}^{(0-loop)}\left(
g_{0,}m_{e0,}x\right) =-ig_{0}:\psi ^{\prime +}\gamma _{4}\gamma _{\mu
}A_{\mu }^{\prime }\psi ^{\prime }:,  \tag{2.5.7}
\end{equation}%
\begin{equation}
\mathcal{H}_{WI}^{(0)}=\mathcal{H}_{WI}^{(0-loop)}\left(
g_{0,}m_{e0,}x\right) =-ig_{0}:\underline{\psi }^{^{\prime }+}\gamma
_{4}\gamma _{\mu }\underline{A}_{\mu }^{\prime }\underline{\psi }^{^{\prime
}}:,  \tag{2.5.8}
\end{equation}%
where\ all field operators in (2.5.5)-(2.5.8)$\ $satisfy free-field equtions
as the conventional $QED$, $\ G_{F}$ and $M_{F}$ \ are respectively
determined by (2.3.10) and (2.3.26), $G_{W}$ and $M_{W}$ \ are respectively
determined by (2.3.15) and (2.3.29), the symbol $T$ denotes that the
products are in time-ordered form. Replacing $g\left( p_{1,}p_{2}\right) $
and $m_{e}\left( p_{1}\right) $ in $\mathcal{H}_{FI}$ and $\mathcal{H}_{WI}$
by $g_{0}$ and $m_{e0}=0,$ reapectively, we obtain $\mathcal{S}%
_{f}^{(0)}\left( g_{0,}m_{e0}\right) $ and $\mathcal{S}_{w}^{(0)}\left(
g_{0,}m_{e0}\right) $ \ from $\mathcal{S}_{f}\left( g_{,}m_{e}\right) $ and $%
\mathcal{S}_{w}\left( g_{,}m_{e}\right) .$ If for arbitrary momenta $p_{1}$
and $p_{2}$ \ we could determine the accurate coupling constants $g\left(
p_{1},p_{2}\right) $ and the masses $m_{e}\left( p\right) ,$ we could
evaluate accurate scattering amplitudes by only tree diagrams determined by $%
S_{f}\left( g,m_{e}\right) $ or $S_{w}\left( g,m_{e}\right) $. But this is
impossible. It should be pointed that\ in the meaning of perturbation
theory, for initial states and final states with given momenta, e.g., the
momemta at the subtraction point, we can give the absolutely precise
coupling coefficients and masses, and from this give absolutely precise $%
\mathcal{L}_{F}$ , $\mathcal{L}_{W},$ $\ \mathcal{H}_{F}^{(0-loop)},$ $%
\mathcal{H}_{W}^{(0-loop)},$ $\mathcal{S}_{f}^{(0)}$ and $\mathcal{S}%
_{w}^{(0)}.$ But for arbitrary initial states and final states, we cannot
give the absolutely precise coupling coefficients, masses, $\mathcal{L}_{F}$
, $\mathcal{L}_{W},$ $\ \mathcal{H}_{F},$ $\mathcal{H}_{W},$ $\mathcal{S}%
_{f} $ and $\mathcal{S}_{w}.$\ In fact the coupling coefficients and masses\
will be corrected by n-loop diagrams,\ hence for arbitrary initial states
and final states we can only give approximate $\mathcal{L}_{F}$ , $\mathcal{L%
}_{W},$ $\ \mathcal{H}_{F},$ $\mathcal{H}_{W},$ $\mathcal{S}_{f}$ and $%
\mathcal{S}_{w}.$Of course, by such $\mathcal{S}_{f}^{(0)}$ and $\mathcal{S}%
_{w}^{(0)}$ we can obtain scattering amplitudes approximate to arbitrary
n-loop diagrams. Thus, the two scattering operators $S_{f}$ and $S_{w}$ are
not practical to evaluate scattering amlitudes. Therefore we can evaluate
Feynman amplitudes by only $S_{f}^{(0)}$ \ and $S_{w}^{(0)}$ and pertubtion
approximation. The scattering operator $S_{f}^{(0)}\left(
g_{0},m_{e0}\right) $ is the same as the scattering operator of the
conventional QED. $S_{w}^{(0)}$ differ from $S_{f}^{(0)}$ only in the sign
before $i$ in form. Hence we can evaluate scattering amplitudes by $%
S_{f}^{(0)}$ \ and $S_{w}^{(0)}$ in the same method as is used by the
conventional QED

In order to simplify $S_{f}^{(0)}$ and $S_{w}^{(0)},$ we must decompose the
time-ordered products appearing in $S_{f}^{(0)}$ and $S_{w}^{(0)}$ into
normal-ordered products. Considering the expansions of $\psi ^{\prime
}\left( x\right) $ and $A_{\mu }^{\prime }\left( x\right) $ to be the same
as those of the free fields of the conventional $QED$ and $\underline{A}%
_{\mu }^{\prime }\left( x\right) $ and $\underline{\psi ^{\prime }}\left(
x\right) $ to have the similar expansions, as the conventional $QED,$ we have%
\begin{eqnarray}
\langle 0 &\mid &T\psi _{\alpha }^{\prime }\left( x_{1}\right) \overline{%
\psi }_{\beta }^{\prime }\left( x_{2}\right) \mid 0\rangle  \notag \\
&=&\frac{-i}{\left( 2\pi \right) ^{4}}\int d^{4}p\frac{\left( m-ip\gamma
\right) _{\alpha \beta }}{p^{2}+m^{2}-i\varepsilon }e^{ip\left(
x_{1}-x_{2}\right) }\equiv S_{Ff}\left( x_{1}-x_{2}\right) ,  \TCItag{2.5.9}
\end{eqnarray}

\begin{eqnarray}
\langle 0 &\mid &TA_{\mu }^{\prime }\left( x_{1}\right) A_{\nu }^{\prime
}\left( x_{2}\right) \mid 0\rangle  \notag \\
&=&\delta _{\mu \nu }\frac{-i}{\left( 2\pi \right) ^{4}}\int d^{4}k\frac{1}{%
k^{2}-i\varepsilon }e^{ik\left( x_{1}-x_{2}\right) }\equiv \delta _{\mu \nu
}D_{Ff}\left( x_{1}-x_{2}\right) ,  \TCItag{2.5.10}
\end{eqnarray}

\begin{eqnarray}
\langle 0 &\mid &T\underline{\psi }_{\alpha }^{\prime }\left( x_{1}\right) 
\underline{\overline{\psi }}_{\beta }^{\prime }(x_{2})\mid 0\rangle  \notag
\\
&=&\frac{-i}{\left( 2\pi \right) ^{4}}\int d^{4}p\frac{\left( m-ip\gamma
\right) _{\alpha \beta }}{p^{2}+m^{2}+i\varepsilon }e^{ip\left(
x_{1}-x_{2}\right) }\equiv S_{Fw}\left( x_{1}-x_{2}\right) ,  \TCItag{2.5.11}
\end{eqnarray}

\begin{eqnarray}
\langle 0 &\mid &T\underline{A}_{\mu }^{\prime }\left( x_{1}\right) 
\underline{A}_{\nu }^{\prime }\left( x_{2}\right) \mid 0\rangle  \notag \\
&=&\delta _{\mu \nu }\frac{i}{\left( 2\pi \right) ^{4}}\int d^{4}k\frac{1}{%
k^{2}+i\varepsilon }e^{ik\left( x_{1}-x_{2}\right) }\equiv \delta _{\mu \nu
}D_{Fw}\left( x_{1}-x_{2}\right) ,  \TCItag{2.5.12}
\end{eqnarray}%
where $\varepsilon \rightarrow 0^{+},$ the integral contours are shown in
Fig.2.$1.$ The other contractions are all equal to zero. Comparing
(2.5.11)-(2.5.12) with (2.5.9)-(2.5.10), we find the differences and the
common properties among the four propagators as follows.

1. The places of the polar points of $D_{Fw\text{ }}$ and $S_{Fw}$ are
opposite to those of $D_{Fw\text{ }}$ and $S_{Fw}$, respectively. The polar
points of $D_{Fw\text{ }}$ and $S_{Fw}$ are in the first and third
quadrants, the polar points of $D_{Ff\text{ }}$ and $S_{Ff\text{ }}$ are in
the second and the fourth quadrants (see Fig.2.1.1 and Fig.2.1.2).

2. Neglecting $\varepsilon ,D_{Ff}=-D_{Fw}$ and $S_{Ff}=S_{Fw}$.

It can be seen that the Wick theorem still applicable to the present theory.
Decomposing the time-ordered products appearing in $(2.5.3)$ and$\ \left(
2.5.4\right) $ into normal-ordered products, we can evaluate S-matrix
elements approximate to tree-diagrams$.$ It is readily proven the $S$-matrix
elements $\langle F_{f}\mid S_{f}^{(0)}\left( g_{0},m_{e0}\right) \mid
F_{i}\rangle $ approximate to tree-diagrams are the same as those obtained
by the conventional QED. From the $S$-matrix elements $\langle F_{f}\mid
S_{f}\left( g_{0},m_{e0}\right) \mid F_{i}\rangle $ and $\langle W_{f}\mid
S_{w}\left( g_{0},m_{e0}\right) \mid W_{i}\rangle $ approximate to
tree-diagrams we can obtain the Feynman rules in the momentum space (see
Fig. 2.2.1 and Fig.2.2.2).

In addition to the rules given by Fig. 2.1 and Fig.2.2.1-2, there is an
overall minus sign corresponding to a closed fermion loop and three
supplimentary Feynman rules (see the following chapter). The rules fully
determine the method evaluating a F-amplitude or a W-amplitude.

\subsection{Summary}

On the basis of the first chapter we discuss quantization for interacting
fields, derive Feynman rules and scattering operators. Since we quantize
fields by the transformation operators replacing creation and annihilation
operators in the conventional QED, it is necessary to replace coupling
constants and masses by coupling operators and mass operators in the present
theory. In contrast with the conventional QED, in the present theory, there
is only one sort of parameters which are all finite and measurable and the
two sorts of corrections originating $S_{f\text{ }}$ and $S_{w}.$ We will
see that the integrands causing divergence in the two sorts of correction
cancel each other out in the following.

\section{One-loop Correction and Supplementary Feynman Rules}

\subsection{Introduction}

We have constructed a new Lagrangian density of free fields and quantized
the free field by the transformation operators in the first chapter, have
discussed quantization for interacting fields and have derived the
scattering operators $S_{f\text{ \hspace{0in}}},$ $S_{w}$ \ and the Feynman
rules in second chapter. In present paper we evaluate one-loop correction to
the coupling coefficients and the masses in detail, derive the method
evaluating n-loop correction.

If $g_{0}$ $\ $and $m_{e0}$ are regarded constants, $\mathcal{L}_{F}$ $\ $and%
$\ \mathcal{L}_{W}$ \ were fully independent. In this case we can only
obtain one new result i.e., $\langle 0\mid H\mid 0\rangle =0.$ But in fact,
after quantization, though there is no coupling between the field operators $%
\psi $ and $A_{\mu }$ in $\mathcal{L}_{F}$ and the field operators 
\underline{$\psi $} and \underline{$A$}$_{\mu }$ in $\mathcal{L}_{W}$, $%
\mathcal{L}_{F}$ $\ $and$\ \mathcal{L}_{W}$ \ are dependent on each other,
since $g_{f}(p_{2},p_{1})$ $\ $and $m_{ef}(p_{2},p_{1})$ \ are determined by 
$\mathcal{L}_{W}$ and $g_{w}(p_{2},p_{1})$ $\ $and $m_{ew}(p_{2},p_{1})$ \
are determined by $\mathcal{L}_{F}$. $\mathcal{S}_{f}$ $\ $and$\ \mathcal{S}%
_{w}$ together determine scattering amplitudes. Thus, all scattering
amplitudes will be convergent and it is unnecessary to introduce
regularization and counterterms.

\subsection{Two sorts of correction}

When the momenta meeting at a single vertex $p_{2}=q_{2},$ $p_{1}=q_{1},$
scattering amplidutes $A_{gf}^{(0-loop)}(q_{2},q_{1}),$ $%
A_{gw}^{(0-loop)}(q_{2},q_{1})$, $A_{mf}^{(0-loop)}(q_{1})$ and $%
A_{mf}^{(0-loop)}(q_{1})$ evaluated by $S_{f}^{(0)}\left(
g_{0},m_{e0}\right) $ and $S_{w}^{(0)}\left( g_{0},m_{e0}\right) $ with tree
diagrams are all accurate (it is possible that there is no real process for
a single vertex). $g_{0}$ and $m_{e0}$ are determined on the basis of
experiments and are accurate. When $p_{2}\neq q_{2},$ $p_{1}\neq q_{1}$ or
for such a Feynman diagram with many vertices, $g_{0}$, $m_{e0}$ and the
scattering amplidutes obtained by $S_{f}^{(0)}$ and $S_{w}^{(0)}$ with tree
diagrams are no longer accurate and must be corrected. According to the
perturbation theory, Feynman diagrams with the same initial state $\mid a_{%
\mathbf{p}_{1}s_{1}}\rangle \mid c_{\mathbf{k}\lambda }\rangle $ and final
state $\mid a_{\mathbf{p}_{2}s_{2}}\rangle ,$ no matter whether or not $%
p_{1}=q_{1\text{ \ }}$and $p_{2}=q_{2},$ ($q_{1}$ and $q_{2}$ are the
momenta at the subtraction point), the tree diagrams and all diagrams with
n-loop must exist and must simultaneously be considered. It is obvious that
corrections must begin with one--loop diagrams. The coupling constants, the
masses and scattering amplitudes corrected to n-loop diagrams can be written
as

\begin{equation}
Z^{\left( n-loop\right) }\left( p_{2},p_{1},k\right)
=Z_{0}+\sum_{i=1}^{n}Z^{\left( i\right) }\left( p_{2},p_{1},k\right) ,\text{%
\hspace{0in} }  \tag{3.2.1}
\end{equation}%
where $Z=g_{f},$ $g_{w},$\hspace{0in} \hspace{0in}$m_{ef},$ \hspace{0in}$%
m_{ew},\;A_{gf},\;A_{mf},\;A_{gw},\;A_{mw},$ the superscript $i$ denotes
only the $ith-loop$ diagrams to be evaluated. Because of the gauge
invariance of the Lagrangian density, there is no correction of the mass of
a free F-photon or W-photon.

We consider that a self-consistent theory should satisfy the following
demands:

1. There is only one sort of physical parameters which are measurable and
finite, e.g., there are only $g_{0}$ and $m_{e0}$ in the present theory,\
and there is no other sort of parameters as the bare mass and the bare
charge in the conventional $QED$\ \ which are unmeasured and divergent.

2. $Z^{\left( n-loop\right) }\left( p_{2},p_{1},k\right) $ and $Z^{\left(
i\right) }\left( p_{2},p_{1},k\right) $ must be finite and must tend to zero
as $p_{2}\rightarrow q_{2},$ $p_{1}\rightarrow q_{1}$ and $k\rightarrow
q^{\prime },$ i.e., 
\begin{equation}
Z^{\left( i\right) }\left( q_{2},q_{1},q^{\prime }\right) =0,\text{ \hspace{%
0in} \hspace{0in} \hspace{0in}  \hspace{0in} \hspace{0in} \hspace{0in} 
\hspace{0in} \hspace{0in} \hspace{0in}}Z^{\left( n-loop\right) }\left(
q_{2},q_{1},q^{\prime }\right) =Z_{0}.  \tag{3.2.2}
\end{equation}

3. $Z^{\left( n-loop\right) }\left( p_{2},p_{1},k\right) $ should be
consistent with the results obtained by the given QED.

According to the conventional $QED,$ the 1-loop corrections to $%
g_{f0}(q_{2},q_{1})$ \ and $m_{ef0}(q)$ \ are all divergent. This is because
there is only one sort of corrections originating $S_{f}.$ In order to
remove divergence, it is necessary to introduce regularization and
counterterms. For instance, in order to remove divergence of the one-loop
correction of the mass of a free electron

\begin{equation*}
\delta m=m_{eff}^{(1)}(q)=\frac{3\alpha }{4\pi ^{2}}m\left( \frac{1}{i\pi
^{2}}\int d^{4}k\frac{1}{(k^{2}+m^{2})^{2}}+\frac{3}{2}\right) \equiv A,
\end{equation*}%
the counterterm $A\overline{\psi }\psi $ must be introduced, here $A$ is a
divergent constant. Thus, there are two sorts of parameters in the
conventional QED, i. e., the physical mass and charge and the so-called bare
mass and bare charge. Both bare mass and bare charge are divergent and
unmeasured. It is noteworthy that in order to determine $A$ we must firstly
carry out integrantion over $k.$ Thus regularization must be introduced$.$

As the conventional $QED,$ it can be proven from the Feynman rules that the
one-loop corrections only originating from $S_{f}^{(0)}\left(
g_{0},m_{e0}\right) $ or $S_{w}^{(0)}\left( g_{0},m_{e0}\right) $ are all
divergent and do not tend to zero as $p_{2}\rightarrow q_{2}$ and $%
p_{1}\rightarrow q_{1}.$ This implies that there must be another sort of
corrections which and the first sort of corrections must cancel each other
out as $p_{2}\rightarrow q_{2},$ $p_{1}\rightarrow q_{1}$ and $k\rightarrow
q^{\prime }=q_{2}-q_{1}$ and the sum of the two corrections must be finite.
In contrast with the conventional $QED,$ in the present theory, there is
only one sort of parameters ( $g_{0}$ and $m_{e0}$) which are finite and
measurable and two sorts of correction originating from $S_{f}$ and $S_{w}.$ 
$\mathcal{S}_{f}$ $\ $and$\ \mathcal{S}_{w}$ together determine $Z^{\left(
n-loop\right) }\left( p_{2},p_{1},k\right) $ and $Z^{\left( i\right) }\left(
p_{2},p_{1},k\right) $ which satisfy the demands above.

We first take one-loop correction as an example to discuss the first sort of
correction. As the conventional QED, there are one-loop corrections $%
A_{gff}^{(1)}(p_{2},p_{1})$ to $A_{gf}^{(0-loop)}(p_{2},p_{1})$ and $%
A_{mff}^{(1)}(p_{2},p_{1})$ to $A_{mf}^{(0-loop)}(p_{2},p_{1})$ originating
from $S_{f}^{(0)}\left( g_{0},m_{e0}\right) $, and one-loop corrections $%
A_{gww}^{(1)}(p_{2},p_{1})$ to $A_{gw}^{(0-loop)}(p_{2},p_{1})$ \ and $%
A_{mww}^{(1)}(p_{2},p_{1})$ \ to $A_{mw}^{(0-loop)}(p_{2},p_{1})$
originating from $S_{w}^{(0)}\left( g_{0},m_{e0}\right) $. From the
definition (2.3.1) we see that $A_{gff}^{(1)}$ firstly corrects the
amplitude $A_{gf}^{(0)}(p_{2},p_{1})$ and thereby corrects $g_{f0},$ hence
the momenta of $A_{gff}^{(1)}(p_{2},p_{1})$ and $g_{ff}^{(1)}(p_{2},p_{1})$
must be the same as those of $A_{gf}^{(0)}(p_{2},p_{1}).$ We can evaluate $%
g_{ff}^{(1)}(p_{2},p_{1})$ by $S_{f}^{\left( 1\right) }\left(
g_{0},m_{e0}\right) $ as the conventional QED. Thus from (2.3.1) we have%
\begin{equation}
g_{ff}^{(1)}=g_{ff}^{(1)}(p_{2},p_{1})=\frac{\langle a_{p_{2}s_{2}}\mid
S_{f}^{(1)}(g_{0},m_{e0})\mid a_{p_{1}s_{1}}c_{k\lambda }\rangle }{\langle
a_{p_{2}s_{2}}\mid S_{f}^{(0)}(g_{0}=1,m_{e0}=0)\mid
a_{p_{1}s_{1}}c_{k\lambda }\rangle },  \tag{3.2.3}
\end{equation}%
where $p$ may be a momentum of an internal line and does not satisfy the
mass shell restriction, $S_{f}^{(1)}$ is the scattering operator approximate
to one-loop. Analogously to $g_{ff}^{(1)},$ from (2.3.17), (2.3.4) and
(2.3.20) we have%
\begin{equation}
m_{eff}^{(1)}(p_{1})=\frac{\langle a_{p_{2}s_{2}}\mid
S_{f}^{(1)}(g_{0},m_{e0})\mid a_{p_{1}s_{1}}\rangle }{\langle
a_{p_{2}s_{2}}\mid S_{f}^{(0)}(g_{0}=1,m_{e0}=0)\mid a_{p_{1}s_{1}}\rangle },
\tag{3.2.4}
\end{equation}%
\begin{equation}
g_{ww}^{(1)}(p_{2},p_{1})=\frac{\langle \underline{a}_{p_{2}s_{2}}\mid
S_{w}^{(1)}(g_{0},m_{e0})\mid \underline{a}_{p_{1}s_{1}}\underline{c}%
_{k\lambda }\rangle }{\langle \underline{a}_{p_{2}s_{2}}\mid
S_{w}^{(0)}(g_{0}=1,m_{e0}=0)\mid \underline{a}_{p_{1}s_{1}}\underline{c}%
_{k\lambda }\rangle },  \tag{3.2.5}
\end{equation}%
\begin{equation}
m_{eww}^{(1)}(p_{1})=\frac{\langle \underline{a}_{p_{2}s_{2}}\mid
S_{w}^{(1)}(g_{0},m_{e0})\mid \underline{a}_{p_{1}s_{1}}\rangle }{\langle 
\underline{a}_{p_{2}s_{2}}\mid S_{w}^{(0)}(g_{0}=1,m_{e0}=0)\mid \underline{a%
}_{p_{1}s_{1}}\rangle }.  \tag{3.2.6}
\end{equation}%
where the denominators of (3.2.3)-(3.2.6) are the same as (2.3.2), (2.3.18),
(2.3.5) and (2.3.21), respectively. (3.2.3)-(3.2.6) are the first sort of
corrections.

Secondly, we take one-loop correction as an example to consider the second
sort of one-loop corrections. It is seen from (2.3.9) that $A_{gww}^{(1)}$
originating from $S_{w}$ not only corrects $A_{gw}^{(0-loop)}$ $\ $and $%
g_{w0},$ but also corrects $g_{f0}$ and $m_{ef0}.$ We denote the sort of
one-loop correction to $g_{f0}$ originating from $S_{w}$ by \ $g_{fw}^{(1)}$%
. In contrast with the first sort of corrections, the corrections of the
second sort are to directly correct the coupling constants and the masses.
Because $g_{0}(q_{2},q_{1})$ and $m_{e0}(q)$ are defined at the subtraction
point, the second sort of corretions must also be defined at the subtraction
point, i.e. $g_{fw}^{(1)}=g_{fw}^{(1)}(q_{2},q_{1})$. Considering%
\begin{equation}
g_{f}^{(1)}(p_{2},p_{1})\equiv
g_{ff}^{(1)}(p_{2},p_{1})+g_{fw}^{(1)}(q_{2},q_{1})  \tag{3.2.7}
\end{equation}%
should be finite, from (2.3.9) and (3.2.5) we define 
\begin{equation}
g_{fw}^{(1)}=g_{fw}^{(1)}(q_{2},q_{1})\equiv -g_{ww}^{(1)}(q_{2},q_{1}), 
\tag{3.2.8}
\end{equation}%
Analogously to $g_{fw}^{(1)},$ from (2.3.25) and (3.2.6) we define 
\begin{equation}
m_{efw}^{(1)}=m_{efw}^{(1)}(q)\equiv -m_{eww}^{(1)}(q),\;\ \ \ \ \
m_{ef}^{(1)}(p)\equiv m_{eff}^{(1)}(p)+m_{efw}^{(1)}(q).  \tag{3.2.9}
\end{equation}%
Analogously to $g_{fw}^{(1)}$ and $m_{efw}^{(1)},$ there are the corrections 
$g_{wf}^{(1)}$ and $m_{ewf}^{(1)}$ originating from $S_{f}$ to $g_{w0}$ and $%
m_{ew0}.$ From (2.3.14), (3.2.3), (2.3.28) and (3.2.4) we define 
\begin{equation}
g_{wf}^{(1)}=g_{wf}^{(1)}(q_{2},q_{1})\equiv -g_{ff}^{(1)}(q_{2},q_{1}),\ \
g_{w}^{(1)}(p_{2},p_{1})=g_{ww}^{(1)}(p_{2},p_{1})+g_{wf}^{(1)}(q_{2},q_{1}),
\tag{3.2.10}
\end{equation}%
\begin{equation}
m_{ewf}^{(1)}=m_{ewf}^{(1)}(q)\equiv -m_{eff}^{(1)}(q),\;\ \ \ \ \ \ \ \ \
m_{ew}^{(1)}(p)\equiv m_{eww}^{(1)}(p)+m_{ewf}^{(1)}(q).  \tag{3.2.11}
\end{equation}%
(3.2.8)-(3.2.11) are the second sort of 1-loop corrections. (3.2.3)-(3.2.11)
can be generalized to n-loop diagrams. Since there must be the two sorts of
corrections simultaneously, both must simultaneously be considered and be
approximated to the same loop-number, i.e., after their integrands are added
up, the corresponding integral is completed. Thus the integrands causing
divergences will cancel each other out, consequently $\langle F_{f}\mid
S_{f}\mid F_{i}$ $\rangle $ and $\langle W_{f}\mid S_{w}\mid W_{i}\rangle $
will be all convergent, and it is unnecessary to introduce regularization
and counterterms. For example, in the new QED, we have (1.1.1). From
(3.2.7)-(3.2.11) we see that if let $g_{fw}^{(1)}=g_{fw}^{(1)}(p_{2},p_{1}),$
the total sum of corrections of the first and second sorts must be zero.
This is equivalent to redefine the point with the momenta $p_{2}$ and $p_{1}$
as the subtraction point and $g_{f}(p_{2},p_{1})$ is regarded as an accurate
value. In fact, $g_{f}(p_{2},p_{1})$ is unknown, and our aim is to evaluate
an approximate $g_{f}(p_{2},p_{1})$ from the accurate $g_{0}(q_{2},q_{1})$
and $m_{e0}(q_{1})$. In meaning of perturbation theory, $g_{0}(q_{2},q_{1})$
as a accurate value already contains $g_{ff}^{(1)}(q_{2},q_{1})$ and $%
g_{fw}^{(1)}(q_{2},q_{1})$, i.e., $g_{f}^{(1)}(q_{2},q_{1})=$ $%
g_{ff}^{(1)}(q_{2},q_{1})+g_{fw}^{(1)}(q_{2},q_{1}).$ From the F-W symmetry
and (3.2.7)-(3.2.11) we can prove $g_{f}^{(1)}(q_{2},q_{1})=0.$ Thus when $%
g_{ff}^{(1)}(q_{2},q_{1})$ become $g_{ff}^{(1)}(p_{2},p_{1}),$ $%
g_{fw}^{(1)}(q_{2},q_{1})$ appears. Anologous analysis holds water for $%
m_{efw}^{(1)}(q)$, $g_{wf}^{(1)}(q_{2},q_{1})$ and $m_{ewf}^{(1)}(q)$ and 
\begin{equation}
g_{w}^{(1)}(q_{2},q_{1})=m_{ew}^{(1)}(q_{1})=m_{ef}^{(1)}(q_{1})=0. 
\tag{3.2.12}
\end{equation}

For a free particle, $p=q$ or $k=q^{\prime }$ \ and $%
m_{ew}^{(1)}(q)=m_{ef}^{(1)}(q)=0.$ When $p\neq q$ , $m_{ef}^{\left(
1\right) }\left( p\right) $ $=m_{ew}^{\left( 1\right) }\left( p\right) \neq
0.$ In this case $m_{ef}^{\left( 1\right) }\left( p\right) $ or $%
m_{ew}^{\left( 1\right) }\left( p\right) $ must join one or two vertexes,
hence we can regard $m_{ef}^{\left( 1\right) }\left( p\right) $ or $%
m_{ew}^{\left( 1\right) }\left( p\right) $ as a part of corrections of the
vertexes. We regard specially a correction to a propagator as the sum of
partial corrections to the two vertexes joining the propagator. The
correction to every vertex joining the propagator is the same, hence this
correction is half of the correction to the propagator.\ On the other hand,
we can also regard correction to an external line as partial correction to a
vertex. We can prove that these results hold water for n-loop diagrams (see
section 3.5). Thus, in present theory there is no correction to an external
line and a propagator and there is only correction to a vertex.

As mentioned above, we see that $\psi ^{\prime }$ and $A_{\mu }^{\prime }$
are the same as $\psi $ and $A_{\mu }$ in the conventional $QED$.
Formulating the new $QED$ by $\psi ^{\prime }$ , $A_{\mu }^{\prime }$, 
\underline{$\psi $}$^{\prime }$ and \underline{$A$}$_{\mu }^{\prime }$ , the
new $QED$ will be the same as the conventional $QED$ in form. The essential
differences between the new $QED$ and the conventional $QED$ are that the
Lagrangian density of the conventional $QED$ is replaced by $\mathcal{L=L}%
_{F}\mathcal{+L}_{W}$ \ and $\mathcal{L}_{F}$ and $\mathcal{L}_{W}$ together
determine correction to a scattering amplitude $\langle F_{f}\mid S_{f}\mid
F_{i}$ $\rangle $ or $\langle W_{f}\mid S_{w}\mid W_{i}\rangle $. Thus,
though $\mathcal{L}_{F}$ $\ $and $\mathcal{L}_{W}$ \ are independent of each
other in classical meanings, after quantization both are dependent on each
other.

\subsection{The first sort of 1-loop corrections}

\subsubsection{The first sort of one-loop  corrections originating  from  $%
S_{f}\left( g_{0},m_{e0}\right) .$}

\ By $S_{f}\left( g_{0},m_{e0}\right) $ ((2.5.3)) we can evaluate the 1-loop
correction to the scattering amplitude and the electromagnetic mass $m_{e0}$
of a free F-electron corresponding to figure 3.1.1-2$^{\left[ 8\right] }$ ,

\begin{eqnarray}
A_{mff}^{\left( 1\right) }\left( q_{1}\right) &=&\langle a_{\mathbf{q}%
_{2}s_{2}}\mid S_{f}^{\left( 1\right) }\left( g_{0},m_{e0}\right) \mid a_{%
\mathbf{q}_{1}s_{1}}\rangle  \notag \\
&=&\delta ^{4}\left( q_{2}-q_{1}\right) \frac{1}{V}\overline{u}_{\mathbf{q}%
_{2}s_{2}}\Sigma _{ff}^{\left( 1\right) }\left( q_{1}\right) u_{\mathbf{q}%
_{1}s_{1}},  \TCItag{3.3.1}
\end{eqnarray}%
\begin{eqnarray}
\Sigma _{ff}^{\left( 1\right) }\left( q_{1}\right) &=&g_{0}^{^{2}}\int d^{4}k%
\frac{-i}{k^{2}-i\varepsilon }r_{\mu }\frac{-i\left[ m-i\left(
q_{1}-k\right) \gamma \right] }{\left( q_{1}-k\right)
^{2}+m^{2}-i\varepsilon }r_{\mu }  \notag \\
&=&-i\left( 2\pi \right) ^{4}A+BS_{Ff}^{-1}\left( q_{1}\right)
+S_{Ff}^{-1}\left( q_{1}\right) \Sigma _{fc}^{\left( 1\right) }\left(
q_{1}\right) S_{Ff}^{-1}\left( q_{1}\right) ,  \TCItag{3.3.2}
\end{eqnarray}

\begin{equation}
A\equiv \frac{\alpha }{2\pi }m\left( \frac{3}{2}D+\frac{9}{4}\right) ,\;\ \
\ D\equiv \frac{1}{i\pi ^{2}}\int d^{4}k\frac{1}{(k^{2}+m^{2})^{2}}, 
\tag{3.3.3}
\end{equation}%
\begin{equation}
B\equiv -\frac{\alpha }{4\pi }\left( D-4\int_{0}^{1}\frac{dx}{x}+\frac{11}{2}%
\right) ,  \tag{3.3.4}
\end{equation}%
\begin{equation}
\overline{u}_{\mathbf{q}_{2}s_{2}}S_{Ff}^{-1}\left( q_{1}\right)
=S_{Ff}^{-1}\left( q_{1}\right) u_{\mathbf{q}_{1}s_{1}}=\Sigma _{fc}^{\left(
1\right) }\left( q_{1}\right) =0.  \tag{3.3.5}
\end{equation}%
From (3.2.4) we have 
\begin{equation}
m_{eff}^{\left( 1\right) }\left( q\right) =\frac{\langle a_{\mathbf{q}%
_{2}s_{2}}\mid S_{f}^{\left( 1\right) }\left( g_{0},m_{e0}\right) \mid a_{%
\mathbf{q}_{1}s_{1}}\rangle }{\langle a_{\mathbf{q}_{2}s_{2}}\mid -i\int
d^{4}x:\overline{\psi ^{\prime }}\psi ^{\prime }:\mid a_{\mathbf{q}%
_{1}s_{1}}\rangle }=A  \tag{3.3.6}
\end{equation}%
where $D$ is divergent and $\alpha =g_{0}^{2}/4\pi .$ The 1-loop correction
of the F-electron propagator with its two vertices shown in figure 3.2.1 is

\begin{equation}
(-g_{0}\gamma _{\mu })S_{Fff}^{\left( 1\right) }\left( p\right)
(-g_{0}\gamma _{\nu })=g_{0}^{2}\gamma _{\mu }S_{Ff}\left( p\right) \Sigma
_{f}^{\left( 1\right) }\left( p\right) S_{Ff}\left( p\right) \gamma _{\nu },
\tag{3.3.7}
\end{equation}%
where the factors $\left( 2\pi \right) ^{4}\delta \left(
p_{2}-p+k_{2}\right) $ and $\left( 2\pi \right) ^{4}\delta \left(
p-p_{1}-k_{1}\right) $ are ignored$,$ $\Sigma _{fc}^{\left( 1\right) }\left(
p\right) $ is finite, $B$ is divergent, and $S_{Ff}^{-1}\left( p\right)
=i\left( 2\pi \right) ^{4}\left( ip\gamma +m\right) .$ But it is not
necessary to evaluate the integral, because the integrand of $D$ \ will be
cancelled out by the other integrand in the total scattering amplitude.
Because a propagator must join its two vertexes, $(3.3.7)$ may also be
regarded as the sum of corrections to the two vertices shown in figure
3.2.1. The contributions of the two diagrams are the same. Thus, we obtain
the first sort of one-loop corrections $\Lambda _{ff1\text{ }}^{\left(
1\right) }$to a F-vertex to be

\begin{equation}
\Lambda _{ff1\mu \text{ }}^{\left( 1\right) L}\left( p\right) \equiv -\frac{1%
}{2}g_{0}\gamma _{\mu }S_{Ff}\left( p\right) \Sigma _{f}^{\left( 1\right)
}\left( p\right) ,  \tag{3.3.8}
\end{equation}%
\begin{equation}
\text{ }\Lambda _{ff1\nu \text{ }}^{\left( 1\right) R}\left( p\right) \equiv
-\frac{1}{2}\Sigma _{f}^{\left( 1\right) }\left( p\right) S_{Ff}\left(
p\right) g_{0}\gamma _{\nu }.  \tag{3.3.9}
\end{equation}%
The diagrams corresponding to $\left( 3.3.8\right) $ \hspace{0in} and $%
\left( 3.3.9\right) $ are shown in figure 3.2.1L and 3.2.1R. We always
regard $\Lambda _{ff1\mu \text{ }}^{\left( 1\right) L}\left( p\right) $ and $%
\Lambda _{ff1\nu \text{ }}^{\left( 1\right) R}\left( p\right) $ as a whole.
Thus, in addition to the Feynman rules in the section $2.5$, from $\left(
3.3.8\right) ,$ \hspace{0in}$\left( 3.3.9\right) $, figure 3.2.1L and figure
3.2.1R \ we obtain the first supplementary Feynman rule.

\textbf{The first supplementary Feynman rule: There is no correction to
propagator, \ there are only correction to mass of a free particle and to a
vertex. The sort of corrections analogous to (3.3.8)-(3.3.9) is the first
sort of corrections to a vertex. \ A correction to a propagator can be
regarded as a sum of the corrections to two vertexes joining the propagator.}

We will see the total correction to the mass of a free particle to be zero.

Operating $\Lambda _{ff1\mu \text{ }}^{\left( 1\right) L}\left( q\right) $
on $u_{\mathbf{q}s}$ and $\Lambda _{ff1\nu \text{ }}^{\left( 1\right)
R}\left( q\right) $ on $\overline{u}_{\mathbf{q}s},$ considering (3.3.5),
from (3.3.8)\hspace{0in} and $\left( 3.3.9\right) $ we have

\begin{equation}
\Lambda _{ff1\mu \text{ }}^{\left( 1\right) L}u_{\mathbf{q}s}=-\frac{1}{2}%
g_{0}\gamma _{\mu }[-i\left( 2\pi \right) ^{4}AS_{Ff}\left( q\right) +B]u_{%
\mathbf{q}s},  \tag{3.3.10}
\end{equation}

\begin{equation}
\overline{u}_{\mathbf{q}s}\Lambda _{ff1\nu \text{ }}^{\left( 1\right) R}=-%
\overline{u}_{\mathbf{q}s}\frac{1}{2}[-i\left( 2\pi \right)
^{4}AS_{Ff}\left( q\right) +B]g_{0}\gamma _{\nu }.  \tag{3.3.11}
\end{equation}%
From (3.3.10) or (3.3.11) we see that the rule above naturally eliminates
the ambiguity of corrections to external lines in the conventional QED$%
^{[8]} $.

Because of the gauge invariance of the Lagrangian density, there is no
correction to the mass of a free F-photon or a W-photon. The one-loop
contribution of the scattering operator $S_{f}\left( g_{0},m_{e0}\right) $
to the F-photon propagator with its two vertices can be represented by the
Feynman diagram shown in figure 3 .3.1-2 Considering the current to be
consevational, by the Feynman rules we have$^{[8]}$

\begin{equation}
(-g_{0}\gamma _{\mu })D_{Fff\mu \nu }^{\left( 1\right) }\left( k\right)
(-g_{0}\gamma _{\nu })=g_{0}\gamma _{\mu }D_{Ff}\left( k\right) \Pi _{f\mu
\nu }^{\left( 1\right) }\left( k\right) D_{Ff}\left( k\right) g_{0}\gamma
_{\nu },  \tag{3.3.12}
\end{equation}

\begin{eqnarray}
\Pi _{f\mu \nu }^{\left( 1\right) }\left( k\right) &=&g_{0}^{2}\int
d^{4}pTr\gamma _{\mu }\frac{m-ip\gamma }{p^{2}+m^{2}-i\varepsilon }\gamma
_{\nu }\frac{m-i\left( p-k\right) \gamma }{\left( p-k\right)
^{2}+m^{2}-i\varepsilon }  \notag \\
&=&\left( \delta _{\mu \nu }-\frac{k_{\mu }k\nu }{k^{2}}\right) \left[
CD_{Ff}^{-1}\left( k\right) +\Pi _{fc}^{\left( 1\right) }\left( k\right)
D_{Ff}^{-2}\left( k\right) \right]  \notag \\
&=&\delta _{\mu \nu }\left[ CD_{Ff}^{-1}\left( k\right) +\Pi _{fc}^{\left(
1\right) }\left( k\right) D_{Ff}^{-2}\left( k\right) \right] , 
\TCItag{3.3.13}
\end{eqnarray}%
where $D_{Ff}^{-1}\left( k\right) =i\left( 2\pi \right) ^{4}k^{2},$%
\begin{equation}
C=-\frac{\alpha }{3\pi }\left( D+\frac{5}{6}\right) .  \tag{3.3.14}
\end{equation}%
$C$ is logarithmically divergent , $\Pi _{fc}^{\left( 1\right) }\left(
k\right) $ is finite and $\Pi _{fc}^{\left( 1\right) }\left( q^{\prime
}\right) =0$.

Analogously to discussion for (3.3.8) and (3.3.9), from $\left(
3.3.12\right) $ and $\left( 3.3.13\right) $ we obtain the second sort of
correction $\Lambda _{ff2\nu \text{ }}^{\left( 1\right) }$ to a F-vertex to
be

\begin{eqnarray}
\Lambda _{ff2\nu \text{ }}^{\left( 1\right) }\left( k\right) &=&-\frac{1}{2}%
g_{0}\gamma _{\mu }D_{Ff}\left( k\right) \Pi _{f\mu \nu }^{\left( 1\right)
}\left( k\right)  \notag \\
&=&-\frac{1}{2}g_{0}\gamma _{\mu }\left( \delta _{\mu \nu }-\frac{k_{\mu
}k\nu }{k^{2}}\right) \left[ C+\Pi _{fc}^{\left( 1\right) }\left( k\right)
D_{Ff}^{-1}\left( k\right) \right] .  \TCItag{3.3.15}
\end{eqnarray}%
Considering $q^{\prime 2}=0$ for external photon lines, from $\left(
3.3.15\right) $ we obtain

\begin{equation}
\Lambda _{ff2\mu \text{ }}^{\left( 1\right) }\left( q^{\prime }\right) \frac{%
1}{\sqrt{2\omega _{\mathbf{q}}V}}e_{\mathbf{q}\mu }^{\lambda }=-\frac{1}{2}%
g_{0}\gamma _{\mu }\delta _{\mu \nu }C\frac{1}{\sqrt{2\omega _{\mathbf{q}}V}}%
e_{\mathbf{q}\nu }^{\lambda }.  \tag{3.3.16}
\end{equation}%
From $\left( 3.3.16\right) $ and (3.2.3) we obtain the second sort of the
one-loop corrections originating from $S_{f}^{\left( 1\right) }\left(
g_{0},m_{e0}\right) $ to the coupling constant to be 
\begin{eqnarray}
g_{ff2}^{\left( 1\right) }\left( p_{2},p_{1}\right)  &=&\frac{\left( 2\pi
\right) ^{4}\delta ^{4}\left( p_{2}-k_{1}-p_{1}\right) V^{-3/2}\overline{u}_{%
\mathbf{p}_{2}s_{2}}\Lambda _{ff2\nu \text{ }}^{\left( 1\right) }\left(
k\right) u_{\mathbf{p}_{1}s_{1}}e_{\mathbf{k}_{1}\nu }^{\lambda }\left(
2\omega _{\mathbf{k}_{1}}\right) ^{-1/2}}{\langle a_{\mathbf{p}%
_{2}s_{2}}\mid -\int d^{4}x:\overline{\psi }^{\prime }\gamma _{\mu }A_{\mu
}^{\prime }\psi ^{\prime }:\mid a_{\mathbf{p}_{1}s_{1}}c_{\mathbf{k}%
_{1}\lambda }\rangle }  \notag \\
&=&\frac{1}{2}g_{0}\left\{ C+\Pi _{fc}^{\left( 1\right) }\left( k\right)
D_{Ff}^{-1}\left( k\right) \right\} ,  \TCItag{3.3.17}
\end{eqnarray}%
\begin{equation}
g_{ff2}^{\left( 1\right) }\left( q_{2},q_{1}\right) =\frac{1}{2}g_{0}C, 
\tag{3.3.18}
\end{equation}%
where the numerator in (3.3.17) corresponds to Fig. 3.4.3A to be the one
part of the scattering amplitude $\langle a_{\mathbf{p}s}\mid S_{f}^{\left(
1\right) }\left( g_{0},m_{e0}\right) \mid a_{\mathbf{p}_{1}s_{1}}\rangle
\mid c_{\mathbf{k}_{1}\lambda }\rangle .$ The first supplementary Feynman
rule still applies to this case.

The third sort of one-loop contribution $A_{gff3}^{\left( 1\right) }$ of the
scattering operator $S_{f}\left( g_{0},m_{e0}\right) $ to the scattering
amplitude $A_{gf}\left( p_{2,}p_{1}\right) $ is represented by the Feynman
diagram shown in figure 3.4.4.A. By the Feynman rules we have$^{[8]}$

\begin{eqnarray}
A_{gff3}^{\left( 1\right) }\left( p_{2,}p_{1}\right)  &=&\langle a_{\mathbf{p%
}_{2}s_{2}}\mid S_{f3}^{\left( 1\right) }\left( g_{0},m_{e0}\right) \mid a_{%
\mathbf{p}_{1}s_{1}}c_{\mathbf{k}\lambda }\rangle   \notag \\
&=&-g_{0}\left( 2\pi \right) ^{4}\delta ^{4}\left( p_{2}-p_{1}-k\right)  
\notag \\
&&\cdot \frac{1}{V}\overline{u}_{\mathbf{p}_{2}s_{2}}\Lambda _{\mu ,ff3\text{
}}^{\left( 1\right) }\left( p_{2,}p_{1}\right) u_{\mathbf{p}_{1}s_{1}}\frac{1%
}{\sqrt{2\omega _{\mathbf{k}}V}}e_{\mathbf{k}\mu }^{\lambda }, 
\TCItag{3.3.19}
\end{eqnarray}

\begin{eqnarray}
\Lambda _{\mu ,ff3\text{ }}^{\left( 1\right) }\left( p_{2,}p_{1}\right)  &=&%
\frac{ig_{0}^{2}}{\left( 2\pi \right) ^{4}}\int \frac{d^{4}k}{%
k^{2}-i\varepsilon }\gamma _{\nu }\frac{m-i\left( p_{2}-k\right) \gamma }{%
\left( p_{2}-k\right) ^{2}+m^{2}-i\varepsilon }  \notag \\
&&\cdot \gamma _{\mu }\frac{m-i\left( p_{1}-k\right) \gamma }{\left(
p_{1}-k\right) ^{2}+m^{2}-i\varepsilon }\gamma _{\nu }  \notag \\
&=&L\gamma _{\mu }+\Lambda _{\mu ,ff3c\text{ }}^{\left( 1\right) }\left(
p_{2,}p_{1}\right) ,  \TCItag{3.3.20}
\end{eqnarray}%
where $L=-B,$ is logarithmically divergent, $\Lambda _{\mu ,ff3\text{ }%
}^{\left( 1\right) }\left( p_{2,}p_{1}\right) $ is finite and $\Lambda _{\mu
,ff3\text{ }}^{\left( 1\right) }\left( q_{2,}q_{1}\right) =0.$ From $\left(
3.3.19\right) $ and $\left( 3.2.3\right) $ we obtain the third sort of the
one-loop corrections originating from $S_{f}^{\left( 1\right) }\left(
g_{0},m_{e0}\right) $ to the coupling constant $g_{f0}$ to be 
\begin{equation}
g_{ff3}^{\left( 1\right) }\left( p_{2},p_{1}\right) =g_{0}\frac{\overline{u}%
_{\mathbf{p}_{2}s_{2}}\Lambda _{\nu ,ff3c\text{ }}^{\left( 1\right) }\left(
p_{2,}p_{1}\right) \gamma _{\nu }u_{\mathbf{p}_{1}s_{1}}e_{\mathbf{k}_{1}\nu
}^{\lambda }}{\overline{u}_{\mathbf{p}_{2}s_{2}}\gamma _{\mu }u_{\mathbf{p}%
_{1}s_{1}}e_{\mathbf{k}_{1}\mu }^{\lambda }}+g_{0}L,  \tag{3.3.21}
\end{equation}%
\begin{equation}
g_{ff3}^{\left( 1\right) }\left( q_{2},q_{1}\right) =g_{0}L.  \tag{3.3.22}
\end{equation}%
(3.3.1)-(3.3.7), (3.3.12)-(3.3.14) and (3.3.19)-(3.3.20) are the same as
those of the conventional $QED$, respectively$^{[8]}$.

\subsubsection{The first sort of one-loop corrections to $g_{w}$\ and $m_{w}$%
}

We can obtain the Feynman diagrams determined by $S_{w}\left(
g_{0},m_{e0}\right) $ corresponding to those determined by $S_{f}\left(
g_{0},m_{e0}\right) $, provided we transform real lines into the dotted
lines corresponding to them. Analogously to computation of $\left(
3.3.1\right) $ -$\left( 3.3.22\right) ,$ and paying attention to the
differences between the vertex factor, the propagators and the external line
factors of $\langle F_{f}\mid S_{f}\left( g_{0},m_{e0}\right) \mid
F_{i}\rangle $ and those of $\langle W_{f}\mid S_{w}\left(
g_{0},m_{e0}\right) \mid W_{i}\rangle $ , we can obtain the results
corresponding to $\left( 3.3.1\right) $ -$\left( 3.3.22\right) .$

\begin{eqnarray}
A_{mww}^{(1)}\left( q_{1}\right) &=&\langle \underline{a}_{\mathbf{q}%
_{2}s_{2}}\mid S_{w}^{\left( 1\right) }\left( g_{0},m_{e0}\right) \mid 
\underline{a}_{\mathbf{q}_{1}s_{1}}\rangle  \notag \\
&=&\delta ^{4}\left( q_{2}-q_{1}\right) \frac{1}{V}\overline{v}_{\mathbf{q}%
_{2}s_{2}}\Sigma _{ww}^{\left( 1\right) }\left( p\right) v_{\mathbf{q}%
_{1}s_{1}},  \TCItag{3.3.23}
\end{eqnarray}

\begin{eqnarray}
\Sigma _{ww}^{\left( 1\right) }\left( q_{1}\right) &=&g_{0}^{2}\int d^{4}k%
\frac{i}{k^{2}+i\varepsilon }r_{\mu }\frac{-i\left[ m-i\left( q_{1}-k\right)
\gamma \right] }{\left( q_{1}-k\right) ^{2}+m^{2}+i\varepsilon }r_{\mu } 
\notag \\
&=&-i\left( 2\pi \right) ^{4}A+BS_{Fw}^{-1}\left( q_{1}\right)
+S_{Fw}^{-1}\left( q_{1}\right) \Sigma _{wc}^{\left( 1\right) }\left(
q_{1}\right) S_{Fw}^{-1}\left( q_{1}\right)  \notag \\
&=&\Sigma _{ff}^{\left( 1\right) }\left( q_{1}\right) ,  \TCItag{3.3.24}
\end{eqnarray}

\begin{equation}
\Sigma _{wc}^{\left( 1\right) }\left( q_{1}\right) =\Sigma _{fc}^{\left(
1\right) }\left( q_{1}\right) =0,\;\ \ \ \ \ \ \ S_{Fw}^{-1}\left(
q_{1}\right) =S_{Ff}^{-1}\left( q_{1}\right) .  \tag{3.3.25}
\end{equation}%
Substituting \hspace{0in}$A_{mww}^{(1)}\left( p_{1}\right) $ into $(3.2.6),$
we obtain the one-loop correction originating from $S_{w}\left(
g_{0},m_{e0}\right) $ to the mass of a free W-electron to be

\begin{equation}
m_{eww}^{\left( 1\right) }\left( q\right) =A.  \tag{3.3.26}
\end{equation}%
The one-loop contribution for the W-electron propagator with its two
vertices originating $S_{w}\left( g_{0},m_{e0}\right) $ is

\begin{equation}
g_{0}\gamma \mu S_{Fw,ww}^{\left( 1\right) }\left( p\right) g_{0}\gamma
_{\nu }=g_{0}^{2}\gamma _{\mu }S_{Fw}\left( p\right) \Sigma _{ww}^{\left(
1\right) }\left( p\right) S_{Fw}\left( p\right) \gamma _{\nu }.  \tag{3.3.27}
\end{equation}%
From $\left( 3.3.27\right) $ we obtain the first sort of one-loop
corrections to a W-vertex to be

\begin{eqnarray}
\Lambda _{ww1\mu \text{ }}^{\left( 1\right) L} &=&\frac{1}{2}g_{0}\gamma \mu
S_{Fw}\left( p\right) \Sigma _{ww}^{\left( 1\right) }\left( p\right)
=-\Lambda _{ff1\mu \text{ }}^{\left( 1\right) L},\text{ \hspace{0in} } 
\TCItag{3.3.28} \\
\Lambda _{ww1\nu \text{ }}^{\left( 1\right) R} &=&\frac{1}{2}\Sigma
_{ww}^{\left( 1\right) }\left( p\right) S_{Fw}\left( p\right) g_{0}\gamma
_{\nu }=-\Lambda _{ff1\nu \text{ }}^{\left( 1\right) R}.  \TCItag{3.3.29}
\end{eqnarray}%
From $\left( 3.3.28\right) -(3.3.29)$ we have

\begin{equation}
\Lambda _{ww1\mu \text{ }}^{\left( 1\right) L}\left( q\right) v_{\mathbf{q}s}%
\frac{1}{\sqrt{V}}=\frac{1}{2}g_{0}\gamma \mu \lbrack -i\left( 2\pi \right)
^{4}AS_{Fw}\left( p\right) +B]v_{\mathbf{q}s}\frac{1}{\sqrt{V}}, 
\tag{3.3.30}
\end{equation}

\begin{equation}
\frac{1}{\sqrt{V}}\overline{v}_{\mathbf{q}s}\Lambda _{ww1\nu \text{ }%
}^{\left( 1\right) R}\left( q\right) =\frac{1}{\sqrt{V}}\overline{v}_{%
\mathbf{q}s}\frac{1}{2}[-i\left( 2\pi \right) ^{4}AS_{Fw}\left( p\right)
+B]g_{0}\gamma _{\nu }.  \tag{3.3.31}
\end{equation}

\hspace{0pt}The one-loop contribution of the scattering operator $%
S_{w}\left( g_{0},m_{e0}\right) $ for the W-photron propagator with its two
vertices is

\begin{equation}
g_{0}\gamma \mu D_{Fw\mu \nu }^{\left( 1\right) }\left( k\right) g_{0}\gamma
_{\nu }=g_{0}\gamma \mu D_{Fw}\left( k\right) \Pi _{w\mu \nu }^{\left(
1\right) }\left( k\right) D_{Fw}\left( k\right) g_{0}\gamma _{\nu }, 
\tag{3.3.32}
\end{equation}

\begin{eqnarray}
\Pi _{w\mu \nu }^{\left( 1\right) }\left( k\right) &=&g_{0}^{^{2}}\int
d^{4}pTr\left\{ r_{\mu }\frac{m-ip}{p^{2}+m^{2}+i\varepsilon }r_{\nu }\frac{%
m-i\left( p-k\right) }{\left( p-k\right) ^{2}+m^{2}+i\varepsilon }\right\} 
\notag \\
&=&\left( \delta _{\mu \nu }-\frac{k_{\mu }k\nu }{k^{2}}\right) \left[
CD_{Fw}^{-1}\left( k\right) +\Pi _{wc}^{\left( 1\right) }\left( k\right)
D_{Fw}^{-2}\left( k\right) \right]  \notag \\
&=&-\Pi _{f\mu \nu }^{\left( 1\right) }\left( k\right) ,  \TCItag{3.3.33}
\end{eqnarray}

\begin{equation}
\Pi _{wc}^{\left( 1\right) }\left( k\right) =-\Pi _{fc}^{\left( 1\right)
}\left( k\right) ,\text{ \hspace{0in} \hspace{0in} \hspace{0in} \hspace{0in} 
\hspace{0in} \hspace{0in} \hspace{0in} \hspace{0in} \hspace{0in} \hspace{0in}
\hspace{0in} \hspace{0in} \hspace{0in}}D_{Fw}^{-1}\left( k\right)
=-D_{Ff}^{-1}\left( k\right) .  \tag{3.3.34}
\end{equation}%
Analogously to $(3.3.15)$, from $\left( 3.3.32\right) $ we obtain the second
sort of corrections $\Lambda _{ww2\text{ }}^{\left( 1\right) }$ for a
W-vertex to be

\begin{eqnarray}
\Lambda _{ww2\nu \text{ }}^{\left( 1\right) }\left( k\right) &=&\frac{1}{2}%
g_{0}\gamma \mu D_{Fw}\left( k\right) \Pi _{w\mu \nu }^{\left( 1\right)
}\left( k\right)  \notag \\
&=&\frac{1}{2}g_{0}\gamma \mu \left( \delta _{\mu \nu }-\frac{k_{\mu }k\nu }{%
k^{2}}\right) \left[ C+\Pi _{wc}^{\left( 1\right) }\left( k\right)
D_{Fw}^{-1}\left( k\right) \right] .  \TCItag{3.3.35}
\end{eqnarray}%
Considering $q^{\prime 2}=0$ for external photron lines, from $\left(
3.3.35\right) $ we have

\begin{equation}
\Lambda _{ww2\mu \text{ }}^{\left( 1\right) }\left( q^{\prime }\right) \frac{%
1}{\sqrt{2\omega _{\mathbf{q}^{\prime }}V}}e_{\mathbf{q}^{\prime }\mu
}^{\lambda }=\frac{1}{2}g_{0}\gamma \mu \left( \delta _{\mu \nu }-\frac{%
q_{\mu }^{\prime }q^{\prime }\nu }{q^{\prime 2}}\right) C\frac{1}{\sqrt{%
2\omega _{\mathbf{q}^{\prime }}V}}e_{\mathbf{q}^{\prime }\nu }^{\lambda }. 
\tag{3.3.36}
\end{equation}%
From $\left( 3.3.35\right) $ \ and $(3.2.5)$ we obtain the second sort of
the one-loop corrections originating from $S_{w}^{\left( 1\right) }\left(
g_{0},m_{e0}\right) $ to the coupling constant to be 
\begin{eqnarray}
g_{ww2}^{\left( 1\right) }\left( p_{2},p_{1}\right)  &=&\frac{\left( 2\pi
\right) ^{4}\delta ^{4}\left( p_{2}-k_{1}-p_{1}\right) V^{-3/2}\overline{v}_{%
\mathbf{p}_{2}s_{2}}\Lambda _{ww2\nu \text{ }}^{\left( 1\right) }\left(
k\right) v_{\mathbf{p}_{1}s_{1}}e_{\mathbf{k}_{1}\nu }^{\lambda }\left(
2\omega _{\mathbf{k}_{1}}\right) ^{-1/2}}{\langle \underline{a}_{\mathbf{p}%
_{2}s_{2}}\mid \int d^{4}x:\underline{\overline{\psi }}^{^{\prime }}\gamma
_{\mu }\underline{A}_{\mu }^{^{\prime }}\underline{\psi }^{^{\prime }}:\mid 
\underline{a}_{\mathbf{p}_{1}s_{1}}\underline{c}_{\mathbf{k}_{1}\lambda
}\rangle }  \notag \\
&=&\frac{1}{2}g_{0}\left\{ C+\Pi _{wc}^{\left( 1\right) }\left( k\right)
D_{Fw}^{-1}\left( k\right) \right\} .  \TCItag{3.3.37}
\end{eqnarray}

\begin{equation}
g_{ww2}^{\left( 1\right) }\left( q_{2},q_{1}\right) =\frac{1}{2}g_{0}C. 
\tag{3.3.38}
\end{equation}

Let $A_{gww3}^{(1)}\left( p_{2,}p_{1}\right) $ be the third sort of one-loop
contributions originating from $S_{w}\left( g_{0},m_{e0}\right) $ for the
scattering amplitude $A_{gw3}^{(1)}\left( p_{2,}p_{1}\right) $. Analogously
to $A_{gff3}^{(1)}\left( p_{2,}p_{1}\right) ,$ by the Feynman rules we have

\begin{eqnarray}
A_{gww3}^{(1)}\left( p_{2,}p_{1}\right)  &=&\langle \underline{a}_{\mathbf{p}%
_{2}s_{2}}\mid S_{w}^{\left( 1\right) }\left( g_{0},m_{e0}\right) \mid 
\underline{a}_{\mathbf{p}_{1}s_{1}}\underline{c}_{\mathbf{k}\lambda }\rangle 
\notag \\
&=&g_{0}\left( 2\pi \right) ^{4}\delta ^{4}\left( p_{2}-p_{1}-k\right) \frac{%
1}{V}  \notag \\
&&\overline{v}_{\mathbf{p}_{2}s_{2}}\Lambda _{\mu ,ww3\text{ }}^{\left(
1\right) }\left( p_{2,}p_{1}\right) v_{\mathbf{p}_{1}s_{1}}\frac{1}{\sqrt{%
2\omega _{\mathbf{k}}V}}e_{\mathbf{k}\mu }^{\lambda },  \TCItag{3.3.39}
\end{eqnarray}

\begin{eqnarray}
&&\Lambda _{\mu ,ww3\text{ }}^{\left( 1\right) }\left( p_{2,}p_{1}\right) 
\notag \\
&=&\frac{g_{0}^{2}}{\left( 2\pi \right) ^{4}}\int d^{4}k\frac{i}{%
k^{2}+i\varepsilon }\gamma _{\nu }\frac{-i\left[ m-i\left( p_{2}-k\right)
\gamma \right] }{\left( p_{2}-k\right) ^{2}+m^{2}+i\varepsilon }\gamma _{\mu
}\frac{-i\left[ m-i\left( p_{1}-k\right) \gamma \right] }{\left(
p_{1}-k\right) ^{2}+m^{2}+i\varepsilon }\gamma _{\nu }  \notag \\
&=&L\gamma _{\mu }+\Lambda _{\mu ,ww3c\text{ }}^{\left( 1\right) }\left(
p_{2,}p_{1}\right) =\Lambda _{\mu ,ff3\text{ }}^{\left( 1\right) }\left(
p_{2,}p_{1}\right) ,\ \Lambda _{\mu ,ww3c\text{ }}^{\left( 1\right)
}=\Lambda _{\mu ,ff3c\text{ }}^{\left( 1\right) }.  \TCItag{3.3.40}
\end{eqnarray}%
From $\left( 3.3.39\right) $ and (3.2.5) we obtain the third sort of the
one-loop corrections originating from $S_{w}^{\left( 1\right) }\left(
g_{0},m_{e0}\right) $ to the coupling constant $g_{w0}$ to be 
\begin{equation}
g_{ww3}^{\left( 1\right) }\left( p_{2},p_{1}\right) =g_{0}\frac{\overline{v}%
_{\mathbf{p}_{2}s_{2}}\Lambda _{\nu ,ww3c\text{ }}^{\left( 1\right) }\left(
p_{2,}p_{1}\right) \gamma _{\nu }v_{\mathbf{p}_{1}s_{1}}e_{\mathbf{k}_{1}\nu
}^{\lambda }}{\overline{v}_{\mathbf{p}_{2}s_{2}}\gamma _{\mu }v_{\mathbf{p}%
_{1}s_{1}}e_{\mathbf{k}_{1}\mu }^{\lambda }}+g_{0}L,  \tag{3.3.41}
\end{equation}%
\begin{equation}
g_{ww3}^{\left( 1\right) }\left( q_{2},q_{1}\right) =g_{0}L.  \tag{3.3.42}
\end{equation}

\subsection{The second sort of one-loop corrections and the total one-loop
corrections}

\subsubsection{The second sort of one-loop corrections}

The second sort of correction is such correction to $g_{f}^{(0)}$ and $%
m_{ef0}$ originating from $S_{w}(g_{0}$, $m_{e0})$ and such correction to $%
g_{w}^{(0)}$ and $m_{ew0}$ originating from $S_{f}(g_{0}$, $m_{e0}).$ The
characters of the second sort of corrections are as follows.

1. The momenta at the vertexes of the first corrections are arbitrary. In
contrast with the first sort of correction, from (3.2.8)-(3.2.11) we see
that the momenta at the vertexes of the second sort of corrections must be
equal to $q_{2},$ $q_{1}$ and $q^{\prime }$.

2. From (3.2.8)-(3.2.11) and (3.2.3)-(3.2.6) we see that the second sort of
corrections is easily evaluated by the first sort of corrections. For
example, from (3.3.6) and (3.3.26), (3.2.9) and (3.2.11), we have

\begin{equation}
m_{efw}^{(1)}(q)=-m_{eww}^{(1)}(q)=-A=m_{ewf}^{(1)}(q)=-m_{eff}^{(1)}(q), 
\tag{3.4.1}
\end{equation}%
from (3.2.8), (3.2.10), (3.3.18), (3.3.22), (3.3.38) and (3.3.42), we can
derive%
\begin{equation}
g_{fw2}^{(1)}(q_{2,}q_{1})=-g_{ww2}^{(1)}(q_{2,}q_{1})=-\frac{1}{2}%
g_{0}C=g_{wf2}^{(1)}(q_{2,}q_{1})=-g_{ff2}^{(1)}(q_{2,}q_{1}),  \tag{3.4.2}
\end{equation}

\begin{equation}
g_{fw3}^{(1)}(q_{2,}q_{1})=-g_{ww3}^{(1)}(q_{2,}q_{1})=-g_{0}L=g_{wf3}^{(1)}(q_{2,}q_{1})=-g_{ff3}^{(1)}(q_{2,}q_{1}).
\tag{3.4.3}
\end{equation}

3. The second sort of corrections directly corrects the coupling constants
and the masses, hence it is equivalent corrections to a vertex. We represent
it by a circle. From (2.3.2), (2.3.5), (2.3.18), (2.3.21), (3.2.3)-(3.2.6)
and (3.2.8)-(3.2.11), we can derive the Feynman rules for a circle as
follows.

\textbf{The second supplimentary Feynman rule:}

\textbf{A. A circle is equivalent to a vertex for the part outside the
circle.}

\textbf{B. The momenta of the external lines of the Feynman diagram inside
the circle are the momenta }$q_{2},$\textbf{\ }$q_{1}$\textbf{\ and }$%
q^{\prime }$\textbf{\ at the subtraction point.}

\textbf{C.The dotted- line circle with two speckles denotes the coefficient}

\begin{equation}
A_{mw}^{\prime -1}\left( \mid \underline{a}_{q_{1}s_{1}}\rangle \rightarrow
\mid \underline{a}_{q_{2}s_{2}}\rangle \right) =\langle \underline{a}%
_{q_{2}s_{2}}\mid -i\int d^{4}x:\underline{\overline{\psi }}^{\prime }%
\underline{\psi }^{\prime }:\mid \underline{a}_{q_{1}s_{1}}\rangle ^{-1}, 
\tag{3.4.4}
\end{equation}%
\textbf{the transition amplitude inside this dotted- line circle is}%
\begin{equation}
\langle \underline{a}_{q_{2}s_{2}}\mid S_{w}^{(1)}\left( g_{0},m_{e0}\right)
\mid \underline{a}_{q_{1}s_{1}}\rangle .  \tag{3.4.5}
\end{equation}%
From (3.4.4) and (3.4.5) we derive the factor $m_{efw}^{(1)}(q)$
corresponding to the dotted-line in figure 3.1.2, figure 3.2.2, figure
3.4.1.B and figure 3.4.2.B to be $-A$ (see (3.4.1)).

\textbf{The real- line circle with two speckles denotes the coefficient }%
\begin{equation}
A_{mf}^{^{\prime }-1}\left( \mid a_{q_{1}s_{1}}\rangle \rightarrow \mid
a_{q_{2}s_{2}}\rangle \right) =\langle a_{q_{2}s_{2}}\mid -i\int d^{4}x:%
\overline{\psi }^{\prime }\psi ^{\prime }:\mid a_{q_{1}s_{1}}\rangle ^{-1}, 
\tag{3.4.6}
\end{equation}%
\textbf{the transition amplitude inside this real- line circle is}%
\begin{equation}
\langle a_{q_{2}s_{2}}\mid S_{f}^{(1)}\left( g_{0},m_{e0}\right) \mid
a_{q_{1}s_{1}}\rangle .  \tag{3.4.7}
\end{equation}%
From (3.4.6) and (3.4.7) we derive the factor $m_{ewf}^{(1)}(q)$
corresponding to the real-line circle in figure 3.2.4, figure 3.4.1.D and
figure 3.4.2.D to be $-A$ (see (3.4.1)).

\textbf{The dotted- line circle with three speckles denotes the coefficient}%
\begin{eqnarray}
&&A_{gw}^{\prime -1}\left( \mid \underline{a}_{q_{1}s_{1}}\underline{c}%
_{q^{\prime }\lambda }\rangle \rightarrow \mid \underline{a}%
_{q_{2}s_{2}}\rangle \right)   \notag \\
&=&\{\langle \underline{a}_{q_{2}s_{2}}\mid \int d^{4}x:\underline{\overline{%
\psi }}^{\prime }\gamma _{\mu }\underline{A^{\prime }}_{\mu }\underline{\psi 
}^{\prime }:\mid \underline{a}_{q_{1}s_{1}}\underline{c}_{q^{\prime }\lambda
}\rangle \}^{-1},  \TCItag{3.4.8}
\end{eqnarray}%
\textbf{the transition amplitude inside this dotted- line circle is}%
\begin{equation}
\langle \underline{a}_{q_{2}s_{2}}\mid S_{w}^{(1)}\left( g_{0},m_{e0}\right)
\mid \underline{a}_{q_{1}s_{1}}\underline{c}_{q^{\prime }\lambda }\rangle . 
\tag{3.4.9}
\end{equation}%
From (3.4.8) and (3.4.9) we derive the factor $g_{fw2}^{(1)}(q_{2},q_{1})$
corresponding to the dotted-line in figure 3.4.3.B to be $-\frac{1}{2}g_{0}C$
(see (3.4.2)) and the factor $g_{fw3}^{(1)}(q_{2},q_{1})$ corresponding to
the dotted-line in figure 3.4.4.B to be $-g_{0}L$ (see (3.4.3)).

\textbf{The real- line circle with three speckles denotes the coefficient }%
\begin{eqnarray}
&&A_{gf}^{^{\prime }-1}\left( \mid a_{q_{1}s_{1}}c_{q^{\prime }\lambda
}\rangle \rightarrow \mid a_{q_{2}s_{2}}\rangle \right)   \notag \\
&=&\{\langle a_{q_{2}s_{2}}\mid \int d^{4}x:-\overline{\psi }^{\prime
}\gamma _{\mu }A_{\mu }^{\prime }\psi ^{\prime }:\mid
a_{q_{1}s_{1}}c_{q^{\prime }\lambda }\rangle \}^{-1},  \TCItag{3.4.10}
\end{eqnarray}%
\textbf{the transition amplitude inside this real- line circle is}%
\begin{equation}
\langle a_{q_{2}s_{2}}\mid S_{f}^{(1)}\left( g_{0},m_{e0}\right) \mid
a_{q_{1}s_{1}}c_{q^{\prime }\lambda }\rangle .  \tag{3.4.11}
\end{equation}%
From (3.4.10) and (3.4.11) we derive the factor $g_{wf2}^{(1)}(q_{2},q_{1})$
to be $-\frac{1}{2}g_{0}C$ (see (3.4.2)) and the factor $%
g_{wf3}^{(1)}(q_{2},q_{1})$ to be $-g_{0}L$ (see (3.4.3)).

4. The first supplementary Feynman rule still applies to the second sort of
corrections. In this case, an explanation for the rule is shown by the
figures 3.2.2 and 3.3.2. The examples are the figures 3.4.1.B, 3.4.1.D and
3.2.4.

Because the two sorts of corrections must simultaneously exist, both must
simultaneously be considered, and both should be corrected to the same
loop-number. For the reason, we define `whole Feynman diagram group with
n-loop'.

\textbf{If one group of Feynman diagrams corresponding to the same initial
state and final state contain complete diagrams with n-loop coming from }$%
S_{w}^{(1)}\left( g_{0},m_{e0}\right) $\textbf{\ and }$S_{f}^{(1)}\left(
g_{0},m_{e0}\right) ,$\textbf{\ the group of Feynman diagrams is called
`whole Feynman diagram group with n-loop'(WFDGNL).}

For example, the figures 3.1.1-2 ; the figures 3.4.1.A-D, the figures
3.4.2.A-D, the figures 3.4.3.A-B and the figures 3.4.4.A-B; or the figures
3.2.1-4, the figures 3.3.1-2 are respectively a WFDG1L.

Taking the figures 3.4.1.A-D as an example, we explain construct of a
WFDG1L. Because there is the contribution $g_{ff1}^{(1)}$ of the figure
3.4.1.A for $g_{f}^{(1)}$, there must be the contribution $g_{fw1}^{(1)}$ of
the figure 3.4.1 C. Because the contribution $g_{ff1}^{(1)}$ originates from 
$m_{eff}^{(1)},$ there must be the contribution originates from $%
m_{efw}^{(1)}$ which is shown in the figure 3.4.1.B. Because the
contribution $g_{fw1}^{(1)}$ originates from $m_{eww}^{(1)}$ , there must be
the contribution originates from $m_{ewf}^{(1)}$ which is shown in the
figure 3.4.1D. By such method we easily construct a WFDG1L. Figure 3.4.1A-D
contain the complete 1-loop corrections to a F-vertex derived from figure
3.4.1.A, hence the figures 3.4.1.A-D form a WFDG1L. Similarly, we can obtain
the WFDG1Ls as shown by the figures 3.4.2, 3.4.3 and 3.4.4. Thus, we see
that because $S_{w}\left( g_{0},m_{e0}\right) $ and $S_{f}\left(
g_{0},m_{e0}\right) $ together determine 1-loop corrections to a F- or
W-amplitude, for an arbitrary scattering amplitude, there must be a WFDGNL
which determines the total n-loop correction to the scattering amplitude.
From this we obtain the following Feynman rule.

\textbf{The third supplementary Feynman rule:}

\textbf{For an arbitrary scattering amplitude, \ we should firstly determine
a WFDGNL corresponding to the amplitude, then sum up the integrands of the
WFDGNL, last carry out the integral. }

The Feynman rules in the second chapter and the three supplimentary Feynman
rules fully determine the method evaluating a scattering amplitude by $S_{w}$
and $S_{f}.$

The properties of a WFDGNL are as follows.

1. The integrands causing divergence in a WFDGNL must cancel each other out,
thereby the Feynman integral of the WFDGNL must convergent. For example, the
correction to $m_{ef0}$ coming from the WFDG1L composed of figure 3.1.1-2 is
(1.1.1); from the Feynman rules$,$ (3.3.17), (3.4.2) we obtain the second
sort of correction $g_{f2}^{(1)}$ to $g_{f\text{ }}^{(0)}$ \ coming from the
WFDG1L Fig. 3.4.3.A-B is%
\begin{eqnarray}
g_{f2}^{(1)} &=&g_{ff2}^{(1)}(p_{2},p_{1})+g_{fw2}^{(1)}(q_{2},q_{1})  \notag
\\
&=&\frac{1}{2}g_{0}\left[ C+\Pi _{wc}^{\left( 1\right) }\left( k\right)
D_{Fw}^{-1}\left( k\right) \right] -\frac{1}{2}g_{0}C  \notag \\
&=&\frac{1}{2}g_{0}\Pi _{wc}^{\left( 1\right) }\left( k\right)
D_{Fw}^{-1}\left( k\right) ,\;\ \ \ \ \ \ \ \ k=p_{2}-p_{1},  \TCItag{3.4.12}
\end{eqnarray}%
from (3.3.21) and (3.4.3) we obtain the third sort of correction $%
g_{f3}^{(1)}$ to $g_{f\text{ }}^{(0)}$ coming from the WFDG1L figure
3.4.4.A-B is

\begin{eqnarray}
g_{f3}^{(1)}(p_{2},p_{1})
&=&g_{ff3}^{(1)}(p_{2},p_{1})+g_{fw3}^{(1)}(q_{2},q_{1})  \notag \\
&=&g_{0}\frac{\overline{u}_{\mathbf{p}_{2}s_{2}}\Lambda _{\nu ,ff3c\text{ }%
}^{\left( 1\right) }\left( p_{2,}p_{1}\right) \gamma _{\nu }u_{\mathbf{p}%
_{1}s_{1}}e_{\mathbf{k}_{1}\nu }^{\lambda }}{\overline{u}_{\mathbf{p}%
_{2}s_{2}}\gamma _{\mu }u_{\mathbf{p}_{1}s_{1}}e_{\mathbf{k}_{1}\mu
}^{\lambda }}+g_{0}L-g_{0}L  \notag \\
&=&g_{0}\frac{\overline{u}_{\mathbf{p}_{2}s_{2}}\Lambda _{\nu ,ff3c\text{ }%
}^{\left( 1\right) }\left( p_{2,}p_{1}\right) \gamma _{\nu }u_{\mathbf{p}%
_{1}s_{1}}e_{\mathbf{k}_{1}\nu }^{\lambda }}{\overline{u}_{\mathbf{p}%
_{2}s_{2}}\gamma _{\mu }u_{\mathbf{p}_{1}s_{1}}e_{\mathbf{k}_{1}\mu
}^{\lambda }}.  \TCItag{3.4.13}
\end{eqnarray}

Analogously to (1.1.1), we should firstly sum up the integrands in a WFDGNL,
then complete the integral. But for shortness' sake, we directly give the
results above.

2. A WFDGNL always as a whole appears in a more complicated diagram or
exists by itsself. For example, the WFDG1L composed of figure 3.1.1-2 is a
part of the WFDG1L composed of figure 3.4.1.A-D. It is possible that a
WFDGNL appearing in a more complicated diagram is no longer a WFDGNL
relative to the more complicated diagram. For example, when the WFDG1L
figure 3.1.1-2 appear in figure 3.4.1.A-B, it is no longer a WFDGNL. But it
is necessary that there is a new WFDGNL which contains the original group of
diagrams, e.g., the WFDG1L composed of figure 3.4.1.A-D contains figure
3.1.1-2.

\subsubsection{The total one-loop corrections to $g_{f0}$}

By the Feynman rules we easily obtain the corrections to $g_{f0}.$ From
(3.4.1) and (3.3.2) we obtain the sum of the two amplitudes $A_{gfAB}^{(1)}$
corresponding to figure 3.4.1.A-B is%
\begin{eqnarray}
A_{gfAB}^{(1)} &=&\left( 2\pi \right) ^{4}\delta ^{4}\left(
p_{2}-p_{1}-k\right) \frac{1}{V}\overline{u}_{\mathbf{p}_{2}s_{2}}\cdot \{%
\frac{1}{2}\Sigma _{f}^{\left( 1\right) }\left( p_{2}\right) S_{Ff}\left(
p_{2}\right) (-g_{0})\gamma _{\nu }  \notag \\
&&+\frac{1}{2}[-i\left( 2\pi \right) ^{4}m_{efw}^{(1)}S_{Ff}\left(
p_{2}\right) (-g_{0})\gamma _{\nu }]\}u_{\mathbf{p}_{1}s_{1}}\frac{1}{\sqrt{%
2\omega _{\mathbf{k}}V}}e_{\mathbf{k}\mu }^{\lambda }  \notag \\
&=&\left( 2\pi \right) ^{4}\delta ^{4}\left( p_{2}-p_{1}-k\right) \frac{1}{V}%
\overline{u}_{\mathbf{p}_{2}s_{2}}  \notag \\
&&\cdot \frac{1}{2}\{B+S_{Ff}^{-1}\left( p_{2}\right) \Sigma _{fc}^{\left(
1\right) }\left( p_{2}\right) \}(-g_{0})\gamma _{\nu }u_{\mathbf{p}_{1}s_{1}}%
\frac{1}{\sqrt{2\omega _{\mathbf{k}}V}}e_{\mathbf{k}\mu }^{\lambda }. 
\TCItag{3.4.14}
\end{eqnarray}%
From (3.4.1), (3.3.31) and the Feynman rules (3.4.6)-(3.4.9) we obtain the
sum of the two amplitudes $A_{gfCD}^{(1)}$\ corresponding to figure 3.4.1C-D
is%
\begin{eqnarray}
A_{gfCD}^{(1)} &=&\langle a_{\mathbf{p}_{2}s_{2}}\mid
-g_{fw1}^{(1)}(q_{2},q_{1})\int d^{4}x:\overline{\psi ^{\prime }}\gamma
_{\nu }A_{\nu }^{\prime }\psi ^{\prime }:\mid a_{\mathbf{p}_{1}s_{1}}c_{%
\mathbf{k}\lambda }\rangle   \notag \\
&=&\left( 2\pi \right) ^{4}\delta ^{4}\left( p_{2}-p_{1}-k\right) \frac{1}{V}%
\overline{u}_{\mathbf{p}_{2}s_{2}}\frac{1}{2}Bg_{0}\gamma _{\nu }u_{\mathbf{p%
}_{1}s_{1}}\frac{1}{\sqrt{2\omega _{\mathbf{k}}V}}e_{\mathbf{k}\mu
}^{\lambda }  \TCItag{3.4.15}
\end{eqnarray}%
where%
\begin{eqnarray}
g_{fw1}^{(1)}(q_{2},q_{1}) &=&-\{\langle \underline{a}_{\mathbf{q}%
_{2}s_{2}}\mid \lbrack S_{w1}^{\left( 1\right) }\left( g_{0},m_{e0}\right)  
\notag \\
+\int d^{4}x &:&\overline{\underline{\psi }^{\prime }}%
m_{ewf}^{(1)}(q_{2})S_{Fw}(q_{2})\gamma _{\nu }g_{0}\underline{A}_{\nu
}^{\prime }\underline{\psi }^{\prime }:]\mid \underline{a}_{\mathbf{q}%
_{1}s_{1}}\underline{c}_{\mathbf{q}\lambda }\rangle \}  \notag \\
\cdot \{\langle \underline{a}_{\mathbf{q}_{2}s_{2}} &\mid &\int d^{4}x:%
\underline{\overline{\psi }}^{\prime }\gamma _{\mu }\underline{A}_{\mu
}^{\prime }\underline{\psi }^{\prime }:\mid \underline{a}_{\mathbf{q}%
_{1}s_{1}}\underline{c}_{\mathbf{q}\lambda }\rangle \}^{-1}  \notag \\
&=&-\frac{1}{2}B  \TCItag{3.4.16}
\end{eqnarray}

From (3.4.14) and (3.4.15) we obtain the transition amplitude $A_{g1R}^{(1)}$
corresponding to the WFDG1L figure 3.4.1.A-D to be%
\begin{eqnarray}
A_{g1R}^{(1)} &=&A_{gf1AB}^{(1)}+A_{gf1CD}^{(1)}  \notag \\
&=&\frac{1}{2}\left( 2\pi \right) ^{4}\delta ^{4}\left( p_{2}-p_{1}-k\right) 
\frac{1}{V}  \notag \\
&&\cdot \overline{u}_{\mathbf{p}_{2}s_{2}}S_{Ff}^{-1}\left( p_{2}\right)
\Sigma _{fc}^{\left( 1\right) }\left( p_{2}\right) g_{0}\gamma _{\nu }u_{%
\mathbf{p}_{1}s_{1}}\frac{1}{\sqrt{2\omega _{\mathbf{k}}V}}e_{\mathbf{k}\nu
}^{\lambda }.  \TCItag{3.4.17}
\end{eqnarray}%
$A_{g1R}^{(1)}$ is convergent. Similarly, for the WFDG1L composed of figure
3.4.2.A-D we have%
\begin{eqnarray}
A_{g1L}^{(1)} &=&A_{gf2AB}^{(1)}+A_{gf2CD}^{(1)}  \notag \\
&=&\frac{1}{2}\left( 2\pi \right) ^{4}\delta ^{4}\left( p_{2}-p_{1}-k\right) 
\notag \\
&&\cdot \frac{1}{V}\overline{u}_{\mathbf{p}_{2}s_{2}}g_{0}\gamma _{\nu
}\Sigma _{fc}^{\left( 1\right) }\left( p_{1}\right) S_{Ff}^{-1}\left(
p_{1}\right) u_{\mathbf{p}_{1}s_{1}}\frac{1}{\sqrt{2\omega _{\mathbf{k}}V}}%
e_{\mathbf{k}\nu }^{\lambda }.  \TCItag{3.4.18}
\end{eqnarray}%
From (3.4.17), (3.4.18) we obtain the first sort of corrections $%
g_{f1}^{(1)}(p_{2},p_{1})$ to $g_{f0}(q_{2},q_{1})$ to be%
\begin{eqnarray}
g_{f1}^{(1)}(p_{2},p_{1}) &=&\frac{A_{g1R}^{(1)}+A_{g1L}^{(1)}}{\langle a_{%
\mathbf{p}_{2}s_{2}}\mid -\int d^{4}x:\overline{\psi }^{\prime }\gamma _{\mu
}A_{\mu }^{\prime }\psi ^{\prime }:\mid a_{\mathbf{p}_{1}s_{1}}c_{\mathbf{k}%
_{1}\lambda }\rangle }  \notag \\
&=&\frac{g_{0}}{2}\overline{u}_{\mathbf{p}_{2}s_{2}}\{\gamma _{\nu }\Sigma
_{fc}^{\left( 1\right) }\left( p_{1}\right) S_{Ff}^{-1}\left( p_{1}\right)
+S_{Ff}^{-1}\left( p_{2}\right) \Sigma _{fc}^{\left( 1\right) }\left(
p_{2}\right) \gamma _{\nu }\}u_{\mathbf{p}_{1}s_{1}}e_{\mathbf{k}\nu
}^{\lambda }  \notag \\
&&\cdot (\overline{u}_{\mathbf{p}_{2}s_{2}}\gamma _{\nu }u_{\mathbf{p}%
_{1}s_{1}}e_{\mathbf{k}\nu }^{\lambda })^{-1}.  \TCItag{3.4.19}
\end{eqnarray}%
Feynman diagrams corresponding to any process must be composed of some
WFDGNLs. Sum of WFDGNLs must be covergent. The sum of WFDGNLs 3.4.1-4 is the
total one-loop correction $g_{f}^{(1)}(p_{2},p_{1})$ to $g_{f0},$ 
\begin{equation}
g_{f}^{(1)}(p_{2},p_{1})=g_{f1}^{(1)}(p_{2},p_{1})+g_{f2}^{(1)}(p_{2},p_{1})+g_{f3}^{(1)}(p_{2},p_{1}).
\tag{3.4.20}
\end{equation}%
From (3.4.12), (3.4.13) and (3.4.19) we see 
\begin{equation}
g_{f}^{(1)}(p_{2},p_{1})\;\ is\;\ finite,\;\ \ \ \ \ \ \ \ \ \
g_{f}^{(1)}(q_{2},q_{1})=0.  \tag{3.4.21}
\end{equation}%
Substituting (3.4.20) into (3.2.1), we have%
\begin{equation}
g_{f}^{(1-loop)}(p_{2},p_{1})=g_{0}+g_{f}^{(1)}(p_{2},p_{1}),\;\ \ \ \
g_{f}^{(1-loop)}(q_{2},q_{1})=g_{0}.  \tag{3.4.22}
\end{equation}%
Analogously (3.4.22), we can prove%
\begin{eqnarray}
g_{w}^{(1-loop)}(p_{2},p_{1})
&=&g_{0}+g_{w}^{(1)}(p_{2},p_{1})=g_{f}^{(1-loop)}(p_{2},p_{1})\equiv
g^{(1-loop)}(p_{2},p_{1}),  \notag \\
g_{w}^{(1-loop)}(q_{2},q_{1}) &=&g_{0}.  \TCItag{3.4.23}
\end{eqnarray}%
From (3.4.1) and (3.2.1) we obtain the total 1-loop correction to $m_{e0}$
to be 
\begin{equation}
m_{ef}^{(1)}(q)=m_{eff}^{(1)}(q)+m_{efw}^{(1)}(q)=m_{eww}^{(1)}(q)+m_{ewf}^{(1)}(q)=m_{ew}^{(1)}(q)\equiv m_{e}^{(1)}(q)=0,
\tag{3.4.24}
\end{equation}%
\begin{eqnarray}
m_{ef}^{(1-loop)}(q) &=&m_{e0}+m_{ef}^{(1)}(q)=m_{e0}+m_{ew}^{(1)}(q)  \notag
\\
&=&m_{ew}^{(1-loop)}(q)\equiv m_{e}^{(1-loop)}(q)=0.  \TCItag{3.4.25}
\end{eqnarray}%
Replacing $g_{0}$ by $g^{(1-loop)}(p_{2},p_{1})$ and $\ m_{e0}$ \ by $%
m_{e}^{(1-loop)}(q)$ \ in  $S_{f}\left( g_{0},m_{e0}\right) $ and $%
S_{w}\left( g_{0},m_{e0}\right) ,$ we obtain%
\begin{equation}
S_{f}^{\left( 1-loop\right) }\equiv S_{f}\left( g_{f}^{\left( 1-loop\right)
}(p_{2},p_{1}),m^{\left( 1-loop\right) }(q)\right) ,\text{ ~}  \tag{3.4.26}
\end{equation}%
\begin{equation}
S_{w}^{\left( 1-loop\right) }\equiv S_{w}\left( g_{f}^{\left( 1-loop\right)
}(p_{2},p_{1}),m^{\left( 1-loop\right) }(q)\right) .  \tag{3.4.27}
\end{equation}%
Because only when $p=q,$ i.e. for a free F- or W-electron, the definitions
of the masses (2.3.17) and (2.3.20) are meaningful, we take $p=q$ in $%
m^{\left( 1-loop\right) }.$ We can prove 
\begin{equation*}
m_{e}^{\left( n-loop\right) }=m_{e}^{\left( n-loop\right) }(q)=0
\end{equation*}%
and when $p\neq q,$ $m_{e}(p)$ \ must be attributed to correction of a
vertex, in the perturbation theory, we may consider only corrections to a
vertex in $S_{f}^{\left( 1-loop\right) }$ or $S_{w}^{\left( 1-loop\right) }.$

All scattering amplitudes corrected to 1-loop diagrams evaluated by (3.4.26)
or (3.4.27) are finite. From the deductive process above we see that in
contrast with the conventional QED, there is only one sort of parameters ( 
\hspace{0in}mass and charge at the subtraction point) which are finite and
measurable, there are the two sorts of corrections originating $S_{f\text{ }%
} $ and $S_{w}$, all Feynman integrals are covergent and it is unnecessary
to introduce regularization and counterterms in the present theory. It is
easily seen that the one-loop corrections derived by the present theory are
the same as those derived by the given QED.

According to the present theory, there is no correction to any external
line, corrections to the mass of a free particle must be zero, and
corrections to a propagator can be attributed to correction to the two
vertexes joining it, hence we may consider only one sort of corrections,
i.e., the sort of correction to a vertex.

\subsection{$n-$loop corrections for the coupling constants and the masses}

The method deriving 
\begin{equation*}
S_f^{\left( 1-loop\right) }\equiv S_f\left( g_f^{\left( 1-loop\right)
},m^{\left( 1-loop\right) }\right) \text{ ~~~~and ~~~~~}S_w^{\left(
1-loop\right) }\equiv S_w\left( g_f^{\left( 1-loop\right) },m^{\left(
1-loop\right) }\right)
\end{equation*}
from $S_f\left( g_0,m_{e0}\right) $ and $S_w\left( g_0,m_{e0}\right) $ can
be generalized to derive $S_f^{\left( \left( n+1\right) -loop\right) }$ and $%
S_w^{\left( \left( n+1\right) -loop\right) }$ from $S_f^{\left(
n-loop\right) }$ and $S_w^{\left( n-loop\right) }.$ The recursion method is
as follows.

Assume $S_{f}\left( g^{\left( n-loop\right) },m_{e}^{\left( n-loop\right)
}\right) $ and $S_{w}\left( g^{\left( n-loop\right) },m_{e}^{\left(
n-loop\right) }\right) $ to be given and 
\begin{eqnarray*}
g_{f}^{\left( n-loop\right) }\left( p_{2,}p_{1}\right) &=&g_{w}^{\left(
n-loop\right) }\left( p_{2,}p_{1}\right) =g^{\left( n-loop\right) }\left(
p_{2,}p_{1}\right) =g_{0}+\sum_{i=1}^{n}g^{\left( i\right) }\left(
p_{2,}p_{1}\right) , \\
g_{f}^{\left( i\right) }\left( p_{2,}p_{1}\right) &=&g_{w}^{\left( i\right)
}\left( p_{2,}p_{1}\right) =g^{\left( i\right) }\left( p_{2,}p_{1}\right)
,\;\ \ \ \ \ \ g^{\left( n-loop\right) }\left( q_{2,}q_{1}\right) =g_{0}, \\
m_{ef}^{\left( n-loop\right) } &=&m_{ew}^{\left( n-loop\right)
}=m_{e}^{\left( n-loop\right) }(q)=\sum_{i=1}^{n}m_{e}^{\left( i\right)
}\left( q\right) =0,
\end{eqnarray*}%
$g^{\left( i\right) }$ and $m_{e}^{\left( i\right) }$ denote $i-loop$
corrections to $g_{0}$ and $m_{e0},$ respectively. We can evaluate $%
m_{eff}^{\left( n+1\right) }\left( p\right) $ $\ $and $m_{ewf}^{\left(
n+1\right) }$ $=-m_{eff}^{\left( n+1\right) }\left( q\right) $ by virtue of $%
S_{f}^{\left( n-loop\right) }$ and $m_{eww}^{\left( n+1\right) }\left(
p\right) $ $\ $and $m_{efw}^{\left( n+1\right) }=-m_{eww}^{\left( n+1\right)
}\left( q\right) $ by virtue of $S_{w}^{\left( n-loop\right) }.$ It can be
proven 
\begin{eqnarray}
m_{ewf}^{\left( n+1\right) } &=&m_{efw}^{\left( n+1\right) },\text{ \hspace{%
0in} \hspace{0in}\hspace{0in} \hspace{0in} \hspace{0in}}  \notag \\
\text{\hspace{0in} }m_{eff}^{\left( n+1\right) }(p)+m_{efw}^{\left(
n+1\right) } &=&m_{eww}^{\left( n+1\right) }(p)+m_{ewf}^{\left( n+1\right)
}\equiv m_{e}^{\prime (n+1)}(p),  \notag \\
m_{eff}^{\prime \left( n+1-loop\right) }(p) &=&m_{eww}^{\prime \left(
n+1-loop\right) }(p)  \notag \\
&=&m_{e}^{\prime \left( n+1-loop\right) }(p)=m_{e}^{\left( n-loop\right)
}(p)+m_{e}^{\prime (n+1)}(p),  \notag \\
m_{e}^{\prime (n+1)}(q) &=&0,\;\ m_{e}^{\prime \left( n+1-loop\right) }(q)=0.
\TCItag{3.5.1}
\end{eqnarray}%
Replacing $m_{e}^{\left( n-loop\right) }$ in $S_{f}^{\left( n-loop\right) }$
and $S_{w}^{\left( n-loop\right) }$ by $m_{e}^{\prime \left( n+1-loop\right)
},$ we obtain

\begin{equation}
S_{f}^{\prime \left( n+1-loop\right) }\equiv S_{f}\left( g^{\left(
n-loop\right) },m_{e}^{\prime \left( n+1-loop\right) }\right) ,  \tag{3.5.2}
\end{equation}

\begin{equation}
S_{w}^{\prime \left( n+1-loop\right) }\equiv S_{w}\left( g^{\left(
n-loop\right) },m_{e}^{\prime \left( n+1-loop\right) }\right) .  \tag{3.5.3}
\end{equation}%
We can evaluate $g_{ff}^{\left( n+1\right) }\left( p_{2,}p_{1}\right) $ $\ $%
and $g_{wf}^{\left( n+1\right) }\equiv $ $-g_{ff}^{\left( n+1\right) }\left(
q_{2},q_{1}\right) $ by $S_{f}^{\prime \left( n+1-loop\right) },$ and $%
g_{ww}^{\left( n+1\right) }\left( p_{2,}p_{1}\right) $ and $g_{fw}^{\left(
n+1\right) }\equiv -g_{ww}^{\left( n+1\right) }\left( q_{2},q_{1}\right) $
by $S_{w}^{\prime \left( n+1-loop\right) }.$ It can be proven 
\begin{eqnarray}
g_{wf}^{\left( n+1\right) } &=&g_{fw}^{\left( n+1\right) },\text{\hspace{0in}
}g_{ww}^{\left( n+1\right) }\left( p_{2,}p_{1}\right) +g_{wf}^{\left(
n+1\right) }  \notag \\
&=&g_{ff}^{\left( n+1\right) }\left( p_{2,}p_{1}\right) +g_{fw}^{\left(
n+1\right) }\equiv g^{\left( n+1\right) }\left( p_{2,}p_{1}\right) ,  \notag
\\
g_{ff}^{\left( n+1-loop\right) }\left( p_{2,}p_{1}\right) &=&g_{ww}^{\left(
n+1-loop\right) }\left( p_{2,}p_{1}\right) =g^{\left( n+1-loop\right)
}\left( p_{2,}p_{1}\right)  \notag \\
&=&g^{\left( n-loop\right) }\left( p_{2,}p_{1}\right) +g^{\left( n+1\right)
}\left( p_{2,}p_{1}\right) ,  \notag \\
g^{\left( n+1\right) }\left( q_{2,}q_{1}\right) &=&0,\;\ \ \ g^{\left(
n+1-loop\right) }\left( q_{2,}q_{1}\right) =g_{0}.  \TCItag{3.5.4}
\end{eqnarray}%
Replacing $g^{\left( n-loop\right) }$ and $m_{e}^{\left( n-loop\right) }$ \
in $S_{f}^{\left( n-loop\right) }$ and $S_{w}^{\left( n-loop\right) }$ by $%
g^{\left( n+1-loop\right) }$ \ and $m_{e}^{\left( n+1-loop\right) },$ we
obtain

\begin{equation}
S_{f}^{\left( n+1-loop\right) }\equiv S_{f}\left( g^{\left( n+1-loop\right)
},m_{e}^{\left( n+1-loop\right) }\right) ,  \tag{3.5.5}
\end{equation}

\begin{equation}
S_{w}^{\left( n+1-loop\right) }\equiv S_{w}\left( g^{\left( n+1-loop\right)
},m_{e}^{\left( n+1-loop\right) }\right) .  \tag{3.5.6}
\end{equation}%
$g^{\left( n+1\right) }\left( p_{2,}p_{1}\right) $ is finite.

It must be noted that all $m_{fw}^{\left( n+1\right) },$ $m_{wf}^{\left(
n+1\right) },$ $g_{wf}^{\left( n+1\right) },$ $g_{fw}^{\left( n+1\right) }$
are independent of momenta $p_{2},$ $p_{1}$ and $k,$ since they are defined
at the subtraction point. All scattering amplitudes approximate to $\left(
n+1\right) -loop$ diagrams evaluated by virtue of $S_{f}^{\left(
(n+1)-loop\right) }$ and $S_{w}^{\left( (n+1)-loop\right) }$ are finite.
Detailed discussion will be given in another paper.

\section{Generalize to the $SU(2)\times U(1)$ electroweak unified model and
interaction between W-matter and F-matter}

\subsection{Generalize to the $SU(2)\times U(1)$ electroweak unified model}

{{On the basis of the idea we can construct a left-right symmetric
electroweak unified model. Taking the lepton part of the G-W-S model as
example, the corresponding Lagrangian density will become 
\begin{equation}
\mathcal{L}=\mathcal{L}_{F}+\mathcal{L}_{W}-V_{F}-V_{W},  \tag{4.1.1}
\end{equation}%
\ 
\begin{eqnarray}
\mathcal{L}_{F} &=&-\overline{\psi }_{L}\gamma _{\mu }(\partial _{\mu
}-i\cdot G_{F}\cdot \frac{\tau ^{j}}{2}A_{\mu }^{j}+\frac{i}{2}\cdot
G_{F}^{\prime }\cdot B_{\mu })\cdot \psi _{L}  \notag \\
&&-\frac{1}{4}F_{\mu \nu }^{j}\cdot F_{\mu \nu }^{j}-\frac{1}{4}F_{\mu \nu
}\cdot F_{\mu \nu }  \notag \\
&&-(\partial _{\mu }\phi ^{+}+i\phi ^{+}\cdot G_{F}\cdot \frac{\tau ^{j}}{2}%
A_{\mu }^{j}+\frac{i}{2}\phi ^{+}\cdot G_{F}^{\prime }\cdot B_{\mu })  \notag
\\
&&\cdot (\partial _{\mu }-i\cdot G_{F}\cdot \frac{\tau ^{j}}{2}A_{\mu }^{j}-%
\frac{i}{2}\cdot G_{F}^{\prime }\cdot B_{\mu })\phi  \notag \\
&&-(\overline{e}_{R}\cdot M_{F}^{+}\cdot \phi ^{+}\psi _{L}+\overline{\psi }%
_{L}\phi \cdot M_{F}\cdot e_{R}),  \TCItag{4.1.2}
\end{eqnarray}%
}}

\begin{equation}
V_{F}=-\phi ^{+}\cdot \Omega _{F}\cdot \phi +\frac{1}{4}\phi ^{+}\phi \cdot
\Lambda _{F}\cdot \phi ^{+}\phi ,  \tag{4.1.3}
\end{equation}%
{{%
\begin{eqnarray}
\mathcal{L}_{W} &=&\overline{\underline{\psi }}_{R}\gamma _{\mu }(\partial
_{\mu }+i\cdot {{G_{W}\cdot }}\frac{\tau ^{j}}{2}\underline{A}_{\mu }^{j}-%
\frac{i}{2}\cdot G{{_{W}^{\prime }\cdot }}\underline{B}_{\mu })\cdot 
\underline{\psi }_{R}  \notag \\
&&-\frac{1}{4}\underline{F}_{\mu \nu }^{j}\cdot \underline{F}_{\mu \nu }^{j}-%
\frac{1}{4}\underline{F}_{\mu \nu }\cdot \underline{F}_{\mu \nu }  \notag \\
&&-(\partial _{\mu }\underline{\phi }^{+}+i\underline{\phi }^{+}\cdot {{%
G_{W}\cdot }}\frac{\tau ^{j}}{2}\underline{A}_{\mu }^{j}+\frac{i}{2}%
\underline{\phi }^{+}\cdot {{G_{W}^{\prime }\cdot }}\underline{B}_{\mu }) 
\notag \\
&&\cdot (\partial _{\mu }-i\cdot {{G_{W}\cdot }}\frac{\tau ^{j}}{2}%
\underline{A}_{\mu }^{j}-\frac{i}{2}\cdot {{G_{W}^{\prime }\cdot }}%
\underline{B}_{\mu })\underline{\phi }  \notag \\
&&+(\underline{\overline{e}}_{L}\cdot M_{W}^{+}\cdot \underline{\phi }^{+}%
\underline{\psi }_{R}+\underline{\overline{\psi }}_{R}\underline{\phi }\cdot
M_{W}\cdot \underline{e}_{L}),  \TCItag{4.1.4}
\end{eqnarray}%
}}

\begin{equation}
V_{W}=-\underline{\phi }^{+}\cdot \Omega _{W}\cdot \underline{\phi }+\frac{1%
}{4}\underline{\phi }^{+}\underline{\phi }\cdot \Lambda _{W}\cdot \underline{%
\phi }^{+}\underline{\phi }  \tag{4.1.5}
\end{equation}

{{\ \ where $G_{F},$ $G_{F}^{\prime },$ }}${{M_{F},}}$ $\Omega _{F},$ $%
\Lambda _{F},$ {{$G_{W},$ $G_{W}^{\prime },$ }}${{M_{W},}}$ $\Omega _{W}$ \
and $\Lambda _{W}${\ are all operators, }

{{\ 
\begin{equation}
\psi _{L}=\left( 
\begin{array}{l}
\upsilon _{e} \\ 
e%
\end{array}%
\right) _{L},\;\;\;e_{R}\;,\;\;\;\;\;\;\;\;\phi =\left( 
\begin{array}{l}
\varphi ^{+} \\ 
\varphi ^{0}%
\end{array}%
\right) ,  \tag{4.1.6}
\end{equation}%
}}%
\begin{equation}
F_{\mu \nu }^{i}=\partial _{\mu }A_{\nu }^{i}-\partial _{\nu }A_{\mu
}^{i}+\varepsilon _{ijk}\cdot G_{F}\cdot A_{\mu }^{j}A_{\nu }^{k}, 
\tag{4.1.7}
\end{equation}%
\begin{equation}
F_{\mu \nu }=\partial _{\mu }B_{\nu }-\partial _{\nu }B_{\mu },  \tag{4.1.8}
\end{equation}

{{%
\begin{equation}
\underline{\psi }_{R}=\left( 
\begin{array}{l}
\underline{\nu }_{e} \\ 
\underline{e}%
\end{array}%
\right) _{R},\;\;\;\;\underline{e}_{L},\;\;\;\;\;\;\;\;\underline{\phi }%
=\left( 
\begin{array}{l}
\underline{\varphi }^{+} \\ 
\underline{\varphi }^{0}%
\end{array}%
\right) ,  \tag{4.1.9}
\end{equation}%
}}

\begin{equation}
\underline{F}_{\mu \nu }^{i}=\partial _{\mu }\underline{A}_{\nu
}^{j}-\partial _{\nu }\underline{A}_{\mu }^{j}+\varepsilon _{ijk}\cdot
G_{W}\cdot \underline{A}_{\nu }^{j}\underline{A}_{\mu }^{k},  \tag{4.1.10}
\end{equation}%
\begin{equation}
\underline{F}_{\mu \nu }=\partial _{\mu }\underline{B}_{\nu }-\partial _{\nu
}\underline{B}_{\mu }.  \tag{4.1.11}
\end{equation}%
\begin{equation}
{{{\langle \underline{\phi }\rangle }_{0}=\left( \mathbf{%
\begin{array}{l}
0 \\ 
\upsilon%
\end{array}%
}\right) ,\;\ \ \ \ \ \ \ \ \langle \phi \rangle _{0}=\left( \mathbf{%
\begin{array}{l}
0 \\ 
\upsilon%
\end{array}%
}\right) .}}  \tag{4.1.12}
\end{equation}%
{\ {Analogously to the new QED, we can prove for the electroweak unified
model that all Feynman integrals are convergent, it is not necessary to
introduce counterterms and regularization, there is only one sort of
parameters which are all finite and measurable (at the subtraction point), {%
there is no such parameters as a bare mass and a bare charge, {and there is
no triangle anomaly. }}T{he energy of the ground state of all fields except
Higgs fields is still equal to zero. }As the standard {\ electroweak unified
model, the expectation valves of the Higgs fields are not equal to zero. In
this case in order to solve the cosmological constant problem we must
consider constribution of Higgs fields.} }In contrast with the other
left-right electroweak unified models,{\ there is no unknown massive
particles in the present model. A new origin of left-right asymmetry is
presented. }From the model we see that the world is left-right symmetrical
on principle since }$\mathcal{L}_{F}+\mathcal{L}_{W}${\ \ is symmetric, but
because both }$\mathcal{L}_{F}$ and $\mathcal{L}_{W}$ \ are asymmtric{, the
world in which we exist is left-right asymmetical in fact. }

\subsection{Interaction between W-matter and F-matter}

There is no interaction between W-particles and F-particles by a given
quantizable field, hence only possibility is that there is repulsion or
gravitation of the two sorts of particles. If the new interaction is
gravitation, it is possible that W-matter is the candidate for dark matter$%
^{[3]}$.

Because there are the two sorts of particles corresponding to $\mathcal{L}%
_{W}$ and $\mathcal{L}_{F}$, the energy-momentum tensor $\mathcal{T}_{\mu
\nu }$ should be written as 
\begin{equation}
\mathcal{T}_{\mu \nu }=\mathcal{T}_{F\mu \nu }+\mathcal{T}_{W\mu \nu }. 
\tag{4.2.1}
\end{equation}%
Correspondingly, the Einstein's equation should also be written as 
\begin{equation}
R_{\mu \nu }-\frac{1}{2}g_{\mu \nu }R+\lambda g_{\mu \nu }=-8\pi G\left( 
\mathcal{T}_{F\mu \nu }+\mathcal{T}_{W\mu \nu }\right) .  \tag{4.2.2}
\end{equation}%
Because there is no other interaction between the F-particles and the
W-particles except the gravitation, we existing in F-world cannot detect the
W-particles by other methods except the gravition. Thus, if W-particles
exist, they must be the dark matter for the F-world. Because the F-world and
the W-world are symmetric, it is possible that the W-matter is 50 per cent
of all matter in the cosmos. Other components of dark matter for the F-world
may possibly be other undetected F-matter. The world in which we exist,
i.e., the F-world, is left-hand world, then the W-world is the right-hand
world. Thus the right-hand world is the dark matter world for the left-hand
world, and vice versa. Detail discussion is given in [3].

If the new interaction is repulsion, it is possible that W-matter is the
origin of universe expansion. A.Einstein presented the concept `dark
energy'. If there is repulsion between W-matter and F-matter, we identify
W-matter with `dark energy'. It is also possible that there is new and more
important relationship between W-particles and F-particles. We will discuss
the problem in another paper in detail.

\section{Conclusions and prospects}

We have presented a new conjecture. According to the conjectures, a particle
can exist in two forms which are symmetric. From this we have presented a
new Lagrangian density and a new quantization method for QED. That the
energy of the vacuum state is equal to zero is naturally obtained. From this
the cosmological constant is easily determined by astronomical observation
values and it is possible to correct nonperturbational methods which depend
on the energy of the ground state in quantum field theory.

We discuss quantization for interacting fields, derive Feynman rules and
scattering operators. Since we quantize fields by the transformation
operators replacing creation and annihilation operators in the conventional
QED, it is necessary to replace coupling constants and masses by coupling
operators and mass operators in the present theory. In contrast with the
conventional QED, in the present theory, there is only one sort of
parameters which are all finite and measurable and the two sorts of
corrections originating $S_{f\text{ }}$ and $S_{w}.$

We have evaluateed the one-loop corrections to the coupling constants and
the masses of an electron which are the same as those derived by the
conventional QED, and have supplemented three new Feynman rules and one
concept of WFDGNL. A complete method to evaluate correction with n-loop
diagrams is given. Feynman diagrams corresponding to any process must be
composed of some WFDGNLs. The integrands causing divergence in a WFDGNL will
cancel each other out, hence correction coming from a WFDGNL must be
covergent and it is unnecessary to introduce regularization and counterterms.

On the same basis as the new QED, we obtain naturally a new $SU(2)\times
U(1) $ electroweak unified model whose $\mathcal{L=L}_{F}\mathcal{+L}_{W}$ ,
here $\mathcal{L}$ is left-right symmetric. Thus the world is left-right
symmetric in{\Large \ }principle, but the part in which we exist is
asymmetric because both $\mathcal{L}_{W}$ and $\mathcal{L}_{F}$ are
asymmetic. This model do not contain any unknown particle with a massive
mass.

A conjecture that there is gravitation between the W-matter and the F-matter
is presented and the W-matter is identified as `dark matter'. It is possible
that the new interaction is the origin of some cosmic phenomena.

It is seen that the concepts of the new QED can also be generalized to $QCD$.

\begin{description}
\item[Caption for figures] 
\end{description}

Fig.2.1-2. Contours for propagators. $p_{0}=p_{0},$ $k_{0};$ $\omega =\sqrt{%
m^{2}+\mathbf{p}^{2}}\mathbf{,}$ $\mid $ $\mathbf{k\mid .}$ Fig.2.1.1.
Contours for $S_{Ff\text{ \hspace{0in}}}$ and $D_{Ff}.$ \hspace{0in}%
Fig.2.1.2. Contours for $S_{Fw\text{ \hspace{0in}}}$ and $D_{Fw}.$

Fig. 2.1-2. The Feynman rules. In the table $P$ is an algebraric sum of the
three 4-momenta meeting at a vertex. Moreover, there is an overall minus
sign corresponding to a closed fermion loop.

Fig.2.2.1A 
\begin{equation*}
-g_{0}\gamma _{\mu }\left( 2\pi \right) ^{4}\delta ^{4}\left( P\right)
\end{equation*}

Fig.2.2.1B 
\begin{equation*}
\frac{-i}{\left( 2\pi \right) ^{4}}\frac{\left( m-ip\gamma \right) _{\alpha
\beta }}{p^{2}+m^{2}-i\varepsilon }e^{ip\left( x_{1}-x_{2}\right) }\text{ 
\hspace{0in} \hspace{0in} \hspace{0in} \hspace{0in} \hspace{0in}}with\text{ 
\hspace{0in} \hspace{0in} \hspace{0in} }\int d^{4}p\text{ }
\end{equation*}

Fig.2.2.1C 
\begin{equation*}
\text{ }\delta _{\mu \nu }\frac{-i}{\left( 2\pi \right) ^{4}}\frac{1}{%
k^{2}-i\varepsilon }e^{ik\left( x_{1}-x_{2}\right) }\text{ \hspace{0in} 
\hspace{0in} \hspace{0in} \hspace{0in} \hspace{0in} }with\text{ \hspace{0in} 
\hspace{0in} \hspace{0in} \hspace{0in}}\int d^{4}k
\end{equation*}

Fig.2.2.1D 
\begin{equation*}
\frac{1}{\sqrt{V}}u_{\mathbf{p}s}\text{ }(ingoing)\text{ \hspace{0in} 
\hspace{0in} \hspace{0in} \hspace{0in} }or\text{ \hspace{0in} \hspace{0in} 
\hspace{0in} \hspace{0in} }\frac{1}{\sqrt{V}}\overline{v}_{\mathbf{p}s}\text{
}(outgoing)
\end{equation*}

Fig.2.2.1E 
\begin{equation*}
\text{ }\frac{1}{\sqrt{V}}\overline{u}_{\mathbf{p}s}\text{ }(outgoing)\text{ 
\hspace{0in} \hspace{0in} \hspace{0in} \hspace{0in} \hspace{0in} }or\text{ 
\hspace{0in} \hspace{0in} \hspace{0in} \hspace{0in} \hspace{0in} \hspace{0in}
\hspace{0in}}\frac{1}{\sqrt{V}}v_{\mathbf{p}s}\text{ }(ingoing)
\end{equation*}

Fig.2.2.1F 
\begin{equation*}
\text{ }\frac{1}{\sqrt{V\omega _{\mathbf{k}}}}e_{\mathbf{k}\mu }^{\lambda
},\lambda =1,2
\end{equation*}

Fig.2.2.2A 
\begin{equation*}
g_{0}\gamma _{\mu }\left( 2\pi \right) ^{4}\delta ^{4}\left( P\right)
\end{equation*}

Fig.2.2.2B

\begin{equation*}
\frac{-i}{\left( 2\pi \right) ^4}\frac{\left( m-ip\gamma \right) _{\alpha
\beta }}{p^2+m^2+i\varepsilon }e^{ip\left( x_1-x_2\right) }\text{ \hspace{0in%
} \hspace{0in} \hspace{0in} \hspace{0in} \hspace{0in}}with\text{ \hspace{0in}
\hspace{0in} \hspace{0in} }\int d^4p
\end{equation*}
\hspace{0in} \hspace{0in} \hspace{0in} \hspace{0in} \hspace{0in} \hspace{0in}
\hspace{0in} \hspace{0in} \hspace{0in}

Fig.2.2.2C 
\begin{equation*}
\text{ }\delta _{\mu \nu }\frac{i}{\left( 2\pi \right) ^{4}}\frac{1}{%
k^{2}+i\varepsilon }e^{ik\left( x_{1}-x_{2}\right) }\text{ \hspace{0in} 
\hspace{0in} \hspace{0in} \hspace{0in} \hspace{0in} \hspace{0in} \hspace{0in}
\hspace{0in} }with\text{ \hspace{0in} \hspace{0in} \hspace{0in} }\int d^{4}k
\end{equation*}

Fig.2.2.2D 
\begin{equation*}
\text{ }\frac{1}{\sqrt{V}}v_{\mathbf{p}s}\text{ }(ingoing)\text{\hspace{0in} 
\hspace{0in} \hspace{0in} \hspace{0in} \hspace{0in} \hspace{0in} \hspace{0in}%
}or\text{ \hspace{0in} \hspace{0in} \hspace{0in} \hspace{0in} \hspace{0in} 
\hspace{0in} \hspace{0in} \hspace{0in} }\frac{1}{\sqrt{V}}\overline{u}_{%
\mathbf{p}s}\text{ }(outgoing)
\end{equation*}

Fig.2.2.2E 
\begin{equation*}
\text{ }\frac{1}{\sqrt{V}}\overline{v}_{\mathbf{p}s}(outgoing)\text{ \hspace{%
0in} \hspace{0in} \hspace{0in} \hspace{0in} \hspace{0in} \hspace{0in} 
\hspace{0in} \hspace{0in} }or\text{ \hspace{0in} \hspace{0in} \hspace{0in} 
\hspace{0in} \hspace{0in} \hspace{0in} \hspace{0in}}\frac{1}{\sqrt{V}}u_{%
\mathbf{p}s}\text{ }(ingoing)
\end{equation*}

Fig.2.2.2F 
\begin{equation*}
\text{ }\frac{1}{\sqrt{V\omega _{\mathbf{k}}}}e_{\mathbf{k}\mu }^{\lambda },%
\text{ \hspace{0in} \hspace{0in} \hspace{0in} \hspace{0in} \hspace{0in}}%
\lambda =1,2.
\end{equation*}

Fig.3.1.1-2. A WFDG1L representing the two sorts of corrections to the
scattering amplitude $A_{m}(\mid a_{\mathbf{q}s}\rangle \rightarrow \mid a_{%
\mathbf{q}s}\rangle ).$ Fig. 3.1.1 represents the contribution of $\hspace{%
0in}$ $m_{eff}^{\left( 1\right) }\left( q\right) ;$ Fig. 3.1.2 represents
the contribution of $m_{efw}^{\left( 1\right) }\left( q\right) ,$ and 
\hspace{0in}the dotted- line circle in Fig. 3.1.2 represents $%
m_{efw}^{\left( 1\right) }\left( q\right) .$

Fig.3.2.1-4. A WFDG1L representing the one-loop corrections to $S_{Ff}\left(
p\right) $ with its two vertices. Fig. 3.2.1L represents the contribution of 
$\Lambda _{ff1\mu \text{ }}^{\left( 1\right) L}$ (corresponding to $%
g_{f,ff1}^{\left( 1\right) }\left( p\right) $). Fig. 3.2.1R represents the
contribution of $\Lambda _{ff1\mu \text{ }}^{\left( 1\right) R}$
(corresponding to $g_{f,ff1}^{\left( 1\right) }\left( p\right) $).

Fig. 3.2.2 represents the contribution of $g_{f,fw1}^{\left( 1\right)
}\left( p\right) $ (equivalent to $m_{efw}^{\left( 1\right) }\left( q\right) 
$), and the dotted- line circle in Fig. 3.2.2 represents $m_{efw}^{\left(
1\right) }\left( q\right) .$

Fig. 3.2.3 represents the contribution of $g_{f,ww1}^{\left( 1\right)
}\left( q_{2},q_{1}\right) ,$ \hspace{0in}the dotted- line circle in Fig.
3.2.3 represents $g_{f,ww1}^{\left( 1\right) }\left( q_{2},q_{1}\right) .$

Fig. 3.2.4 represents the contribution of $g_{f,wf1}^{\left( 1\right)
}\left( q_{2},q_{1}\right) ,$ \hspace{0in}the dotted- line circle in Fig.
3.2.4 represents $g_{f,wf1}^{\left( 1\right) }\left( q_{2},q_{1}\right) .$

\hspace{0in}Fig. 3.3.1-2. A WFDG1L representing the one-loop corrections to $%
D_{Ff}\left( k\right) $ with its two vertices. Fig. 3.3.1 represents the
contribution of $\hspace{0in}g_{ff2}^{\left( 1\right) }\left( k\right) .$

Fig. 3.3.2 represents the contribution of $g_{fw2}^{\left( 1\right) }\left(
q_{2},q_{1}\right) $, and the dotted- line circle in Fig. 3.3.2 represents $%
g_{fw2}^{\left( 1\right) }\left( q_{2},q_{1}\right) .$

Fig. 3.4.1-4. \hspace{0in}The total one-loop corrections to the \ scattering
\ amplitude $\ \ \ \ A_{g}\left( \mid a_{\mathbf{p}_{1}s_{1}}c_{\mathbf{k}%
\lambda }\rangle \rightarrow \mid a_{\mathbf{p}_{2}s_{2}}\rangle \right) .$

Fig. 3.4.1 is a WFDG1L representing the contributions of the first sort.
Fig. 3.4.1A-D represent the contribution of $g_{f,ff1}^{\left( 1\right)
}\left( p_{2}\right) ,$ $g_{f,fw1}^{\left( 1\right) }\left( p_{2}\right) ,$ $%
g_{f,ww1}^{\left( 1\right) }\left( q_{2},q_{1}\right) $ and $%
g_{f,wf1}^{\left( 1\right) }\left( q_{2},q_{1}\right) $ in turn$,$ the
dotted- line circle in Fig. 3.4.1B, C, D represents $g_{f,fw1}^{\left(
1\right) }\left( q\right) ,$ $g_{f,ww1}^{\left( 1\right) }\left(
q_{2},q_{1}\right) $ and $g_{f,wf1}^{\left( 1\right) }\left(
q_{2},q_{1}\right) $ in turn.

Fig. 3.4.2 is a WFDG1L representing the contributions of the first sort.
Fig. 3.4.2A-D represent $g_{f,ff1}^{\left( 1\right) }\left( p_{1}\right) ,$ $%
g_{f,fw1}^{\left( 1\right) }\left( p_{1}\right) ,$ $g_{f,ww1}^{\left(
1\right) }\left( q_{2},q_{1}\right) $ and $g_{f,wf1}^{\left( 1\right)
}\left( q_{2},q_{1}\right) $ in turn$,$ the dotted- line circle in Fig.
3.4.2B, C, D represents $g_{f,fw1}^{\left( 1\right) }\left( q\right) ,$ $%
g_{f,ww1}^{\left( 1\right) }\left( q_{2},q_{1}\right) $ and $%
g_{f,wf1}^{\left( 1\right) }\left( q_{2},q_{1}\right) $ in turn.

Fig. 3.4.3 is a WFDG1L representing the contributions of the second sort.
Fig. 3.4.3A-B represent the contributions of $g_{ff2}^{\left( 1\right)
}\left( k\right) $ and $g_{fw2}^{\left( 1\right) }\left( q_{2},q_{1}\right) $
in turn, the dotted- line circle in Fig. 3.4.3B represents $g_{fw2}^{\left(
1\right) }\left( q_{2},q_{1}\right) $.

Fig. 3.4.4 is a WFDG1L representing the contributions of the third sort.
Fig. 3.4.4A-B represent the contributions of $g_{ff3}^{\left( 1\right)
}\left( p_{2},p_{1}\right) $ and $g_{fw3}^{\left( 1\right) }\left(
q_{2},q_{1}\right) $ in turn, the dotted- line circle in Fig. 3.4.4B
represents $g_{fw3}^{\left( 1\right) }\left( q_{2},q_{1}\right) $.

\begin{description}
\item[Appendix A] 
\end{description}

All physics quanties of the vacuum state are zero, hence we have 
\begin{eqnarray}
&\mid &\alpha _{\mathbf{p}s}\rangle =\mid 0\rangle \mid \alpha _{\mathbf{p}%
s}\rangle \ =\mid \alpha _{\mathbf{p}s}\rangle \mid 0\rangle ,\; 
\TCItag{A.1a} \\
\ \langle \alpha _{\mathbf{p}s} &\mid &=\ \langle \alpha _{\mathbf{p}s}\mid
\langle 0\mid =\langle 0\mid \ \langle \alpha _{\mathbf{p}s}\mid , 
\TCItag{A.1b} \\
&\mid &0\rangle =\mid 0\rangle \mid 0\rangle .  \TCItag{A.1c}
\end{eqnarray}%
Thus $\mid n_{\mathbf{k}\lambda }\rangle $ and $\mid \underline{n}_{\mathbf{k%
}\lambda }\rangle $ can also be represented by 
\begin{eqnarray}
&\mid &n_{\mathbf{k}\lambda }\rangle =\frac{1}{\sqrt{n!}}\underset{n}{%
\underbrace{\mid c_{\mathbf{k}\lambda }\eqslantgtr \mid 0\rangle \cdots \mid
c_{\mathbf{k}\lambda }\eqslantgtr \mid 0\rangle }},  \TCItag{A.2a} \\
&\mid &\underline{n}_{\mathbf{k}\lambda }\rangle =\frac{1}{\sqrt{n!}}%
\underset{n}{\underbrace{\mid \underline{c}_{\mathbf{k}\lambda }\eqslantgtr
\mid 0\rangle \cdots \mid \underline{c}_{\mathbf{k}\lambda }\eqslantgtr \mid
0\rangle }}.  \TCItag{A.2b}
\end{eqnarray}%
The inner product of two states must be defined as a number. According to
(A.2), only when the number of single-particle bras of a state is the same
as the number of single-particle kets of other state, is it possible that
the inner product of the two states is not equal to zero. It can be seen
from (A.1) that we always have the two numbers to be the same by suitably
supplement the number of $\mid 0\rangle $ or $\langle 0\mid $. Hence in
order to carry out the inner of two states, we should first have the two
numbers are equal to each other. We define the inner product of two sates to
be such a product obtained by the way. For example, 
\begin{eqnarray}
\langle \alpha _{\mathbf{p}s} &\mid &\langle \alpha _{\mathbf{p}s}\mid \cdot
\mid \alpha _{\mathbf{p}s}\rangle \equiv \langle \alpha _{\mathbf{p}s}\mid
\langle \alpha _{\mathbf{p}s}\mid \cdot \mid \alpha _{\mathbf{p}s}\rangle
\mid 0\rangle  \notag \\
&=&\langle \alpha _{\mathbf{p}s}\mid \cdot \mid 0\rangle =0.  \TCItag{A.3}
\end{eqnarray}%
It is obvious according to the definition that for an orthonormal set, the
inner product of two different states must be zero. This result is the same
as those obtained by (1.3.19).\ 

\begin{description}
\item[Appendix B] 
\end{description}

A possible definitions $\underline{I}_{\mathbf{p}}$ etc. are as follows.

\begin{eqnarray}
\underline{I}_{\mathbf{p}} &\equiv &\frac{1}{\sqrt{2}}(\mid \underline{a}_{%
\mathbf{p}}\rangle \xi +\eta ^{+}\langle \underline{b}_{-\mathbf{p}}\mid ), 
\notag \\
\underline{I}_{\mathbf{p}}^{+} &\equiv &\frac{1}{\sqrt{2}}(\xi ^{+}\langle 
\underline{a}_{\mathbf{p}}\mid +\mid \underline{b}_{-\mathbf{p}}\rangle \eta
),  \TCItag{B.1}
\end{eqnarray}%
\begin{equation}
\underline{J}_{\mathbf{k}}\equiv \frac{1}{\sqrt{2}}(\mid \underline{c}_{%
\mathbf{k}}\rangle e^{i\varphi }+e^{-i\varphi }\langle \underline{c}_{-%
\mathbf{k}}\mid )  \tag{B.2}
\end{equation}%
\begin{equation}
\underline{J}_{\mathbf{k}}^{+}\equiv \frac{1}{\sqrt{2}}(\langle \underline{c}%
_{\mathbf{k}}\mid e^{-i\varphi }+e^{i\varphi }\mid \underline{c}_{-\mathbf{k}%
}\rangle )  \tag{B.3}
\end{equation}%
\ \ \ \ \ \ \ \ \ 
\begin{eqnarray}
I_{\mathbf{p}} &\equiv &\frac{1}{\sqrt{2}}(\mid b_{\mathbf{p}}\rangle \eta
+\xi ^{+}\langle a_{-\mathbf{p}}\mid ),  \notag \\
I_{\mathbf{p}}^{+} &\equiv &\frac{1}{\sqrt{2}}(\eta ^{+}\langle b_{\mathbf{p}%
}\mid +\mid a_{-\mathbf{p}}\rangle \xi ),  \TCItag{B.4}
\end{eqnarray}%
\begin{equation}
J_{\mathbf{k}}\equiv \frac{1}{\sqrt{2}}(\mid c_{\mathbf{k}}\rangle
e^{i\varphi }+e^{-i\varphi }\langle c_{-\mathbf{k}}\mid ),  \tag{B.5}
\end{equation}%
\begin{equation}
J_{\mathbf{k}}^{+}\equiv \frac{1}{\sqrt{2}}(\langle c_{\mathbf{k}}\mid
e^{-i\varphi }+e^{i\varphi }\mid c_{-\mathbf{k}}\rangle ),  \tag{B.6}
\end{equation}%
where all $\xi $, $\xi ^{+}$, $\eta $ and $\eta ^{+}$ are Grassman numbers,
and $2\pi \geq \varphi \geq 0.$ We define the inner products of the base
vectors as follows%
\begin{equation}
Z_{\mathbf{p}}\cdot Z_{\mathbf{p}^{\prime }}^{\prime }\equiv Tr\int (d\xi
d\xi ^{+}+d\eta d\eta ^{+})Z_{\mathbf{p}}Z_{\mathbf{p}^{\prime }}^{\prime
}=Z_{\mathbf{p}^{\prime }}^{\prime }\cdot Z_{\mathbf{p}},  \tag{B.7}
\end{equation}%
\begin{equation}
Y_{\mathbf{k}}\cdot Y_{\mathbf{k}^{\prime }}^{^{\prime }}\equiv \frac{1}{%
2\pi }Tr\int_{0}^{2\pi }d\varphi Y_{\mathbf{k}}Y_{\mathbf{k}^{\prime
}}^{^{\prime }}=Y_{\mathbf{k}^{\prime }}^{^{\prime }}\cdot Y_{\mathbf{k}}, 
\tag{B.8}
\end{equation}%
where $Z_{\mathbf{p}}$, $Z_{\mathbf{p}^{\prime }}^{\prime }=I_{\mathbf{p}}$
, $I_{\mathbf{p}}^{+},$ $\underline{I}_{\mathbf{p}}$ and $\underline{I}_{%
\mathbf{p}}^{+},$ $Y_{\mathbf{k}},$ $Y_{\mathbf{k}^{\prime }}^{^{\prime
}}=J_{\mathbf{k}},$ $J_{\mathbf{k}}^{+},$ $\underline{J}_{\mathbf{k}}$ and $%
\underline{J}_{\mathbf{k}}^{+}$. We define 
\begin{equation}
Tr\langle \alpha _{\mathbf{p}}\mid \cdot \mid \alpha _{\mathbf{p}^{\prime
}}^{\prime }\rangle =-Tr\mid \alpha _{\mathbf{p}^{\prime }}^{\prime }\rangle
\langle \alpha _{\mathbf{p}}\mid =\delta _{\alpha \alpha ^{\prime }}\delta _{%
\mathbf{pp}^{\prime }},  \tag{B.9}
\end{equation}%
\begin{equation}
Tr\langle \gamma _{\mathbf{k}}\mid \cdot \mid \gamma _{\mathbf{k}^{\prime
}}^{\prime }\rangle =Tr\mid \gamma _{\mathbf{k}^{\prime }}^{\prime }\rangle
\langle \gamma _{\mathbf{k}}\mid =\delta _{\gamma \gamma ^{\prime }}\delta _{%
\mathbf{kk}^{\prime }}.  \tag{B.10}
\end{equation}%
From (B.1)-(B.10) we can obtain (1.3.26)-(1.3.28), e.g.,%
\begin{eqnarray}
I_{\mathbf{p}}^{+}\cdot I_{\mathbf{p}^{\prime }} &=&Tr\int (d\xi d\xi
^{+}+d\eta d\eta ^{+})  \notag \\
\frac{1}{2}\{\xi ^{+}\xi \langle \underline{a}_{\mathbf{p}} &\mid &\cdot
\mid \underline{a}_{\mathbf{p}^{\prime }}\rangle +\xi ^{+}\eta ^{+}\langle 
\underline{a}_{\mathbf{p}}\mid \langle \underline{b}_{-\mathbf{p}^{\prime
}}\mid  \notag \\
+\eta \xi &\mid &\underline{b}_{-\mathbf{p}}\rangle \mid \underline{a}_{%
\mathbf{p}^{\prime }}\rangle +\eta \eta ^{+}\mid \underline{b}_{-\mathbf{p}%
}\rangle \langle \underline{b}_{-\mathbf{p}^{\prime }}\mid \}  \notag \\
&=&\frac{1}{2}Tr\{\langle \underline{a}_{\mathbf{p}}\mid \underline{a}_{%
\mathbf{p}^{\prime }}\rangle -\mid \underline{b}_{-\mathbf{p}}\rangle
\langle \underline{b}_{-\mathbf{p}^{\prime }}\mid \}  \notag \\
&=&\delta _{\mathbf{pp}^{\prime }}=I_{\mathbf{p}^{\prime }}\cdot I_{\mathbf{p%
}}^{+},  \TCItag{B.11}
\end{eqnarray}%
\begin{eqnarray}
\underline{J}_{\mathbf{k}}\cdot \underline{J}_{\mathbf{k}^{\prime
}}^{^{\prime }} &\equiv &\frac{1}{2\pi }Tr\int_{0}^{2\pi }d\varphi \frac{1}{2%
}\{\langle c_{\mathbf{k}}\mid c_{\mathbf{k}^{\prime }}\rangle +e^{i2\varphi
}\mid c_{-\mathbf{k}}\rangle \mid c_{\mathbf{k}^{\prime }}\rangle  \notag \\
+\langle c_{\mathbf{k}} &\mid &\langle c_{-\mathbf{k}^{\prime }}\mid
e^{-i2\varphi }+\mid c_{-\mathbf{k}}\rangle \langle c_{-\mathbf{k}^{\prime
}}\mid \}  \notag \\
&=&\frac{1}{2}Tr\{\langle c_{\mathbf{k}}\mid c_{\mathbf{k}^{\prime }}\rangle
+\mid c_{-\mathbf{k}}\rangle \langle c_{-\mathbf{k}^{\prime }}\mid \}=\delta
_{\mathbf{kk}^{\prime }}.  \TCItag{B.12}
\end{eqnarray}%
$Z_{\mathbf{p}}$ and $Y_{\mathbf{k}}$ are always regarded as a whole, hence
when an operator or a state is multiplied by $Z_{\mathbf{p}}$ or $Y_{\mathbf{%
k}},$ we must first accomplish third inner, e.g.,

\begin{equation}
Z_{\mathbf{p}}\cdot \eqslantless \sigma \mid =Tr\int (d\xi d\xi ^{+}+d\eta
d\eta ^{+})Z_{\mathbf{p}}\eqslantless \sigma \mid =0.  \tag{B.13}
\end{equation}%
Analogously to (B.13), we have%
\begin{eqnarray}
&\eqslantless &\sigma \mid \cdot Z_{\mathbf{p}}=\langle \sigma \mid \cdot Z_{%
\mathbf{p}}=Z_{\mathbf{p}}\cdot \langle \sigma \mid   \notag \\
&=&Y_{\mathbf{k}}\cdot \eqslantless \sigma \mid =\eqslantless \sigma \mid
\cdot Y_{\mathbf{k}}=\langle \sigma \mid \cdot Y_{\mathbf{k}}=Y_{\mathbf{k}%
}\cdot \langle \sigma \mid =0.  \TCItag{B.14}
\end{eqnarray}%
Thus we have%
\begin{equation}
\lbrack \eqslantless \sigma \mid ,Z_{\mathbf{p}}]=0,\;\ \ \ \ \ \ \ \ \ \ [\
Y_{\mathbf{k}},\eqslantless \sigma \mid ]=0\ .\ \ \ \   \tag{B.15}
\end{equation}%
According to the definitions(B.1)-(B.10), $\underline{J}_{\mathbf{k}}$ and $%
\underline{J}_{\mathbf{k}}^{+}$ etc. form respectively vector spaces. We
call such a space which takes $\underline{J}_{\mathbf{k}}$ as a base vector
the $\underline{J}_{\mathbf{k}}-space,$ etc.. 

\begin{description}
\item[Appendix C] 
\end{description}

The equal-time anticommutation or commutation relations of such operators as 
$\eqslantless a_{\mathbf{p}s}(t)\mid $ are the same as (1.316)-(1.3.17).
From (1.3.16)-(1.3.17) and (1.3.28) or (B.15) we have 
\begin{eqnarray}
\lbrack \underline{I}_{\mathbf{p}} &\eqslantless &a_{\mathbf{p}s}(t)\mid
,\mid a_{\mathbf{p}_{2}s_{2}}\left( t\right) \eqslantgtr \eqslantless a_{%
\mathbf{p}_{1}s_{1}}\left( t\right) \mid ]  \notag \\
&=&\underline{I}_{\mathbf{p}}\eqslantless a_{\mathbf{p}_{1}s_{1}}\left(
t\right) \mid \delta _{\mathbf{pp}_{2}}\delta _{ss_{2}},  \TCItag{C.1}
\end{eqnarray}%
\begin{eqnarray}
\lbrack &\mid &b_{\mathbf{p}s}(t)\eqslantgtr \underline{I}_{(\mathbf{-p)}%
},\mid b_{(-\mathbf{p}_{1})s_{1}}\left( t\right) \eqslantgtr \eqslantless
b_{(-\mathbf{p}_{2})s_{2}}\left( t\right) \mid ]  \notag \\
&=&-\delta _{\mathbf{p(-p}_{2})}\delta _{ss_{2}}\mid b_{(-\mathbf{p}%
_{1})s_{1}}\left( t\right) \eqslantgtr \underline{I}_{(\mathbf{-p)}}, 
\TCItag{C.2}
\end{eqnarray}%
\begin{eqnarray}
\lbrack &\mid &b_{\mathbf{p}s}(t)\eqslantgtr \underline{I}_{(\mathbf{-p)}%
},\eqslantless b_{(-\mathbf{p}_{2})s_{2}}\left( t\right) \mid \eqslantless
a_{\mathbf{p}_{1}s_{1}}\left( t\right) \mid ]  \notag \\
&=&\underline{I}_{(\mathbf{-p)}}\eqslantless a_{\mathbf{p}_{1}s_{1}}\left(
t\right) \mid \delta _{\mathbf{p(-p}_{2})}\delta _{ss_{2}},  \TCItag{C.3}
\end{eqnarray}%
\begin{eqnarray}
\lbrack \underline{I}_{\mathbf{p}} &\eqslantless &a_{\mathbf{p}s}(t)\mid
,\mid a_{\mathbf{p}_{2}s_{2}}\left( t\right) \eqslantgtr \mid b_{(-\mathbf{p}%
_{1})s_{1}}\left( t\right) \eqslantgtr ]  \notag \\
&=&\delta _{\mathbf{pp}_{2}}\delta _{ss_{2}}\mid b_{(-\mathbf{p}%
_{1})s_{1}}\left( t\right) \eqslantgtr \underline{I}_{\mathbf{p}}, 
\TCItag{C.4}
\end{eqnarray}%
\begin{eqnarray}
\lbrack \underline{I}_{\mathbf{p}} &\eqslantless &a_{\mathbf{p}s}(t)\mid
,\mid b_{(-\mathbf{p}_{1})s_{1}}\left( t\right) \eqslantgtr \eqslantless
b_{(-\mathbf{p}_{2})s_{2}}\left( t\right) \mid ]  \notag \\
&=&[\mid b_{\mathbf{p}s}(t)\eqslantgtr \underline{I}_{(\mathbf{-p)}},\mid a_{%
\mathbf{p}_{2}s_{2}}\left( t\right) \eqslantgtr \eqslantless a_{\mathbf{p}%
_{1}s_{1}}\left( t\right) \mid ]  \notag \\
&=&[\underline{I}_{\mathbf{p}}\eqslantless a_{\mathbf{p}s}(t)\mid
,\eqslantless b_{(-\mathbf{p}_{2})s_{2}}\left( t\right) \mid \eqslantless a_{%
\mathbf{p}_{1}s_{1}}\left( t\right) \mid ]  \notag \\
&=&[\mid b_{\mathbf{p}s}(t)\eqslantgtr \underline{I}_{(\mathbf{-p)}},\mid a_{%
\mathbf{p}_{2}s_{2}}\left( t\right) \eqslantgtr \mid b_{(-\mathbf{p}%
_{1})s_{1}}\left( t\right) \eqslantgtr ]=0.  \TCItag{C.5}
\end{eqnarray}%
\begin{equation}
\sum_{s}(u_{\mathbf{p}s}u_{\mathbf{p}s}^{+}+v_{(-\mathbf{p)}s}v_{(-\mathbf{p)%
}s}^{+})=1  \tag{C.6}
\end{equation}%
Substituting (2.4.1)-(2.4.3), (C.1)-(C.6) and 
\begin{equation}
\psi \left( x\right) =\frac{1}{\sqrt{V}}\sum_{\mathbf{p}s}\left( \underline{I%
}_{\mathbf{p}}\eqslantless a_{\mathbf{p}s}(t)\mid u_{\mathbf{p}s}e^{i\mathbf{%
px}}+\mid b_{\mathbf{p}s}(t)\eqslantgtr \underline{I}_{(\mathbf{-p)}}v_{%
\mathbf{p}s}e^{-i\mathbf{px}}\right) ,  \tag{C.7}
\end{equation}%
into (2.2.18), i.e., 
\begin{equation}
\dot{\psi}\left( x\right) =-i[\psi \left( x\right) ,H]=-i[\psi \left(
x\right) ,H_{F}],  \tag{C.8}
\end{equation}%
we obtain (2.2.14).

From (2.2.20) we have%
\begin{equation}
\lbrack \underline{j}_{\mathbf{k}}\eqslantless c_{\mathbf{k}\lambda }(t)\mid
,\mid \overline{c}_{(-\mathbf{k}^{\prime }\lambda ^{\prime })}\left(
t\right) \eqslantgtr ]=\underline{j}_{\mathbf{k}}\delta _{\mathbf{k(-k}%
^{\prime })}\delta _{\lambda \lambda ^{\prime }},  \tag{C.9}
\end{equation}%
\begin{equation}
\lbrack \mid \overline{c}_{\mathbf{k}\lambda }(t)\eqslantgtr \underline{j}%
_{(-\mathbf{k}^{\prime }\mathbf{)}},\eqslantless c_{\mathbf{k}^{\prime
}\lambda ^{\prime }}(t)\mid ]=-\underline{j}_{(-\mathbf{k}^{\prime }\mathbf{)%
}}\delta _{\mathbf{k(-k}^{\prime })}\delta _{\lambda \lambda ^{\prime }}, 
\tag{C.10}
\end{equation}%
\begin{eqnarray}
\lbrack \underline{j}_{\mathbf{k}} &\eqslantless &c_{\mathbf{k}\lambda
}(t)\mid ,\eqslantless c_{\mathbf{k}^{\prime }\lambda ^{\prime }}(t)\mid ] 
\notag \\
&=&[\mid \overline{c}_{\mathbf{k}\lambda }(t)\eqslantgtr \underline{j}_{(-%
\mathbf{k}^{\prime }\mathbf{)}},\mid \overline{c}_{(-\mathbf{k}^{\prime
}\lambda ^{\prime })}\left( t\right) \eqslantgtr ]=0.  \TCItag{C.11}
\end{eqnarray}%
\begin{eqnarray}
\lbrack \underline{j}_{\mathbf{k}} &\eqslantless &c_{\mathbf{k}\lambda
}(t)\mid ,\mid \overline{c}_{\mathbf{k}^{\prime }\lambda ^{\prime }}\left(
t\right) \eqslantgtr \eqslantless c_{\mathbf{k}^{\prime }\lambda ^{\prime
}}\left( t\right) \mid ]  \notag \\
&=&\underline{j}_{\mathbf{k}}\eqslantless c_{\mathbf{k}^{\prime }\lambda
^{\prime }}\left( t\right) \mid \underline{j}_{\mathbf{k}}\delta _{\mathbf{kk%
}^{\prime }}\delta _{\lambda \lambda ^{\prime }}  \TCItag{C.12}
\end{eqnarray}%
\begin{eqnarray}
\lbrack &\mid &\overline{c}_{\mathbf{k}\lambda }(t)\eqslantgtr \underline{j}%
_{(-\mathbf{k}^{\prime }\mathbf{)}},\mid \overline{c}_{\mathbf{k}^{\prime
}\lambda ^{\prime }}\left( t\right) \eqslantgtr \eqslantless c_{\mathbf{k}%
^{\prime }\lambda ^{\prime }}\left( t\right) \mid ]  \notag \\
&=&-\underline{j}_{(-\mathbf{k}^{\prime }\mathbf{)}}\mid \overline{c}_{%
\mathbf{k}^{\prime }\lambda ^{\prime }}\left( t\right) \eqslantgtr \delta _{%
\mathbf{kk}^{\prime }}\delta _{\lambda \lambda ^{\prime }}  \TCItag{C.13}
\end{eqnarray}%
\begin{equation}
\sum_{\lambda =1}^{4}e_{\mathbf{k}\mu }^{\lambda }e_{\mathbf{k}\nu
}^{\lambda }=\delta _{\mu \nu }  \tag{C.14}
\end{equation}%
Substituting (2.4.1), (2.4.3), (C.9)-(C.13) and 
\begin{eqnarray}
A_{\mu }\left( x\right) &=&\frac{1}{\sqrt{V}}\sum_{\mathbf{k}}\frac{1}{\sqrt{%
2\omega _{\mathbf{k}}}}\sum_{\lambda =1}^{4}(e_{\mathbf{k}\mu }^{\lambda }%
\underline{j}_{\mathbf{k}}\eqslantless c_{\mathbf{k}\lambda }(t)\mid e^{i%
\mathbf{kx}}  \notag \\
+ &\mid &\overline{c}_{\mathbf{k}\lambda }(t)\eqslantgtr \underline{j}_{%
\mathbf{(-k)}}e^{-i\mathbf{kx}}),  \TCItag{C.15}
\end{eqnarray}%
\begin{eqnarray}
\pi _{\mu } &=&\dot{A}_{\mu }\left( x\right) \equiv \frac{-i}{\sqrt{V}}\sum_{%
\mathbf{k}}\sqrt{\frac{\omega _{\mathbf{k}}}{2}}\sum_{\lambda =4}^{4}e_{%
\mathbf{k}\mu }^{\lambda }(\underline{j}_{\mathbf{k}}\eqslantless c_{\mathbf{%
k}\lambda }(t)\mid e^{i\mathbf{kx}}  \notag \\
- &\mid &\overline{c}_{\mathbf{k}\lambda }(t)\eqslantgtr \underline{j}_{(-%
\mathbf{k)}}e^{-i\mathbf{kx}}),  \TCItag{C.16}
\end{eqnarray}%
into (2.2.18), i.e., 
\begin{equation}
\dot{A}_{\mu }\left( x\right) =-i[A_{\mu }\left( x\right) ,H_{F}],\;\dot{\pi}%
_{\mu }\ =\ddot{A}_{\mu }\ =-i[\pi _{\mu }\left( x\right) ,H_{F}],\  
\tag{C.17}
\end{equation}%
we obtain (2.2.15). It is seen that (2.2.18) is consistent with
(2.2.14)-(2.2.15). Analogously, from (2.4.4)-(2.4.6) we can prove (2.2.19)
to be consistent with (2.2.16)-(2.2.17).

\begin{description}
\item[Acknowledgement] 
\end{description}

I am very grateful to professor Zhan-yao Zhao for best support and professor
Zhao-yan Wu for helpful discussions.

[2] Shi-Hao Chen, Quantum Field Theory Without Divergence, hep-th/0203220.

[3] Shi-Hao Chen, A Possible Candidate for Dark Matter, hep-th/0103234.

[4] David Lurie, Particles and Fields,1968. New York .London. Sydney.

[5] Yu Yun qiang, An Introduction to General Reletivity (Second Edition),
1997, Peking University, Beijing.

[6] T.Kinoshita and Y. Nambu, Phys. Rev. 94, 598 (1954); A.L. Fetter and
J.D.Walecka, Quantum Theory of Many-Particle Systems (McGraw-Hill, New York,
1971).

[7] Shau-Jin Chang and Jon A. Wright, Phys. Rev. D, 12, 1595 (1975).

[8] Liao-Fu Luo, Quantum Field Theory, Jiangsu science and technology
publishing house, 1990. Hong-yuan Zhu, Quantum Field Theory, Science
publishing house, 1960.


\begin{thebibliography}{9}
\bibitem{References} G.Scharf, Finite Quantum Electrodynamics, Second
Edition, Springer- Verlag, 1995.
\end{thebibliography}
\end{document}